\documentclass[12pt]{article}
\pdfoutput=1

\usepackage{jheppub}

\usepackage{amsfonts}
\usepackage{amsmath,amssymb,amsthm}
\usepackage{graphicx}
\usepackage{color}
\usepackage{setspace}
\usepackage{array}
\usepackage{bbm} 

\newcommand{\p}{\partial}

\newcommand{\N}{\mathcal{N}}
\newcommand{\ti}{\widetilde}

\newcommand{\be}{\begin{equation}}
\newcommand{\ee}{\end{equation}}
\newcommand{\bea}{\begin{eqnarray}}
\newcommand{\eea}{\end{eqnarray}}
\newcommand{\ba}{\begin{aligned}}
\newcommand{\ea}{\end{aligned}}

\newcommand{\lp}{\left(}
\newcommand{\rp}{\right)}
\newcommand{\tr}{\textrm{Tr} \,}

\newcommand{\vev}[1]{\left\langle #1 \right\rangle}

\newcommand{\rap}[2]
{\setbox1=\hbox{#1}%
\setbox2=\hbox to\wd1{\hss #2\hss}%
\mbox{\rlap{\box1}\box2}}

\newcommand{\bC}{\ensuremath{\mathbb{C}}}

\newcommand{\bP}{\ensuremath{\mathbb{P}}}

\newcommand{\bR}{\ensuremath{\mathbb{R}}}

\newcommand{\bZ}{\ensuremath{\mathbb{Z}}}

\newcommand{\scC}{\ensuremath{\mathcal{C}}}

\newcommand{\scH}{\ensuremath{\mathcal{H}}}

\newcommand{\scM}{\ensuremath{\mathcal{M}}}

\newcommand{\scO}{\ensuremath{\mathcal{O}}}
\newcommand{\scP}{\ensuremath{\mathcal{P}}}

\newcommand{\scR}{\ensuremath{\mathcal{R}}}

\numberwithin{equation}{section}       

\title{Ring Relations and Mirror Map from Branes}

\author{Benjamin Assel}
\affiliation{CERN, Theoretical Physics Department, 
 CH-1211 Geneva 23, Switzerland}

\abstract{We study the space of vacua of three-dimensional $\N=4$ theories from a novel approach building on the type IIB brane realization of the theory and
in which the insertion of local chiral operators in the path integral is obtained from integrating out light modes in appropriate brane setups.
 Most of our analysis focuses on abelian quiver theories which can be realized as the low-energy theory of D3-D5-NS5 brane arrays. Their space of vacua contains a Higgs branch, parametrized by the vevs of half-BPS meson operators, and a Coulomb branch, parametrized by the vevs of half-BPS monopole operators.  We show that the Higgs operators are inserted by adding F1 strings and D3 branes, while Coulomb operators are inserted by adding D1 strings and D3 branes, with specific orientations, to the initial brane setup of the theory. This approach has two main advantages. First the ring relations describing the Higgs and Coulomb branches can be derived by looking at specific brane setups with multiple interpretations in terms of operator insertions. This provides a new derivation of the Coulomb branch quantum relations. Secondly the map between the Higgs and Coulomb operators of mirror dual theories can be derived in a trivial way from IIB S-duality.}

\emailAdd{benjamin.assel@gmail.com}

\begin{document}

\vspace*{-2cm} 
\begin{flushright}
{\tt  CERN-TH-2017-018 } 
\end{flushright}

\maketitle

\section{Introduction and Discussion}
\label{sec:Introduction}

Three-dimensional $\N=4$ Yang-Mills gauge theories (i.e. with eight Poincar\'e supercharges) are fully characterized by a choice of gauge group $G$, associated to a vector multiplet, and a pseudo-real representation $\mathcal{R}$ in which the hyper-multiplet matter fields transform. This data fixes uniquely the Lagrangian of the theory. Their space of vacua splits into several subspaces or ``branches", each of which is a product of hyper-K\"ahler manifolds, with two branches playing a special role. The Higgs branch, which is free of quantum corrections \cite{Intriligator:1996ex}, is parametrized by the vacuum expectation values (vevs) of the scalars in the hyper-multiplet, subject to a triplet of D-term constraints, and modulo gauge transformations.
In a chosen complex structure the Higgs branch can be described as a complex algebraic variety, with singularities, parametrized by the vevs of gauge invariant operators, which are chiral with respect to a certain $\N=2$ subalgebra, subject to algebraic relations (inherited from the chiral ring relations).
The Coulomb branch is parametrized by the vevs of half-BPS monopole operators which are chiral with respect to another $\N=2$ subalgebra, and which form a ring with relations arising from non-trivial quantum effects \cite{Seiberg:1996nz,Borokhov:2002cg}. The monopole operators are defined in the quantum theory by imposing in the path integral formulation a Dirac monopole singularity for the gauge field at a point in Euclidean space and ``dressing" it with a polynomial in the vector multiplet complex scalars. A special case are monopoles with zero magnetic charges which are simply gauge invariant combinations of the vector multiplet complex scalars.\footnote{In addition there can be mixed branches which will not be studied in this paper, to keep the presentation simple.}

While there is a relatively clear path to study the Higgs branch from the classical Lagrangian of the theory, the Coulomb branch is more difficult to access, since the ring relations between monopole operators do not follow from a superpotential, but from the quantum dynamics of the theory. In abelian theories the Coulomb branch metric receives corrections only at one-loop and can be computed directly \cite{deBoer:1996mp, deBoer:1996ck}. It is also possible to rely on mirror symmetry \cite{Intriligator:1996ex}, which exchanges the Higgs and Coulomb branches of mirror dual theories. 
Much progress has been made recently on deriving the ring relations of the Coulomb branch of quiver gauge theories from different approaches. 
One approach uses the Coulomb branch Hilbert series \cite{Hanany:2011db,Cremonesi:2013lqa,Cremonesi:2014uva,Hanany:2016ezz,Cheng:2017got} \footnote{See \cite{Cremonesi:2017jrk} for a recent review on 3d (and 4d) Hilbert series.}, which is a protected index counting chiral monopole operators refined with fugacities keeping track of their charges under Cartan R-symmetry and global topological symmetries. From the resumed series one is able to extract a set of generators and to read the Coulomb branch relations, up to coefficients which must be determined by other methods. Non-abelian quiver theories of ADE types with unitary or ortho-symplectic gauge nodes have been studied using this method. 
A different construction was proposed by Bullimore, Dimofte and Gaiotto in \cite{Bullimore:2015lsa} to derive the Coulomb branch relations of non-abelian quiver theories. The construction is based on the embedding on the non-abelian CB (Coulomb branch) chiral ring into the CB chiral ring of the low-energy abelian theory which exists at generic points on the Coulomb branch. Each monopole operator of the non-abelian theory is mapped to a gauge invariant polynomial of abelian monopole operators and the non-abelian relations can be extracted from ``abelianized" relations involving the operators of the abelian theory.
A mathematical approach to 3d $\N=4$ Coulomb branches was also proposed in \cite{Nakajima:2015txa, Braverman:2016wma,Braverman:2016pwk}.
Despite this spectacular progress it remains often cumbersome to extract the ring relations in terms of a minimal basis of generators in a systematic way. 

In this paper we propose an alternative approach to study the Coulomb branch and the Higgs branch of 3d $\N=4$ theories using in a new way the realization of the theories as the low-energy theory of D3 branes stretched between NS5 and D5 branes in type IIB string theory. This elegant brane realization introduced by Hanany and Witten \cite{Hanany:1996ie}  is particularly useful to study mirror symmetry which is realized as S-duality in IIB string theory, leaving the D3 branes invariant and exchanging the D5 and NS5 branes. Using the brane construction and the action of S-duality the Higgs and Coulomb branches of large classes of quiver theories were identified as moduli spaces of solutions of Nahm's equations, namely intersections of (the closure of) nilpotent orbits and Slodowy slices \cite{Gaiotto:2008sa,Gaiotto:2008ak}.\footnote{See also \cite{Cabrera:2016vvv} for an application of the brane formalism to study an inclusion relation for nilpotent orbits associated to minimal singularities and related to higgsings of the quiver theories.} Here we use the brane picture in a spirit closer to \cite{Assel:2015oxa}, where half-BPS loop operators were realized with F1 or D1 string arrays added to the brane configuration of the theory. The path integral insertion of a loop operator was then understood as arising from integrating out the light modes on the strings in the brane configuration. This approach proved to be very useful in understanding the action of mirror symmetry on loop operators, using IIB S-duality.
The idea that we develop is that local half-BPS operators can be engineered in a similar way, by adding extra ingredients in the brane realization of the 3d theory and integrating out light modes. From these new brane setups we are able to extract the chiral ring relations and the mirror map between Higgs branch and Coulomb branch operators.

For simplicity we focus our analysis on quiver theories with abelian gauge nodes, except in the last section of the paper where we derive some preliminary results for non-abelian theories. 
In the brane picture the D3 branes are stretched between NS5 branes and span a finite interval in one direction. The three dimensional theory arises in low-energy limit of the D3 branes worldvolume theory. Using previous results in the literature and some simple arguments, we show that the chiral meson operators, or HB (Higgs branch) operators, are inserted from F1 strings stretched between pairs of D5 branes and ending on the D3 segments along the finite direction (e.g.~Figure \ref{SemiF1}), or by D3 branes, that we call D3', intersecting the initial D3 segments at points. For the CB (Coulomb branch) operators we show that chiral monopole operators  are inserted from D1 strings stretched between pairs of NS5 branes and ending on the D3 segments (e.g.~Figure \ref{SemiD1}-a), while scalar operators are realized by D3 branes, that we call D3'',  intersecting the initial D3 segments at a point (e.g.~Figure \ref{SemiD1}-c). The specific orientations of the F1, D3', D1 and D3'' preserve four out of the eight supercharges\footnote{The F1 and D3' preserve four supercharges. The D1 and D3'' preserve four supercharges. The two sets have two common supercharges.}, as appropriate to realize the insertion of half-BPS operators, and are given in Table \ref{tab:orientationsAll} \footnote{There are actually many choices of brane orientations, related by $SO(3)$ rotations, corresponding to different choices of complex structures/$\N=2$ sub-algebras under which the operators are chiral.}.
In a few places in our derivation we have to rely on indirect arguments or motivated assumptions, which are ultimately validated by the global consistency of the emerging picture.

Having understood how to realize the insertion of the operators forming bases of the Higgs branch and Coulomb branch, we proceed to studying the brane setups related to ring relations. For the Higgs branch there are two types of setups to consider. The F-term relations (or complex D-term equations) follow from identifying brane setups related by D3' brane moves along the interval direction, involving Hanany-Witten F1 string creation effects as a D3' passes through a D5 (e.g.~Figure \ref{D3pMove}). The other HB relations, which are trivial in the sense that they involve only recombining products of hyper-multiplet scalars in different products of mesons, can be related to multiple interpretations of a given brane setup with F1 strings (e.g.~Figure \ref{TwoSemiF1s}). The different interpretations can be associated to different recombinations of the F1 strings across D3 and D5 branes. 
For the Coulomb branch there is a single type of brane setups to consider, involving D1 branes (e.g.~Figure \ref{FullD1}), and the CB relations follow from interpreting each setup in two different ways: either as several semi D1 branes ending on D3s inserting a product of monopole operators, or as D1 branes crossing D3s. In this latter case we are able to integrate out the low-energy zero-dimensional theory arising from open string light modes and to show that the resulting insertions are polynomials in the complex scalars. There are also relations following from D1 brane recombinations across NS5 branes.
This approach provides a new derivation of the quantum Coulomb branch relations.
To be precise the relations that we extract from brane setups are not directly the exact ring relations, we therefore call them {\it pre-relations}. The exact relations are obtained by allowing the cancellation of operators which appear multiplicatively on both sides of a pre-relation ($AC = BC \, \Rightarrow A=B$).
At the end of our analysis we are left with a small set of rules from which one can extract the HB and CB relations in abelian quiver theories from a few brane setups. The dictionary between brane setups and operator insertions is summarized in Appendix \ref{app:Dictionary}. 
It might be worth emphasizing that, to our knowledge, this brane approach provides the first derivation of the Coulomb branch relations in a large class of abelian quiver theories\footnote{A direct derivation of the Coulomb branch relation in $\N=4$ SQED using CFT methods was given in \cite{Borokhov:2002cg}.} which does not rely dualities.

This brane approach is also particularly useful to study mirror symmetry. Acting with S-duality on the type IIB brane setups responsible for the insertions of the chiral operators, one finds the map between HB and CB operators of mirror dual theories with no effort. Under S-duality the F1 strings are simply mapped to D1 strings, and the D3' branes are mapped to D3'' branes.  It also immediately follows from IIB S-duality that the ``trivial" HB ring relations are mapped to the quantum CB ring relations.

We study non-abelian theories in the last section of the paper and provide a derivation of the abelianized relations, postulated in \cite{Bullimore:2015lsa}, in the SQCD theory from a set of simple brane setups. We also comment on the a possible analogue approach to deriving the Higgs branch relations in non-abelian theories. 

The approach presented in this paper could be extended to study the moduli space of vacua of other $\N=4$ theories. The generalization to non-abelian theories is not straightforward, in particular in the analysis of the ring relations beyond what we found in this paper, however it should be possible to find the brane setups inserting the chiral operators and the mirror map.  It would also be interesting to study the moduli space of $\N=4$ Chern-Simons theories with matter which also have a brane realization in IIB string theory (with $(1,\pm 1)$ 5branes). Considering brane setups realizing 3d theories with only $\N=2$ supersymmetry or operator insertions preserving a smaller amount of supersymmetry might be interesting directions to investigate. The direct computation of correlators of Higgs branch operators from supersymmetric localization in \cite{Dedushenko:2016jxl} is also a very fruitful approach. It could be extended to the study of correlators of monopole operators and should reproduce the CB ring relations.
We hope to report on some of these topics in future publications.

After a brief review on the Higgs and Coulomb branches of 3d $\N=4$ theories in Section \ref{sec:ModSpaceReview}, we study the brane configurations in type IIB responsible for the insertion of half BPS local operators in Section \ref{sec:BraneRealization}, identifying the F1 string and D3' branes as the objects inserting HB operators, and the D1 and D3'' branes as the object inserting CB operators. 
We then start the discussion in Section \ref{sec:TSU2} with the analysis of the $T[SU(2)]$ theory. We identify the brane setups inserting specific HB and CB operators and work out the ring relations from other brane setups. We explain how the mirror map of operators follows from S-duality ($T[SU(2)]$ is a self-mirror theory).
In Section \ref{sec:AbelGen} we extend the discussion to other abelian theories. We study in detail the cases of SQED and its mirror dual abelian quiver theory. At this point all the rules for operator insertions and ring relation readings for (linear) abelian quiver theories are derived. We illustrate our method in another example at the end of the section. In Section \ref{sec:OtherOp} we discuss briefly the brane setups inserting chiral operators which do not belong to the bases of the chiral rings, such as monopole operators with non-minimal magnetic charges, or products of mesons. Finally in Section \ref{sec:NATheories} we provide a preliminary analysis of the non-abelian theories, by recovering the SQCD abelianized relations of \cite{Bullimore:2015lsa} from a brane setup. Some computations are relegated to Appendix \ref{app:SQEDHBRel}. The dictionary between brane setups and HB/CB operator insertions and  the general mirror map  are presented in Appendix \ref{app:Dictionary}.

\section{Light review on $\N=4$ theories and their moduli space of vacua}
\label{sec:ModSpaceReview}

A three-dimensional Yang-Mills theory with $\N=4$ supersymmetry with gauge group $G$ has an algebra-valued vector multiplet, whose bosonic fields are a vector $A^{\mu}$ and three real scalars $\phi_1,\phi_2,\phi_3$, transforming in the adjoint representation of $G$. The matter fields come in hyper-multiplets, whose bosonic fields are two complex scalars $Q,\ti Q$ transforming in complex conjugate representations $\scR, \overline{\scR}$ respectively \footnote{This is not the most general situation. In general a hyper-multiplet has a complex scalar transforming in a pseudo-real representation $\scR_{\rm pr}$ of $G$. In this paper we focus on theories with $\scR_{\rm pr}= \scR \oplus \overline{\scR}$.}. In term of these data, the Lagrangian of the theory in uniquely fixed. The couplings of the theory are the Yang-Mills couplings $g_{\rm YM}^{(i)}$ for each semi-simple factor in the gauge group. The couplings $g_{\rm YM}^{(i)}{}^2$ have the dimension of a mass, implying that the theories are asymptotically free and strongly coupled at low energies.

The R-symmetry group is $SU(2)_C \times SU(2)_H$, with $(\phi_1,\phi_2,\phi_3)$ transforming in the $({\bf 3},{\bf 1})$ and $(Q, \ti Q^{\dagger})$ transforming in the $({\bf 1},{\bf 2})$. There also exists {\it twisted} vector and hyper- multiplets, for which the roles of $SU(2)_C$ and $SU(2)_H$ are exchanged.

\medskip

The moduli space of vacua of these theories is parameterized by the vevs of scalar operators, including monopole operators. It is given by a union of subspaces (branches) of the form $\scC_{n} \times \scH_n$, where, in broad lines, $\scC_n$ is parametrized by the vev of a certain subset of the vector multiplet scalars and monopole operators, while $\scH_n$ is parametrized by the vev of a subset of the hyper-multiplet scalars. These subspaces intersect on lower-dimensional loci and the total moduli space is $\scM = \cup_{n} (\scC_{n} \times \scH_n )$. There are two particular subspaces: the {\it Coulomb branch} $\scM_C \simeq \scC \times \{0\}$, where all hyper-multiplet scalar vevs are zero, and the {\it Higgs branch} $\scM_H \simeq \{0\} \times \scH $, where all vector multiplet scalar vevs and monopole vevs are zero. The other subspaces are called {\it mixed branches}.
The R-symmetry group $SU(2)_C$ acts non-trivially on the Coulomb branch $\scM_C$, as well as on the subspaces $\scC_n$, and acts trivially on $\scM_H$ and the subspaces $\scH_n$. Conversely the R-symmetry group $SU(2)_H$ acts non-trivially on the Higgs branch $\scM_H$, as well as on the subspaces $\scH_n$, and acts trivially on $\scM_C$ and the subspaces $\scC_n$. We will focus our discussion on the Coulomb and Higgs branches, the extension to the mixed branches being straightforward.
The Coulomb and Higgs branches are hyper-K\"ahler spaces, with the three complex structures transforming as triplets of $SU(2)_C$ and $SU(2)_H$ respectively. 

The Higgs branch is protected from quantum corrections and can be studied from the classical Lagrangian. It is parametrized by the vevs of the $4N$ real hyper-multiplet scalars, with $N=$dim$(\mathcal{R})$, satisfying the triplet of D-term equations, and modulo gauge transformations. This defines the Higgs branch as the hyper-K\"ahler quotient $\bR^{4N}///G$.  At a generic point on the Higgs branch the gauge group is completely broken and the low-energy theory is that of free hyper-multiplets. Often one describes the Higgs branch in an equivalent but simpler way, as a holomorphic quotient $\bC^{2N}/G_{\bC}$, by choosing a $\N=2$ subalgebra of the $\N=4$  super-algebra and parametrizing the Higgs branch by the vevs of the $2N$ complex scalars which are chiral with respect to this subalgebra, imposing the $\N=2$ F-term equation (or complex D-term equation) and quotienting by complex gauge transformations. The gauge invariant combinations of the chiral scalar operators form a ring, the Higgs branch chiral ring, and their vevs are holomorphic functions on the Higgs branch moduli space $\scM_H$. There is a $\bC\bP^1$ worth of choices of $\N=2$ subalgebras, each corresponding to a choice of complex structure on $\scM_H$. 

At a generic point on the Coulomb branch the gauge group is broken to a maximal torus $U(1)^{\text{rank} (G)}$ and the matter fields are massive. The deep infrared theory is then described by free abelian vector multiplets\footnote{The abelian vector multiplets can be dualized to twisted hyper-multiplets}. Classically the Coulomb branch is the symmetric product of rank$(G)$ $U(1)$ Coulomb branches $(\bR^3 \times S^1)^{\text{rank}(G)}/W_G$, where the factors $\bR^3\times S^1$ are parametrized by the Cartan components of the scalars $\phi_1, \phi_2, \phi_3$ and by the dual photons (compact scalar dual to the Cartan vector fields), and $W_G$ is the Weyl group of $G$, permuting the Cartan elements. The Coulomb branch receives quantum corrections which modify the geometry and in particular the topology of the classical description. A construction was presented in \cite{Bullimore:2015lsa} to describe the Coulomb branch as a complex algebraic variety. It involves the choice of complex structure on $\scM_C$, or equivalently the choice of an $\N=2$ subalgebra of the $\N=4$ theory. As for the Higgs branch, there is a $\bC\bP^1$ worth of choices of complex structures (since $\scM_C$ is hyper-K\"ahler). The Coulomb branch is then parametrized by monopole operators which are chiral with respect to the $\N=2$ subalgebra, subject to ring relations. 
The chiral (half-BPS) monopoles  are defined in the quantum theory by the prescription in the path integral of a Dirac monopole singularity for the Cartan elements of the vector field and a corresponding singularity for one of the real scalars in the vector multiplet. The monopole singularity breaks the gauge group to a subgroup $G'$. To complete the definition such an operator is dressed with a $G'$-invariant polynomial of the complex scalar (which combines the two remaining vector multiplet scalars) valued in $G'$. This defines a monopole operator $V_{A,p}$ labeled by a vector of monopole charges $A \in \bZ^{\text{rank} (G)}$ and a $G'$ invariant polynomial $p$. When $A=0$, the monopole operator is simply a gauge invariant polynomial in the vector multiplet complex scalars.
Linear combinations of the chiral monopole vevs define holomorphic functions on $\scM_C$. The ring of monopole operators can be called Coulomb branch chiral ring. Unlike the Coulomb branch metric, the ring relations are independent of the gauge coupling constants. 
They can be computed using the method of \cite{Bullimore:2015lsa}. For abelian theories, which is our primary interest in this paper, the ring relations are explicitly given in \cite{Bullimore:2015lsa}.

\medskip

In the UV description, the theories have global symmetries $U(1)^{\text{rank}(G)} \times G_H$, where the $U(1)$ factor acts by shifting the dual photons and $G_H$ is the flavor symmetry acting on the hyper-multiplets. In the infrared theory, the global symmetry group is enhanced to $G_C \times G_H$, where $G_C$ acts on monopole operators. Therefore the group $G_C$ acts on the Coulomb branch, while the group $G_H$ acts on the Higgs branch.

\medskip

The theories admit supersymmetry preserving deformations by mass terms,  with triplets of (real) mass parameters $m_1,m_2,m_3$, and by FI terms, with triplets of (real) FI parameters $\xi_1,\xi_2,\xi_3$. These deformations can be understood as arising from weakly gauging the Cartan subgroup of the $G_H$ and $G_C$ symmetry respectively. The triplets are then identified with the vevs of scalars in background vector multiplets. A triplet $(m_1,m_2,m_3)$ is the background of a regular vector multiplet and therefore transforms as a triplet of $SU(2)_C$, but a triplet $(\xi_1,\xi_2,\xi_3)$ is the background of a  twisted vector multiplet and therefore transforms as a triplet of $SU(2)_H$ instead. The mass deformations lift  part (or all) of the Higgs branch and modify the geometry on the Coulomb branch. In particular, in a chosen complex structure, a complex combination $m_{\bC}$ appear as a deformation parameter in the CB relations. The FI deformations lift part (or all) of the Coulomb branch and affect the geometry on the Higgs branch. In a chosen complex structure, a complex combination $\xi_{\bC}$ appear as a deformation parameter in the HB relations.

\medskip

\begin{figure}[h!]
\centering
\includegraphics[scale=0.8]{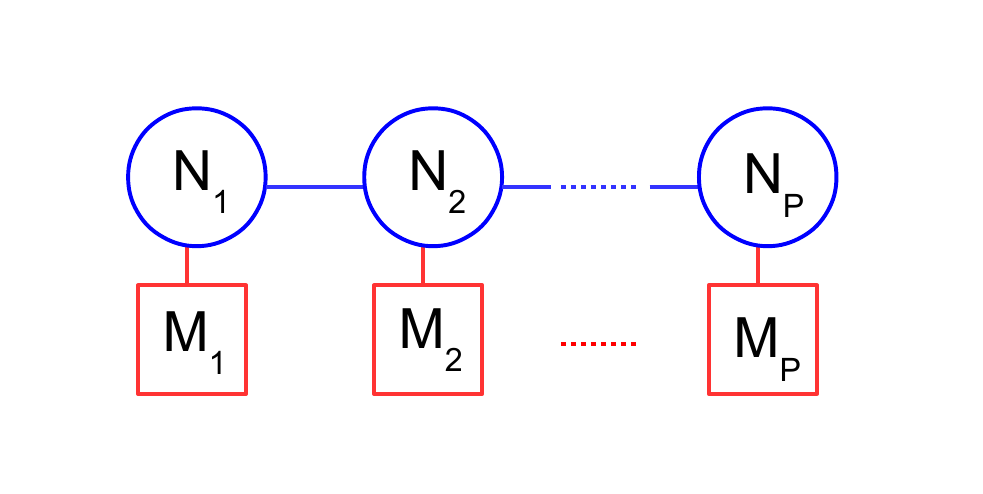} 
\vskip -0.5cm
\caption{\footnotesize General linear quiver.}
\label{linquiv}
\end{figure}

In the following we will study exclusively linear quiver gauge theories with unitary nodes (and for most of the discussion only abelian theories). This means that the gauge group is of the form $G = \prod_{i=1}^P U(N_i)$ and that the matter content comprises only fundamental and bifundamental hyper-multiplets. A fundamental hyper-multiplet of the $U(N_i)$ node transforms in the representation ${\bf N_i} \oplus \overline{\bf N_i}$. A bifundamental hyper-multiplet of the $U(N_i)\times U(N_j)$ nodes transforms in the representation $({\bf N_i},\overline{\bf N_j}) \oplus (\overline{\bf N_i}, {\bf N_j})$.
Such quivers are conveniently described by quiver diagrams, where circles denote gauge nodes, squares denote flavor nodes, and links between two circles, or between a circle and a square, denote bifundamental hyper-multiplets. In this language a bifundamental hyper-multiplet for gauge-flavor nodes $U(N)_{\rm gauge} \times U(M)_{\rm flavor}$ is the same as $M$ fundamental hypermultiplets of the $U(N)_{\rm gauge}$ gauge node.
Moreover we will focus on quiver theories of A type, namely linear quivers. In this case the matter content comprises one bifundamental hyper-multiplets for each pair of nodes $U(N_i) \times U(N_{i+1})$ and arbitrary numbers of fundamental hyper-multiplets. The generic quiver diagram is shown in Figure \ref{linquiv}.

\section{The brane realization of half-BPS local operator insertions}
\label{sec:BraneRealization}

The linear quiver theories described by the general quiver diagram of Figure \ref{linquiv} arise as the low-energy theory of brane arrays in type IIB string theory \cite{Hanany:1996ie}. The configurations involve D3 branes stretched between NS5 branes and intersecting D5 branes, with the orientations given in Table \ref{tab:orientationsQuiver}. 
\begin{table}[h]
\begin{center}
\begin{tabular}{|c||c|c|c|c|c|c|c|c|c|c|}
  \hline
      & 0 & 1 & 2 & 3 & 4 & 5 & 6 & 7 & 8 & 9 \\  \hline
  D3  & X & X & X & X &   &   &   &   &   &   \\
  D5  & X & X & X &   & X & X & X &   &   &   \\
  NS5 & X & X & X &   &   &   &   & X & X & X \\  \hline
\end{tabular}
\caption{\footnotesize Brane array realizing 3d $\N=4$ A-type quiver theories.}
\label{tab:orientationsQuiver}
\end{center}
\end{table}

\begin{figure}[h!]
\centering
\includegraphics[scale=0.75]{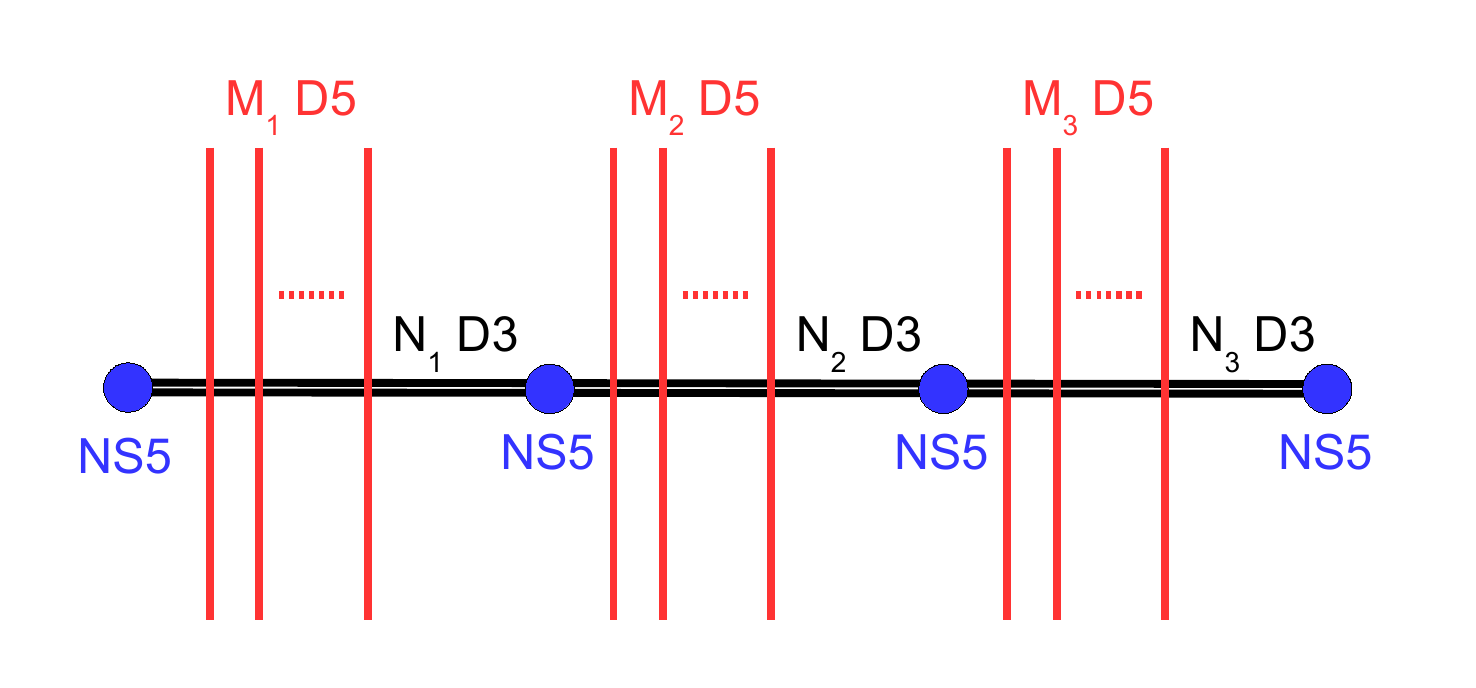} 
\vskip -0.5cm
\caption{\footnotesize Brane configuration associated to a general linear quiver with three nodes.}
\label{linquivbrane}
\end{figure}

A $U(N_i)$ gauge node is associated to $N_i$ D3 branes stretched between two NS5s. The fundamental hypermultiplets are the light modes of  D5-D3 open strings. The bifundamental hyper-mulitplets are the light modes of D3-D3 open strings stretched across NS5 branes.
The 3d quiver theory is the low-energy theory on the D3-brane worldvolume, along the $x^{0,1,2}$ directions common to all branes. The general construction is illustrated in Figure \ref{linquivbrane}, which realizes the generic linear quiver theory of Figure \ref{linquiv} in the three nodes case ($P=3$). 

\subsection{Branes and strings inserting half-BPS local operators}
\label{ssec:BranesLocalOp}

We have reviewed how a 3d $\N=4$ quiver theory can be engineered by a brane configuration in type IIB string theory. In various studies, including studies of 4d $\N=4$ Super-Yang-Mills and 4d $\N=2$ theories \cite{Gomis:2006im, Gomis:2016ljm}, it was found that the insertion of BPS loop operators and surface defects have a realization in terms of brane configurations in string or M theory, in the sense that integrating out the light degrees of freedom of the brane configuration results in the insertion of the loop or surface operator in the path integral of the theory. For three-dimensional $\N=4$ theories the brane configurations associated to the insertion of half-BPS Wilson loops and Vortex loops were constructed in \cite{Assel:2015oxa}.  It is natural to assume that also the local half-BPS operators have a realization in terms of certain brane setups.  

To realize the insertion of half-BPS local operators, namely local operators preserving  four out of the eight supercharges of the 3d theory, we need to include in the setup extra branes and/or strings with two properties: 
\begin{itemize}
\item Their orientation must preserve four type IIB supercharges. In particular extra D-branes must (at least) be oriented such that the numbers of ND (Neumann-Dirichlet) directions with the D3s and with the D5s in the configurations are multiples of four. For extra fundamental strings (F1s), one can consider the S-dual brane configurations, where they become D1s, to apply this criteria.
\item Their intersection with the D3s must be in a point or along the direction $x^3$, in which the D3s have a finite extent. This ensures that the extra brane or string sits at a point in the space $x^{0,1,2}$, which supports the low-energy 3d theory, and therefore inserts a local operator in the 3d theory.
\end{itemize}

These constraints select four possible extra ingredients, which are fundamental strings (F1s), D1-branes and two types of D3-branes, that we denote D3' and D3'', with the orientations given in Table \ref{tab:orientationsAll}.
\begin{table}[h]
\begin{center}
\begin{tabular}{|c||c|c|c|c|c|c|c|c|c|c|}
  \hline
      & 0 & 1 & 2 & 3 & 4 & 5 & 6 & 7 & 8 & 9 \\ \hline
  D3  & X & X & X & X &   &   &   &   &   &   \\
  D5  & X & X & X &   & X & X & X &   &   &   \\
  NS5 & X & X & X &   &   &   &   & X & X & X \\  \hline
  F1 &  &  &  & X  & X  &   &   &  &  &  \\ 
  D3' &  &  &  &   & X  &   &   & X & X & X \\  \hline
  D1 &  &  &  & X  &   &   &   & X &  &  \\ 
  D3'' &  &  &  &   & X  & X  & X  & X &  &  \\  \hline
\end{tabular}
\caption{\footnotesize Brane array for 3d $\N=4$ theories and half-BPS local operators.}
\label{tab:orientationsAll}
\end{center}
\end{table}
There are actually other possible choices of orientation, corresponding to rotations in $(x^4,x^5,x^6)$ and $(x^7,x^8,x^9)$, and preserving different supercharges. These choices of orientation are mapped to the choices of $\N=2$ subalgebra in the $\N=4$ theory reviewed in Section \ref{sec:ModSpaceReview}. We will stick to the orientations displayed in Table \ref{tab:orientationsAll} in our discussion.
Note that these extra branes or strings are extended only along spacial directions, as appropriate to include  point-like operators. 

The selection criteria applied above to preserve supersymmetry are necessary but may not be sufficient. To confirm that these configurations really preserve four supercharges, one needs to check that the projection on the IIB supercharges imposed by each additional brane in the D3-D5-NS5 setup is satisfied by four supercharges. Let us denote by $\epsilon, \ti\epsilon$ the two 32-components Weyl spinors parametrizing the supersymmetries of IIB string theory. We introduce $\Gamma^{A}$, $A=0, 1, \cdots, 9$, the ten-dimensional gamma matrices and $\Gamma^{11}= -\Gamma^0 \Gamma^1 \cdots \Gamma^9$ the chirality matrix. We work here in $(-++ \cdots +)$ signature \footnote{We perform the analysis of brane supersymmetries in the more familiar Lorentzian space. However the field theory discussion is based on the Euclidean theory, so one could adapt the analysis by implementing a Wick rotation in the brane setup.}. The Weyl spinors have the same chirality $\Gamma^{11}\epsilon = \epsilon$ and $\Gamma^{11}\ti\epsilon = \ti\epsilon$. 
The projections imposed by each brane on the spinors $\epsilon, \ti\epsilon$ are the following
\begin{alignat}{3}
& \underline{D3 :} & \qquad   \epsilon &= \Gamma^{0123} \ti\epsilon \,,  \cr
& \underline{D5 :} & \qquad   \epsilon &= \Gamma^{012456} \ti\epsilon \,,  \cr
& \underline{NS5 :} & \qquad   \epsilon &= \Gamma^{012789} \epsilon \,,  \quad  \ti\epsilon = -\Gamma^{012789} \ti\epsilon \,, \cr
& \underline{F1 :} & \qquad   \epsilon &= -i \Gamma^{34} \epsilon \,,  \quad \ti\epsilon = i \Gamma^{34} \ti\epsilon \,, \cr
& \underline{D1 :} & \qquad   \epsilon &= i \Gamma^{37} \ti\epsilon \,,  \cr
& \underline{D3' :} &  \qquad   \epsilon &= i \Gamma^{4789} \ti\epsilon \,, \cr
& \underline{D3'' :} & \qquad   \epsilon &= i \Gamma^{4567} \ti\epsilon \,,
\end{alignat}
where we defined $\Gamma^{A_1 A_2 \cdots A_n} = \Gamma^{A_1}\Gamma^{A_2} \cdots \Gamma^{A_n}$, for $A_1 \neq A_2 \neq \cdots \neq A_n$.
The projections associated to Euclidean branes or strings differ from the projections associated to Lorentzian branes by an extra factor of $i$, which follows from a Wick rotation. The NS5 brane projections were given in \cite{Hanany:1996ie}. The projection due to the fundamental string can be worked out from the Green-Schwarz super-string action in light-cone gauge: in type IIB string theory there are twice sixteen supercharges, the two sets satisfying opposite 2d chirality projections on the worldsheet (which in this case is extended along the $x^3$ and $x^4$ directions) and forming the two ten-dimensional spinors.

Solving the system of equations for each configuration\footnote{We used an explicit representation of the Gamma matrices and solved the systems of equations in Mathematica.}, one finds that the (D3-D5-NS5)-F1 setup and the (D3-D5-NS5)-D3' setup preserve the same four supercharges, say $Q_1,Q_2,Q_3,Q_4$, and that the (D3-D5-NS5)-D1 setup and the (D3-D5-NS5)-D3'' setup also preserve four identical supercharges, with two common supercharges with the previous setups, say $Q_1,Q_2,Q_5,Q_6$.
This shows that the setup with both F1s and D3's preserve four supercharges, that the setup with both D1s and D3''s also preserve four supercharges, and that a setup with F1, D1, D3' and D3''  still preserves two supercharges $Q_1,Q_2$. We will restrict to studying setups with F1s and D3's, or D1s and D3''s.

In the next section we will find that the insertion of Higgs and Coulomb branch operators are realized by F1-D3' and D1-D3'' setups respectively.
Not surprizingly, the F1-D3' setups are mapped to  D1-D3'' setups under S-duality of IIB string theory.

\medskip

\begin{figure}[th]
\centering
\includegraphics[scale=0.75]{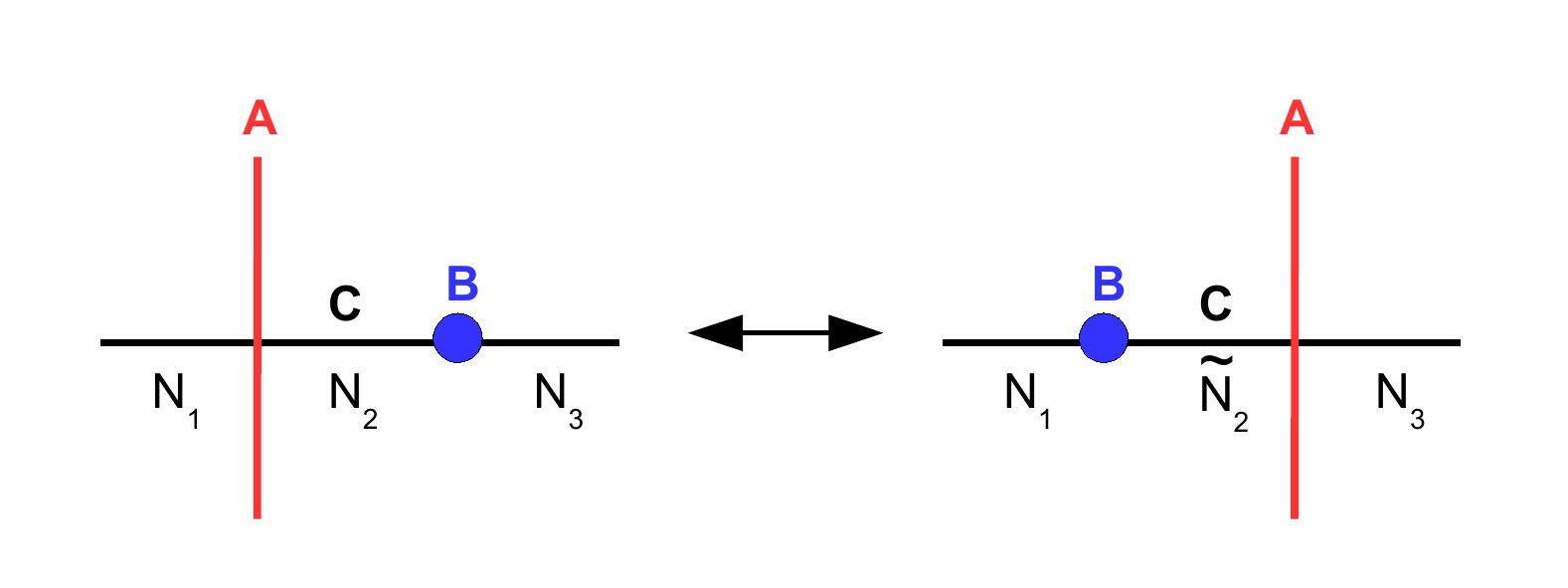}
\vspace{-0.5cm}
\caption{\footnotesize{Hanany-Witten brane creation effect. The triplet of branes $(A,B,C)$ can be either $(D5,NS5,D3)$, or $(D3',D5,F1)$, or $(D3'',NS5,D1)$. As the $A$ and $B$ branes pass across each other the number of $C$ branes stretched between them goes from $N_2$ to $\ti N_2 = N_1 + N_3 - N_2 +1$.}}
\label{HWmove}
\end{figure}

To conclude this discussion on the brane setups, it is important to discuss the Hanany-Witten (HW) brane creation effects \cite{Hanany:1996ie} arising in the above configurations. The Hanany-Witten effect in the D3-D5-NS5 setup is the phenomenon of D3 brane creation, stretched between a D5 and an NS5 branes, as the two 5-branes pass through each-other. Consider a configuration with a D5 on the letf and an NS5 on the right in the $x^3$ direction, with $N_D$ and $N_{NS}$ the net numbers\footnote{The net number is the number ending on the left minus the number ending on the right in the $x^3$ direction.} of D3s ending on the D5 and NS5 respectively. After moving the two 5-branes across each other, the new net numbers $N'_D, N'_{NS}$ are given by $N'_D = N_D -1$, $N'_{NS} = N_{NS} +1$, due to the creation of one D3 brane stretched between them.
Using T-dualities and S-duality of string theories, one can generate various dual setups where this brane/string creation effect happens \cite{Bachas:1997ui, Bachas:1997kn}, with different explanations of the effect in different setups.
In the setups described above we have three such HW effects:
\begin{itemize}
\item the D3-D5-NS5 system, with D3 brane creation, arising in the initial brane setup realizing the quiver theories;
\item the F1-D3'-D5 system, with F1 string creation stretched between the D3' and the D5;
\item the D1-D3''-NS5 system, with D1 brane creation stretched between the D3'' and the NS5.
\end{itemize}
The  two latter situations are related by S-duality.
These HW effects are depicted in Figure \ref{HWmove}. They will play a role in our analysis. There is no other brane creation effects in the setups that we study.

\medskip

In HW configurations with a triplet of branes $(A,B,C)$ and $C$ brane creation effect, the lowest mode on a $C$ brane stretched between an $A$ and a $B$ brane is fermionic. This implies the so-called {\it s-rule}, stating that there can be at most one $C$ brane stretched between an $A$ and a $B$ brane at low energies.

\section{Analysis in the $T[SU(2)]$ theory}
\label{sec:TSU2}

\begin{figure}[th]
\centering
\includegraphics[scale=0.75]{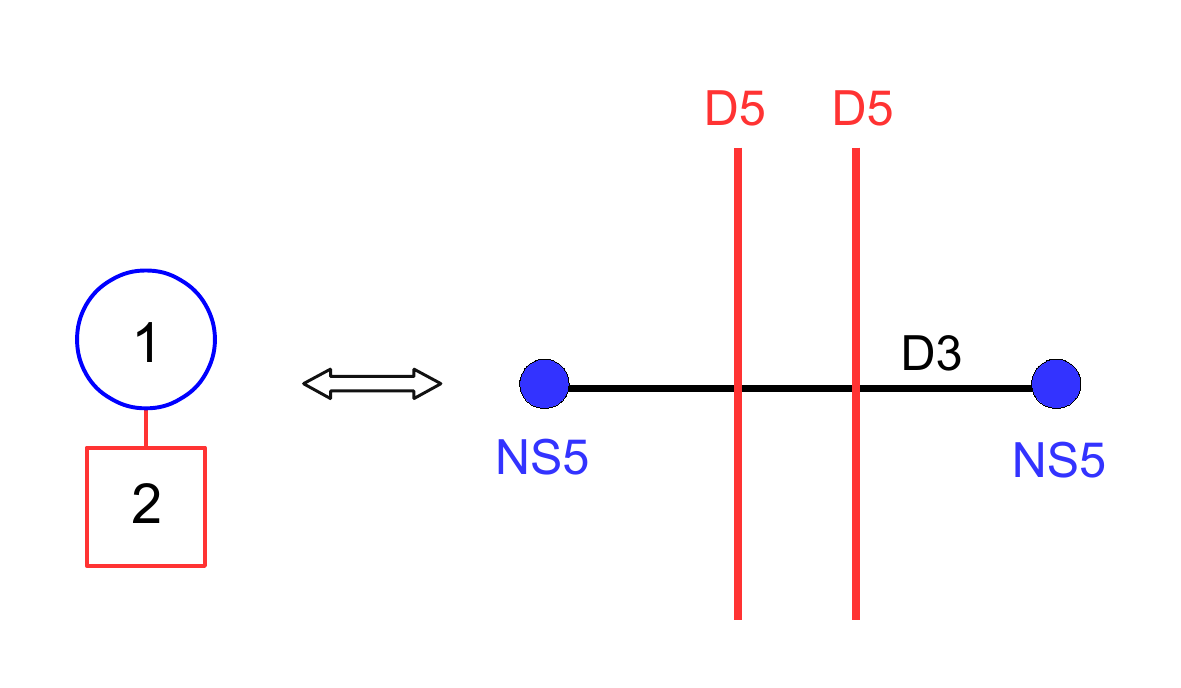}
\vspace{-0.5cm}
\caption{\footnotesize{Quiver and brane configuration for the $T[SU(2)]$ theory.}}
\label{TSU2}
\end{figure}

To start with we consider in this section the simple setup of a $U(1)$ gauge theory with two fundamental hyper-multiplets with complex scalars $(Q^\alpha, \ti Q_\alpha)$, $\alpha=1,2$, also called $T[SU(2)]$ theory.
The brane realization of this theory is shown in Figure \ref{TSU2}. It has two D5s intersecting a single D3. We denote D5$_{(1)}$ the D5 on the left in the figure and D5$_{(2)}$ the D5 on the right.

\subsection{Higgs branch operators}
\label{ssec:HBop}

\begin{figure}[th]
\centering
\includegraphics[scale=0.75]{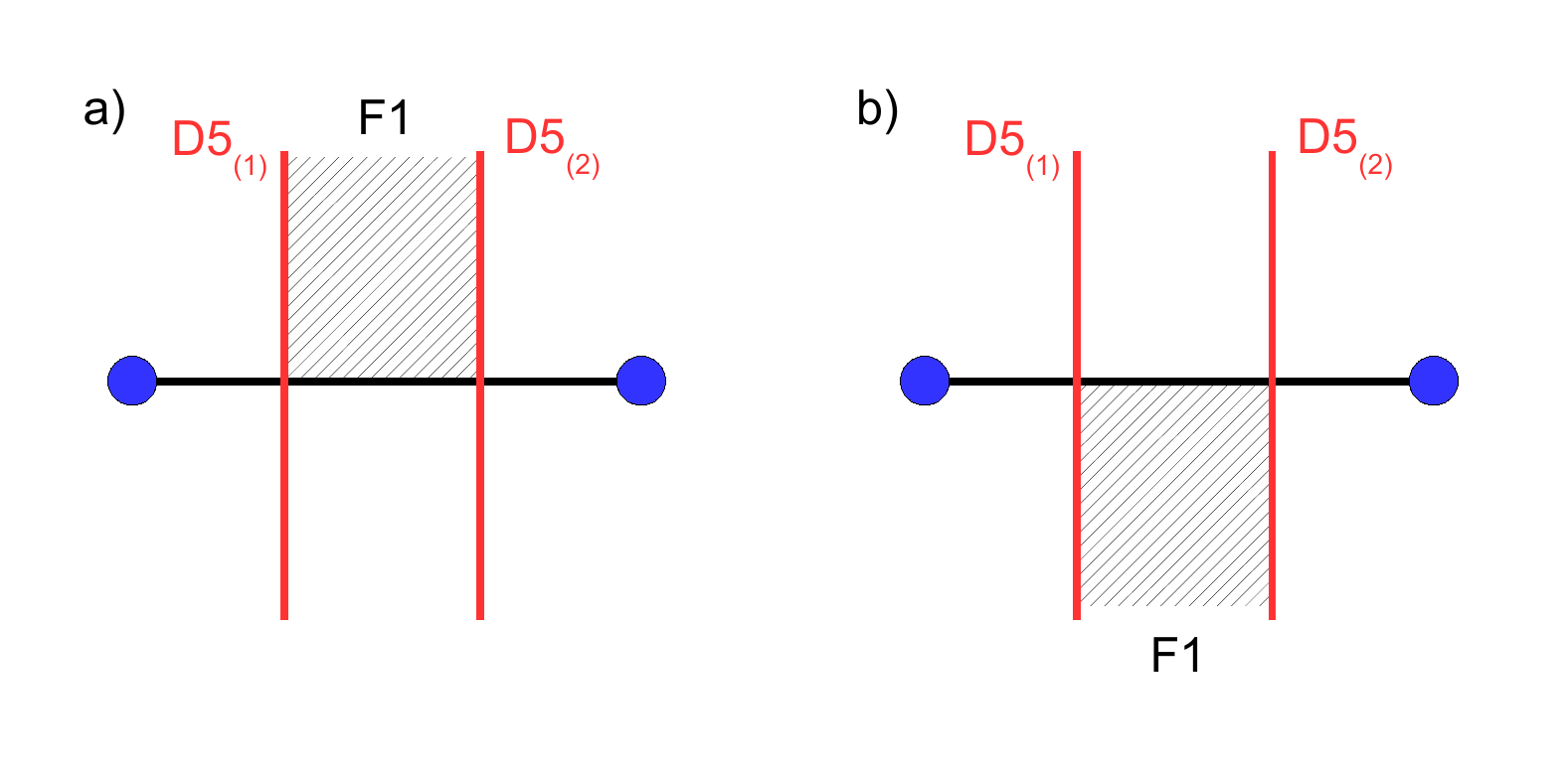}
\vspace{-1cm}
\caption{\footnotesize{Brane configurations for the insertion of the meson operators (a) $Z^2{}_1$ and (b) $-Z^1{}_2$.}}
\label{SemiF1}
\end{figure}

\begin{figure}[th]
\centering
\includegraphics[scale=0.75]{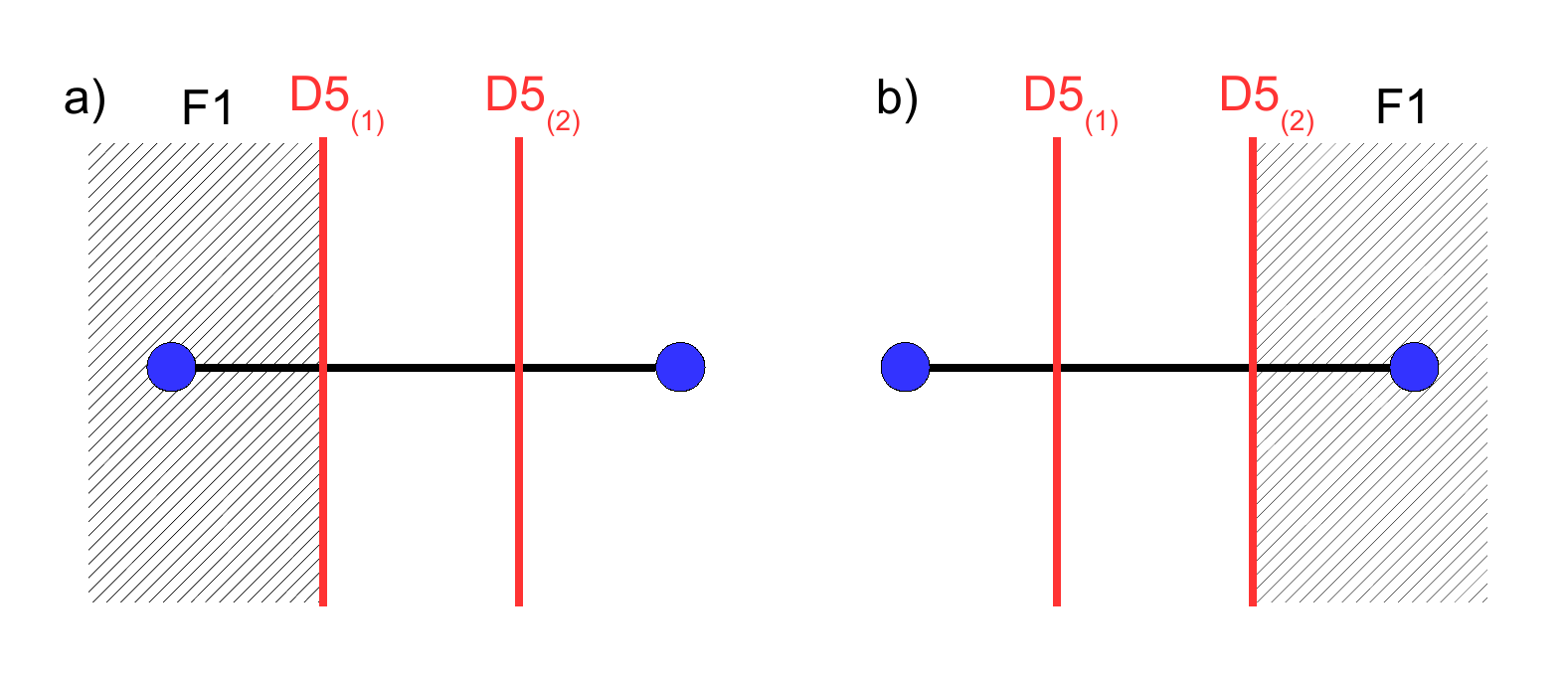}
\vspace{-0.5cm}
\caption{\footnotesize{Brane configurations for the insertion of the meson operators (a) $-Z^1{}_1$  and (b) $Z^2{}_2$.}}
\label{SemiF1_2}
\end{figure}

 The operators whose vevs parametrize the Higgs branch, or HB operators, are the mesons
\be
Z^\alpha{}_\beta = \ti Q_\beta Q^\alpha \,, \quad \alpha,\beta=1,2 \,,
\ee
which satisfy by definition the relation
\be
\det Z \equiv Z^1{}_1 Z^2{}_2 - Z^2{}_1 Z^1{}_2  = 0 \,,
\label{HBRel1}
\ee
and are subject to the F-term constraint
\be
\tr Z \equiv Z^1{}_1 + Z^2{}_2 = 0 \,.
\label{HBRel2}
\ee
We claim that the meson operator insertions are realized by the following brane setups:
\begin{itemize}
\item  The operators $Z^2{}_1$ and $-Z^1{}_2$ are realized with a single semi-infinite F1 string stretched between the two D5s and ending on the D3 from above and from below respectively, as shown in Figure \ref{SemiF1}. 
\item The  operator $-Z^1{}_1$ is realized with one infinite F1 string extended on the left of the configuration and ending on D5$_{(1)}$, as shown in Figure \ref{SemiF1_2}-a. The meson operator $Z^2{}_2$ is realized with one infinite F1 string extended on the right of the configuration and ending on D5$_{(2)}$, as shown in Figure \ref{SemiF1_2}-b. 
\end{itemize}
We will not provide a direct derivation of these results from string perturbation theory. Instead we will rely on known results and consistency arguments leading to the proposals and the observation that they are consistent with expectations. 

First we recall the result in \cite{Gomis:2006im} that a configuration with a semi-infinite F1 string ending on a stack of $N$ D3s is responsible for the insertion a half-BPS Wilson loop in the fundamental representation of $U(N)$, in the four-dimensional low-energy theory on the D3s \footnote{The simplest brane configuration studied in \cite{Gomis:2006im} has a string stretched between a stack of $N$ D3s and a single D3 at a large distance. Here we think of the extra D3 brane as being sent to infinity, leaving only a semi-infinite F1 string  ending on the stack of $N$ D3s.}. 

Let us be more precise. With D3s along $x^{0,1,2,3}$, the low-energy theory on the D3s is the four-dimensional $\N=4$ $U(N)$ SYM theory. The bosonic fields are the 4d gauge field $A_\mu$ and the six real scalars $\phi_i$, $i=4, ...,9$, corresponding to the D3 motions along the $x^{4,5,6,7,8,9}$ directions, all valued in the $\mathfrak{u}(N)$ algebra. 

A semi-infinite  F1 string spanning the directions $x^{3}$ and $x^{4}>0$, ending on the D3s at $x^4=0$, inserts in the path integral of the SYM theory the half-BPS Wilson loop in the fundamental representation\footnote{The factors of $i$ differ from \cite{Gomis:2006im} due to the fact that the string is extended along two space directions here, instead of space and time there.}
\be
W_{\rm fund} = \tr_{\rm fund} \, \scP \exp \lp  \int_{-\infty}^{\infty} dx^3 \, ( i A_3 + \phi_4) \rp \,, 
\ee
where $\tr_{\rm fund}$ is the trace in the fundamental representation and $A_3$ is the component of the gauge field along the direction $x^3$. When there is a single D3 brane, the string inserts the Wilson loop $\scP \exp  \int_{-\infty}^{\infty} dx^3 \, ( i A_3 + \phi_4) $ in the abelian $\N=4$ SYM theory.

Similarly we can consider a stack of $K$ D5s along the directions $x^{0,1,2,4,5,6}$ in the presence of a semi-infinite F1 string along the directions $x^{3}<0$ and $x^{4}$, ending on the D5s at $x^3=0$. The computation in \cite{Gomis:2006im} carries over to this situation, concluding that this setup corresponds to the insertion of a half-BPS Wilson loop in the fundamental representation of the 6d $U(K)$ SYM theory living on the D5s. In the abelian case, $K=1$, the Wilson loop is 
\be
W_{D5} = \exp \lp  \int_{-\infty}^{\infty} dx^4 \, ( i A^{(D5)}_4 + \phi^{(D5)}_3) \rp \,,
\ee
where $A^{(D5)}_4$ is the component of the 6d gauge field along $x^4$ and $\phi^{(D5)}_3$ is the 6d scalar field associated to motion of the D5 along $x^3$.

\begin{figure}[th]
\centering
\includegraphics[scale=0.75]{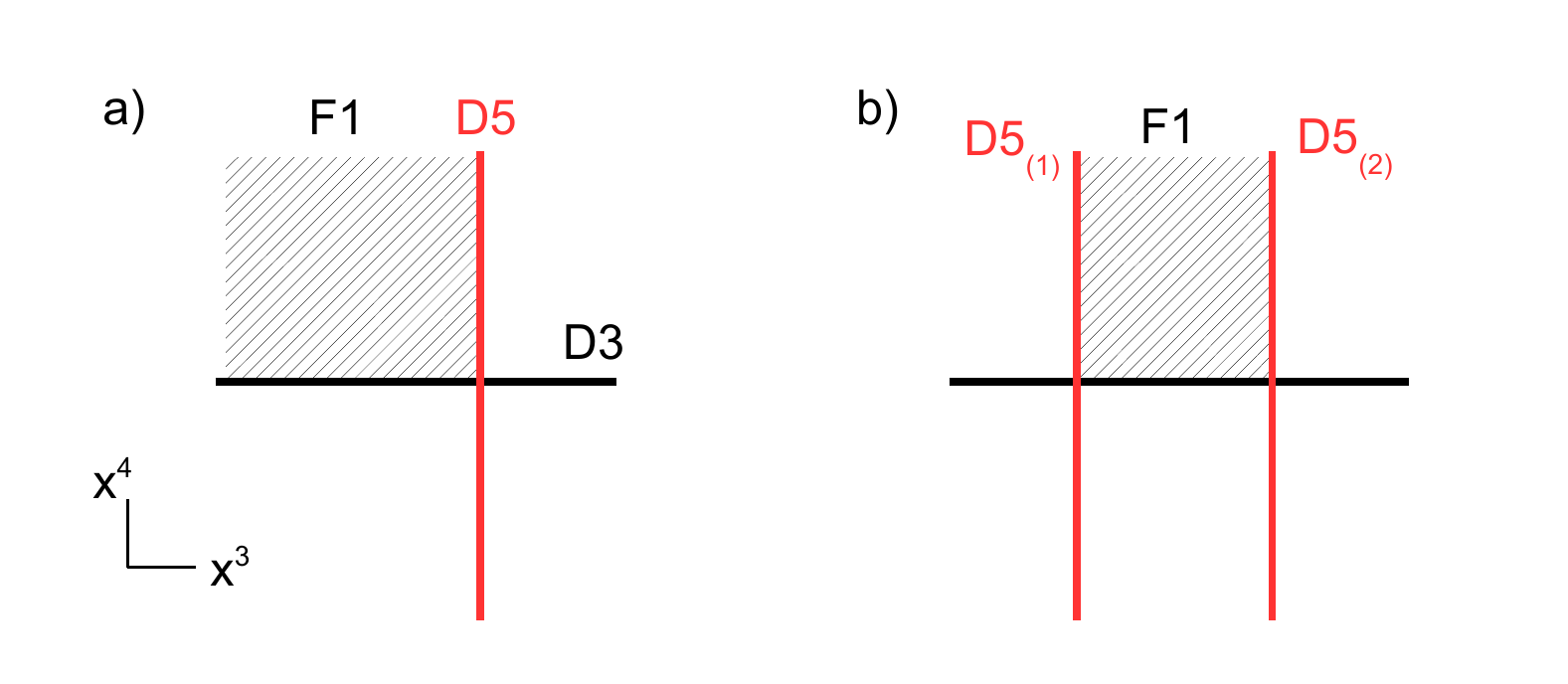}
\vspace{-0.5cm}
\caption{\footnotesize{a) F1 string with a D3-D5 corner. b) F1 string with two D3-D5 corners.}}
\label{CornerF1}
\end{figure}

The situation that we want to consider is slightly more complicated. We have both a D3 and a D5 brane, intersecting, say, at $(x^3,x^4) = (0,0)$, and we stretch an F1 string along the $x^{3,4}$ directions such that it ends both on the D3 and on the D5, and has a D3-D5 corner. So the string spans, say, $x^3 < 0$ and $x^4 > 0$, as shown in Figure \ref{CornerF1}-a. 
At low-energies, we infer from the above analysis that this setup corresponds to the insertion of the product of Wilson loops
\be
\exp \lp \int_{-\infty}^{0} dx^3 \, ( i A_3 + \phi_4) \rp \, \exp \lp  \int_{0}^{\infty} dx^4 \, ( i A^{(D5)}_4 + \phi^{(D5)}_3) \rp \,,
\ee
where the 6d fields can be considered frozen to a background in the low-energy limit.
However this cannot be the whole story, since this operator is not gauge invariant under the 4d or 6d gauge transformations:  the integration contours of the Wilson loops have a common boundary at the point $P =(0,0)$ where the F1, D3 and D5 meet. To restore gauge invariance,  the insertion of an extra local field at the point $P$ is needed, which transforms with charges $(-1,1)$ under $U(1)_{D3} \times U(1)_{D5}$. The minimal extra insertion needed is with a bifundamental scalar, so we propose that the complete operator insertion is
\be
\exp \lp \int_{-\infty}^{0} dx^3 \, ( i A_3 + \phi_4) \rp \, Q(P)  \,  \exp \lp  \int_{0}^{\infty} dx^4 \, ( i A^{(D5)}_4 + \phi^{(D5)}_3) \rp \,,
\ee
where $Q(P)$ is the hypermultiplet complex scalar $Q$ of charge $(-1,1)$ under $U(1)_{D3} \times U(1)_{D5}$ living at the 3d intersection of the D3 and D5 branes, evaluated at $P$. This operator is gauge invariant and preserves half of the supersymmetries of the D5-D3 configuration. 

If one consider the analogous situation of an F1 string stretched along $x^3>0$ and $x^4>0$ instead, the corresponding insertion would be
\be
 \exp \lp  - \int_{0}^{\infty} dx^4 \, ( i A^{(D5)}_4 + \phi^{(D5)}_3) \rp \, \ti Q(P)  \,   \exp \lp \int_{0}^{\infty} dx^3 \, ( i A_3 + \phi_4) \rp \,,
\ee
where the change of sign (or charge) for the Wilson loop in the D5 theory is due to the fact that the F1 string now ends on the other side of the D5 brane. Here $\ti Q(P)$ is the hypermultiplet complex scalar $\ti Q$ of charge $(1,-1)$ under $U(1)_{D3} \times U(1)_{D5}$ living at the 3d intersection of the D3 and D5 branes, evaluated at $P$, and preserving the same four supercharges as the preceeding operator involving $Q$.

Finally we consider the situation when there is one D3 brane and two D5 branes, sitting at $x^3=0$ and $x^3=L>0$, with an F1 string stretched along $x^4>0$ and the finite interval $0 < x^3 < L$, ending on both D5s, as shown in Figure \ref{CornerF1}-b. The same considerations as above lead us to the conclusion that the half-BPS  operator insertion is
\be\ba
 & \exp \lp  - \int_{0}^{\infty} dx^4 \,  ( i A^{(D5_1)}_4 + \phi^{(D5_1)}_3)\rp   \,  \ti Q_1(P_1) \, \exp \lp \int_{0}^{L} dx^3 \, ( i A_3 + \phi_4) \rp \cr
 & \qquad  \, Q^2(P_2) \,   \exp \lp  \int_{0}^{\infty} dx^4 \, ( i A^{(D5_2)}_4 + \phi^{(D5_2)}_3) \rp \,,
 \label{InsertionF14d}
\ea\ee
where $\ti Q_1$ is the 3d scalar of charge $(-1,1)$ under $U(1)_{D3} \times U(1)_{D5_{(1)}}$ and $Q^2$ is  the 3d scalar of charge $(-1,1)$ under $U(1)_{D3} \times U(1)_{D5_{(2)}}$. The points $P_1$ and $P_2$ are the intersections of the F1, D3 and D5$_1$, and F1, D3 and D5$_2$, respectively. They have the same positions in all coordinates, except $x^3$, with $x^3(P_1)=0$ and $x^3(P_2)=L$.

This configuration, with one D3, two D5s at $x^3=0$ and $x^3=L>0$, and an F1 string stretched between them, is embedded in a setup where the D3 spans a finite interval in the $x^3$ direction, ending on two NS5s at positions $x^3_L < 0$ and $x^3_R > L$ (Figure \ref{SemiF1}-a). 
The boundary conditions on the 4d fields living on the D3 at $x^3_L$ and $x^3_R$ have been studied in \cite{Gaiotto:2008sa}. They are of Neumann type for the vector field. We have\footnote{The scalars $\phi_{4,5,6}$ can actually be fixed to non-zero constants  at the boundaries, but we take these constants to be vanishing for simplicity.}
\be
A_3 = 0 \,, \quad  \phi_{4,5,6} = 0 \,, \quad \p_3 \phi_{7,8,9} = 0 \,, \quad \p_3 A_{\mu} = 0 \,, \ \mu = 0,1,2\,.
\label{NSBdyCond}
\ee
At low energies the 6d fields are non-dynamical and the 4d fields obey the constraints \eqref{NSBdyCond} on the whole interval $x^3_L< x^3 <x^3_R$. The operator \eqref{InsertionF14d} then reduces to
\be
\ti Q_1(P_1) Q^2(P_2) \overline{W}_{D5_1} W_{D5_2} \,,
\ee
where $\overline{W}_{D5_i}$ and $W_{D5_i}$ denote the flavor Wilson loops of charge $-1$ and $+1$ under the $U(1)_{D5_i}$ respectively. 
In the 3d effective theory the points $P_1$ and $P_2$ are identified, $P_1 \sim P_2 \equiv x \in \bR^{3}$. The flavor Wilson loops will play no role in our discussion, so we take them to be trivial (equal to one). We conclude that the brane configuration inserts the local meson operator
\be
\ti Q_1(x) Q^2(x) = Z^{2}{}_1 (x) \,.
\ee
The same analysis for the configuration with a semi-infinite F1 string spanning $0<x^3<L$ and $x^4 <0$ (figure \ref{SemiF1}-b), where the Wilson loop in the D3 worldvolume theory has opposite charge, leads to the insertion of the half-BPS meson operator
\be
-Q^1(x) \ti Q_2(x) = -Z^{1}{}_2 (x) \,.
\ee
The minus sign inserted here does not follow from our analysis, which is not sensitive to overall constants. We fix it by consistency with the analysis of ring relations that is presented in later sections.

The brane setup with an F1 string extended along $x^3<0$ and $x^4$, ending on $D5_{(1)}$ on its left, as shown in Figure \ref{SemiF1}-a, can be understood as having two D3-D5$_{(1)}$ corners, one above and one below the D3, inserting the meson operator
\be
-Q^1(x) \ti Q_1(x) = -Z^{1}{}_1 (x) \,.
\ee
Again the minus sign is fixed by consistency with later analysis.

The brane setup with an F1 string extended along $x^3>L$ and $x^4$, ending on $D5_{(2)}$ on its right, as shown in Figure \ref{SemiF1}-b, can be understood as having two D3-D5$_{(2)}$ corners, one above and one below the D3, inserting the meson operator
\be
Q^2(x) \ti Q_2(x) = Z^{2}{}_2 (x) \,.
\ee
This interpretation requires the F1 string to lie at the same $x^{5,6}$ positions as the D3, so that it breaks in two pieces, both ending on the D3.

We recover all the insertions proposed at the beginning of the section. So far the D3' branes did not play a role in the HB operator insertions. They will enter into play when we study the HB relations and mirror symmetry.

\subsection{Coulomb branch operators}
\label{ssec:CBop}

\begin{figure}[th]
\centering
\includegraphics[scale=0.72]{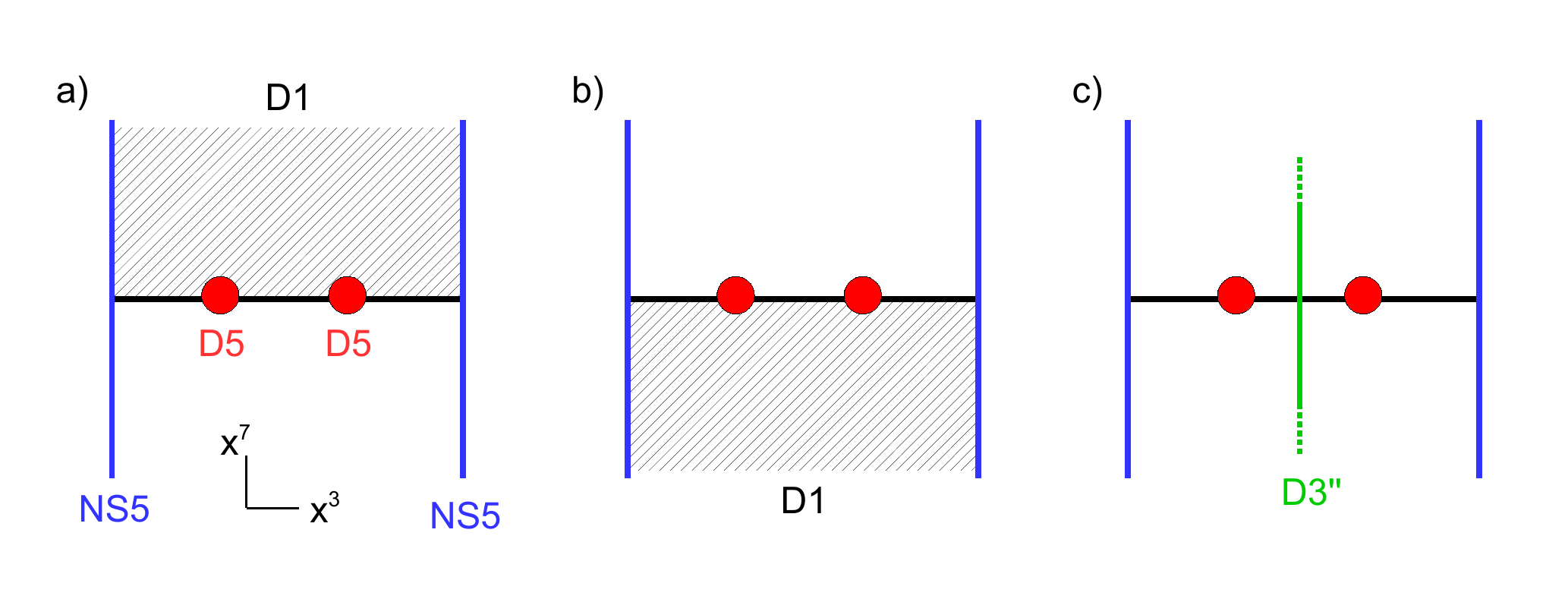}
\vspace{-0.5cm}
\caption{\footnotesize{a) Semi infinite D1 ending on the D3 from above, inserting the monopole operator $u^+$. b) Semi infinite D1 ending on the D3 from below, inserting the monopole operator $-u^-$. c) D3" brane crossing the D3 at an arbitrary position in $x^3$ between the two NS5s, inserting the scalar operator $\varphi$.}}
\label{SemiD1}
\end{figure}

We now turn to the Coulomb branch operators, or CB operators, of the $T[SU(2)]$ theory. A basis generating the CB chiral ring is given by three half-BPS (or chiral) scalar operators: the complex scalar $\varphi$ and the abelian monopole operators $u^{\pm}$ of monopole charge $\pm1$. They are subject to the CB quantum relation
\be
u^+ u^- = - \varphi^2 \,.
\label{TSU2CBRel}
\ee
To realize the path integral insertion of the CB operators, we propose that the following brane setups:
\begin{itemize}
\item The $u^+$ and $-u^-$ \footnote{The minus sign here is purely conventional (it could be eliminated by redefining $u^-$). We introduce it in order to get a simple mirror map with HB operators.} operator insertions are realized by adding a semi-infinite D1 brane stretched between the two NS5s and ending on the D3 from above and from below respectively, as shown in Figure \ref{SemiD1}-a and -b. 
\item The $\varphi$ operator insertion is realized by adding a D3" brane between the two NS5s, intersecting the D3 at a point, as shown in Figure \ref{SemiD1}-c. 
\end{itemize}

To argue for the realization of monopole operator insertions, we start from the brane configuration with only a D3-brane and a semi-infinite D1, along $x^3$ and $x^7>0$, ending on it at $x^7=0$. At low energies on the worldvolume of the D3 lives the abelian 4d $\N=4$ SYM theory. It is well-known \cite{Diaconescu:1996rk} that a D1 brane stretched between two separated D3 branes can be associated to the presence of an $SU(2)$ half-BPS monopole of charge $(1,-1)$ under the $U(1)^2$ Cartan subalgebra, in the D3 worldvolume YM theory. The limit when one of the D3 brane is sent at infinity corresponds to the limit of infinitely small size of the monopole soliton, leaving  a half BPS monopole line operator, called 't Hooft loop, in the 4d theory, with charge $+1$ (or $-1$). We pick the convention that the charge is $+1$ when the D1 ends on the D3 from above.
This 't Hooft loop is defined  as a Dirac monopole singularity of the abelian gauge field at each point on the $x^3$ line (here in Euclidean space),
\be
F=  \star d \bigg( \frac{1}{2r} \bigg) \,, \quad  r = \sqrt{(x^0)^2 + (x^1)^2 +(x^2)^2 } \,,
\ee
and a corresponding singularity for the real scalar $\phi^7$ corresponding to motions of the D3 along $x^7$, and ensuring that the configuration preserves half of the supersymmetries.

In our situation the D3 and D1 span only a finite interval along $x^3$ since they end on the two NS5 branes. This means that the 4d theory with a unit charge 't Hooft loop lives on an interval with Neumann boundary conditions \eqref{NSBdyCond}. These boundary conditions are compatible with the 't Hooft loop singularity. At low energies, the theory becomes effectively three-dimensional and the loop operator becomes a local operator inserting a charge one Dirac monopole singularity and the corresponding singularity for the real scalar $\sigma \sim \phi^7$, preserving half of the supersymmetries. This is precisely the insertion of the $u^+$ monopole operator advertized above. 

The insertion of the $u^-$ monopole operator from a D1 ending on the D3 from below follows from the same argument, with the negative charge 't Hooft loop insertion in 4d.

\medskip

The last configuration involves a D3" crossing the D3 at an arbitrary point along $x^3$ between the two NS5s (Figure \ref{SemiD1}-c).
We can analyse the spectrum of light string modes at the intersection point between D3 and the D3". These branes have eight Neumann-Dirichlet directions, therefore the light string excitations corresponds to a single zero-dimensional fermion $\chi$, with minimal coupling to the two 4d bulk SYM theories, preserving eight supercharges. These couplings can be worked out from the dimensional reduction to zero dimensions of the (8,0) Lagrangian of a 2d Weyl spinor, leading to
\be
S_{\rm 0d} \sim \bar\chi \big(\Delta\phi^8 + i \Delta\phi^9 \big)\big|_P \chi \,,
\ee
where $\Delta\phi^8 + i \Delta\phi^9$ can be understood as arising from the reduction of a 2d gauge field in Euclidean space\footnote{The two scalars are constant values of the 2d gauge field along the two space directions. In Lorentzian 2d space the combination appearing would be $\Delta\phi^8 +  \Delta\phi^9$, but after Wick rotation this becomes $\Delta\phi^8 + i \Delta\phi^9$.}, and are identified in the brane picture with the relative motion between the D3 and D3'' in the directions $x^8$ and $x^9$ transverse to both branes. The scalar fields are evaluated at the point $P$ corresponding to the location of the D3-D3" intersection. In the low energy limit the 4d SYM theory on the D3" is non-dynamical and the vector multiplet complex scalars are set to background values that we take to be vanishing. The scalar combination $\Delta\phi^8 + i \Delta\phi^9 = \phi^8 + i \phi^9$ then corresponds to the 3d vector multiplet complex scalar $\varphi$. 
Integrating out the complex fermion $\chi$ in the path integral leads to the insertion of the local operator $\phi^8 + i \phi^9 \sim \varphi$ at the point $P$. 
\be
\int d\bar\chi d\chi \exp[\bar\chi \big(\phi^8 + i \phi^9 \big)\big|_P  \chi ] = \big(\phi^8 + i \phi^9 \big)\big|_P \sim \varphi(P) \,.
\ee
We conclude that the D3" brane is associated to the insertion of the local operator $\varphi$ in the low-energy limit.
Here we have assumed that the D3" interactions with the other branes in the configuration are irrelevant at low energies.

\subsection{Ring relations}
\label{ssec:Relations}

From the brane realizations of the Higgs and Coulomb branches operators of $T[SU(2)]$ described in the previous section, we can deduce the relations they obey. 
\medskip

\begin{figure}[th]
\centering
\includegraphics[scale=0.75]{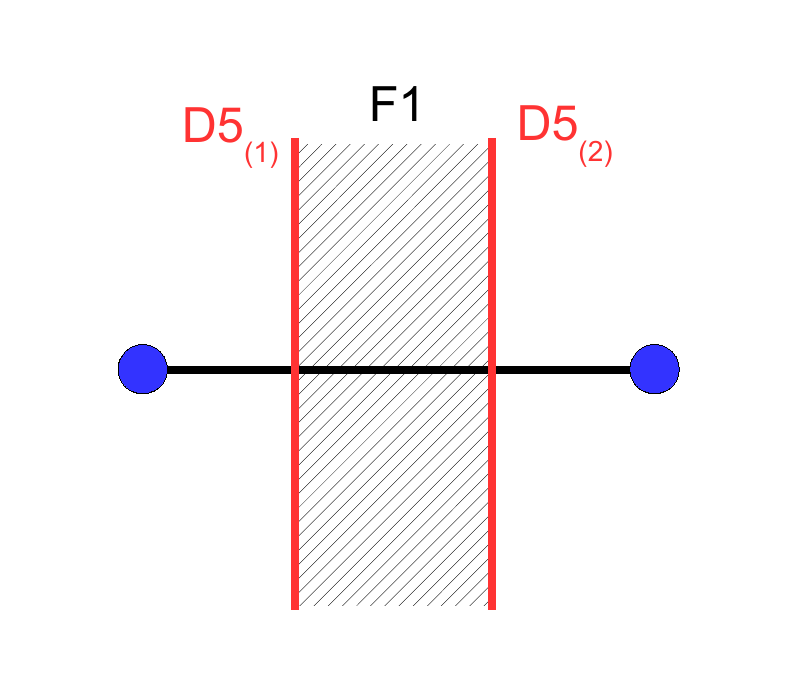}
\vspace{-0.5cm}
\caption{\footnotesize{Setup for the operator $- Z^2{}_1 Z^1{}_2$ or equivalently $-Z^1{}_1 Z^2{}_2$.}}
\label{TwoSemiF1s}
\end{figure}
First we explain how to find the relations of the Higgs branch, \eqref{HBRel1} and \eqref{HBRel2}, which we repeat here for convenience:
\be
\det Z \equiv Z^1{}_1 Z^2{}_2 - Z^2{}_1 Z^1{}_2  = 0 \,, \quad \tr Z \equiv Z^1{}_1 + Z^2{}_2 = 0 \,.
\label{HBRelTot}
\ee
The first relation follow trivially from the definition of the meson operators in terms of the hyper-multiplet scalars. However we will see that this trivial relation is mapped under mirror symmetry to a non-trivial quantum relation on the Coulomb branch of the dual $T[SU(2)]$ theory. Therefore, to provide a unified picture between Higgs branch and Coulomb branch, we wish to find a way to recover the relation from the brane picture (as we will do for the CB relation). For that purpose we consider the setup with an infinite F1 string stretched between the two D5s and intersecting the D3, as in Figure \ref{TwoSemiF1s}. This brane configuration can be interpreted in two ways. 
First we can see it as two semi-infinite F1s ending on the D3. This leads, according to the discussion above, to the insertion of the meson operators $Z^2{}_1$ and $-Z^1{}_2$, which means the insertion of the product $-Z^2{}_1Z^1{}_2$. 
Alternatively it can be seen as a single infinite F1 string, crossing the D3, in which case the insertion of meson operators can be associated to the two D5-D3 intersections with a full F1 ending on the D5. Re-using the arguments of section \ref{ssec:HBop} one finds that the F1 ending onright side of D5$_{(1)}$ inserts the operator $Z^1{}_1$ and the same F1 ending on the left side of D5$_{(2)}$ inserts the operator $-Z^2{}_2$. In section \ref{ssec:HBop} we had a configuration with infinite F1 extended to the left or to the right and realizing $-Z^1{}_1$ and $Z^2{}_2$ separately. Here we have a single F1 string respondible for the insertion the product of the two operators $-Z^1{}_1 Z^2{}_2$.
The first Higgs branch relation $\det Z =0$ follows from the identification of the two operator insertions: $-Z^2{}_1Z^1{}_2 = -Z^1{}_1 Z^2{}_2$. 
These two ``readings" of the same brane setup may seem artificial when one thinks in terms of (gauge non-invariant) scalars $Q^\alpha, \ti Q_{\alpha}$ insertions. The purpose of this discussion is to extract the rules to read ring relations directly in terms of gauge invariant operators. We will then show that parallel rules apply to Coulomb branch operators, making mirror symmetry transparent.

\smallskip

\begin{figure}[th]
\centering
\includegraphics[scale=0.73]{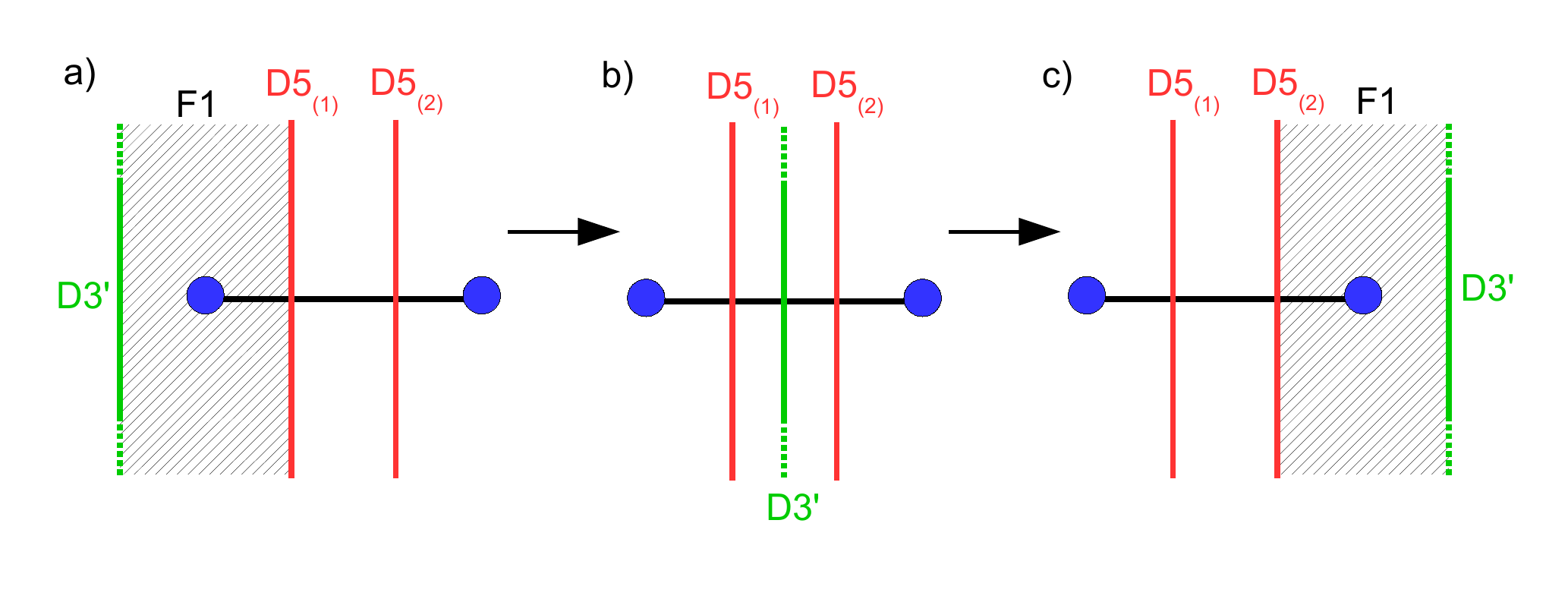}
\vspace{-1cm}
\caption{\footnotesize{a) Setup realizing $-Z^1{}_1$, with a D3' on the left of the configuration. b) Setup after moving the D3' inbetween the D5s. c) Setup after moving the D3' to the right, realizing $Z^2{}_2$.}}
\label{D3pMove}
\end{figure}

The second relation $\tr Z =0$ follows from the F-term constraint in the gauge theory. We find it by first considering the brane setup for the operator $-Z^1{}_1$ (Figure \ref{SemiF1_2}-a), with an infinite F1 extended to the left and ending on D5$_{(1)}$. We can then let the string end on a D3' brane far away to the left, without changing the operator insertion (and preserving the same supersymmetry), as in Figure \ref{D3pMove}-a. It was a crucial point in \cite{Hanany:1996ie} to argue that the brane moves along the $x^3$ direction leave invariant the low-energy physics on the D3 branes, as long as branes of the same type do not cross each other (e.g.~a D5 does not cross another D5). This property allows to derive 3d mirror symmetry from IIB string theory using S-duality and 5-brane moves along $x^3$. Here we use this property to move freely the D3' along $x^3$, across the brane configuration, all the way to a far region on the right. As it passes through the two D5s, HW F1-creation effects occur, as explained in section \ref{ssec:BranesLocalOp}: after crossing D5$_{(1)}$ the D3' stands inbetween the two D5s with no F1 ending on it and after crossing D5$_{(2)}$, there is an F1 string stretched between D5$_{(2)}$ and the D3'. This process is depicted in Figure \ref{D3pMove}. The end configuration is the one inserting the operator $Z^2{}_2$. Identifying the initial and final brane setups, we obtain the second relation $-Z^1{}_1= Z^2{}_2$. 
Therefore the F-term relation follows from a D3' brane move and involves the HW effects.

One may wonder what is the operator insertion corresponding to the brane setup with the D3' brane standing between the two D5s and crossing the D3 segment as in the middle figure in Figure \ref{D3pMove}. It is a nice and instructive exercise to study this configuration. The operator insertion in this case is obtained after integrating out the light modes of the D3'-D3 open strings. This brane system is almost identical to the D3-D3" intersection studied above. The D3' and D3 branes intersect at a point in space and the light open string modes correspond to a single zero-dimensional fermion $\chi$, whose zero dimensional action is
\be
S_{0d} \sim \bar\chi (\phi^5  + i \phi^6)\big|_P \chi \,,
\ee
where $P$ is the intersection point between the D3 and D3' and $\phi^5  + i \phi^6$ corresponds to the relative motion between the two branes in the direction $x^5 +i x^6$ and is identified simply with the D3 brane position along $x^5 +i x^6$ by placing the D3' at $(\phi^5  + i \phi^6)_{D3'}=0$. The scalar $\phi^5 + i \phi^6$ then corresponds to a complex scalar in the 4d SYM theory on the D3 brane, which belongs to the 3d hyper-multiplet with complex scalars ($A_3 + i \phi^4, \phi^5  + i \phi^6)$ under the embedding of the 3d $\N=4$ super-algebra into the 4d $\N=4$ super-algebra. In the infrared limit the boundary conditions imposed by the NS5 branes and the D5 branes, studied in \cite{Gaiotto:2008sa} are such that the complex scalar $\phi^5  + i \phi^6 \equiv \Phi$ vanishes at the NS5 brane positions, say at $x^3=0$ and $x^3=L>0$, $\Phi(0)=\Phi(L)=0$, and to obey the equation\footnote{We thank Davide Gaiotto for informative discussions on this point.}
\be
\p_3 \Phi + \delta(x^3_{(1)})Z^1_{1} + \delta(x^3_{(2)})Z^2_{2} = 0 \,,
\ee
with $x^3_{(\alpha)}$ the position of D5$_{(\alpha)}$. This problem admits a solution only if $Z^1_{1} + Z^2_{2} = 0$, which is another way to recover the F-term constraint, and the profile of the scalar $\Phi$ is then
\be
\Phi(x^3) = \left\lbrace
\begin{array}{cc}
0 & \,, \quad 0 < x^3 < x^3_{(1)} \,, \\
-Z^1_{1} & \,, \  x^3_{(1)} <  x^3 < x^3_{(2)} \,, \\
-Z^1_{1} - Z^2_{2} = 0 & \,, \quad  x^3_{(2)} <  x^3 < L \,.
\end{array}
\right.
\ee
The action for the 0d fermion $\chi$ is then
\be
S_{0d} \sim \bar\chi (-Z^1_{1})\big|_P \chi \,,
\ee
and produces, upon integrating out the fermion, the insertion of the meson operator $-Z^1_{1}(P) = Z^2_{2}(P)$ in complete agreement with the previous analysis. In this different point of view, the F-term constraint does not follow from identifying brane setups (which is our preferred point of view in this paper) but instead from solving the equations on the scalar $\Phi$ which becomes non-dynamical in the infrared limit. 

\medskip

\begin{figure}[th]
\centering
\includegraphics[scale=0.75]{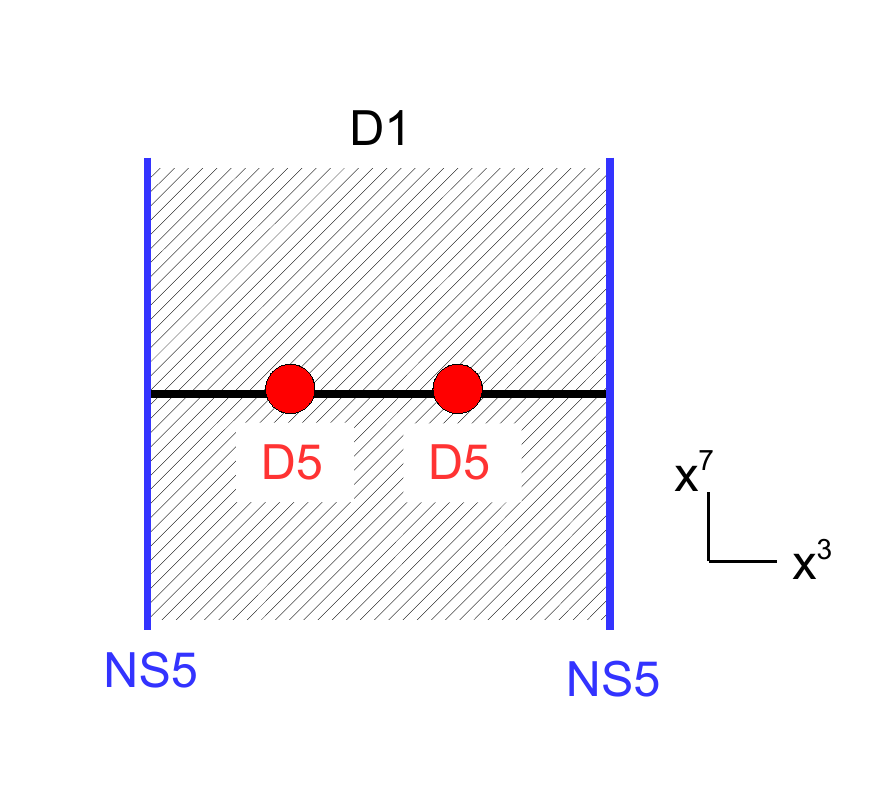}
\vspace{-1cm}
\caption{\footnotesize{A D1 brane stretched between the two NS5s, crossing the D3 brane.}}
\label{FullD1}
\end{figure}

We now move on to the Coulomb branch relations. There is actually a single relation for the $T[SU(2)]$ theory, found in \cite{Borokhov:2002cg}, given by
\be
u^{+} u^- = -\varphi^2 \,.
\label{CBRelBis}
\ee
This non-trivial quantum relation can be recovered from a rather simple analysis of the brane realizations inserting the Coulomb branch operators. The relevant brane setup has a infinite D1 brane stretched between the two NS5s and crossing the D3, as in Figure \ref{FullD1}. We can regard this configurations in two ways, that we detail now.

First we can see it as two semi-infinite D1s ending on the D3 from above and below respectively. According to the rules described before, this corresponds to the insertion of the product of two monopole operators $-u^+ u^-$. 

On the other hand we can regard the configuration as a single D1 stretched between the two NS5s and crossing the D3 and study the low-energy spectrum of the open strings stretched between the various branes. 
In the infrared limit, when the dynamic of the theory in the $x^3$ direction is frozen, the D1-D1 open string modes give rise to an abelian vector multiplet of 1d $\N=(4,4)$ supersymmetry\footnote{This is the dimensional reduction of a 6d $\N=(1,0)$ vector multiplet. It has five real scalars.} living on a line along the $x^7$ direction and intersecting the three-dimensional theory living on the D3 at a point.

The D1-D3 open strings give rise to a zero-dimensional hyper-multiplet \footnote{This is the dimensional reduction of a 6d hyper-multiplet.}  transforming in the bifundamental representation of $U(1)_{\rm 1d}\times U(1)_{\rm 3d}$, where $U(1)_{\rm 1d}$ denotes the abelian gauge symmetry of the theory on the D1 and $U(1)_{\rm 3d}$ that of the theory living on the D3. The hyper-multiplet is coupled to 0d $\N=(4,0)$ vector multiplets embedded in the 1d and 3d vector multiplets.  The coupling to the bosonic 1d and 3d fields is through a complex mass coupling with mass parameter $\varphi' - \varphi$, where $\varphi'$ combines two out of the five real scalars of the 1d vector multiplet. The difference $\varphi' - \varphi$ corresponds to the distance between the D1 and D3 branes along $x^8+ix^9$. In the configuration that we study, we impose that the D1 crosses the D3, which means that $\varphi'=\varphi$, making the hyper-multiplet massless. To regularize the path integral we need to remove the zero modes of this massless hypermultiplet from the path integral. A more detailed analysis of this hyper-multiplet theory will be given in Section \ref{sec:NATheories}.

The D1-D5$_{(\alpha)}$ open strings give rise to a single zero-dimensional fermion living at the intersection of the branes, with complex mass $\varphi - \varphi^{(D5)}$, associated to the distance between the D1 and the D5  along the $x^{8}+ix^{9}$ direction\footnote{The D1 is at the same position $\varphi$ as the D3 in $x^8 + i x^9$.} (this is similar to the D3-D3'' intersection studied above). Here the D5 is at the origin $\varphi^{(D5)}=0$, so that the 0d action is
\be
S_{\rm 0d ,(\alpha)} \sim \overline{\chi}_{(\alpha)} \varphi \chi_{(\alpha)} \,, \quad \alpha=1,2 \,.
\ee
In addition there can be cubic interactions between the D1-D3, D3-D5 and D5-D1 open string modes. We will assume that they do not affect the integration over the zero-dimensional and one-dimensional fields.

Integrating out the 1d vector multiplet and the 0d hyper-multiplet yields a factor independent of the 3d fields, so we neglect it. Integrating out the fermion $\chi_{\alpha}$ yields a factor $\varphi(P_{\alpha})$, where $P_{\alpha}$ is the intersection point of the D1 and D5$_{(\alpha)}$ branes. The total insertion after integrating out the zero-dimensional fields is then
\be
\varphi(P_{1})\varphi(P_{2}) \simeq \varphi(P)^2 \,, 
\ee
since the points $P_1$ and $P_2$ are indistinguishable in the low-energy limit $P_1 \simeq P_2 \equiv P$.

Identifying the two ``readings" of the brane setup, we obtain the relation \eqref{CBRelBis}.
Here we see that no brane move was necessary to get the relation. 

\smallskip

Finally we note that we made an implicit assumption in our analysis, which is that the presence of the D5 branes do not affect the monopole operator insertion in a configuration with a single semi-infinite D1 ending on the D3 (Figure \ref{SemiD1}-a,-b). 
Our analysis of the CB relation indicates that there is no effect due to these strings when the D1 is dissolved in the D3 to insert a monopole operator. On the contrary, when the D1 is flat (crossing the D3) the D1-D5 string modes play a crucial role in the operator insertion, as we discussed. One way to think about this phenomenon is that in the configuration inserting a monopole operator the D1 forms a spike ending on the D3. Then there is no distinction between D1-D5 strings and D3-D5 strings and therefore no additional light modes due to D1-D5 strings.

\subsection{Mirror symmetry}
\label{ssec:MirrorSym}

Three-dimensional $\N=4$ theories are subject to an infrared duality called mirror symmetry. The main statement is that pairs of mirror dual theories flow in the same (strongly coupled) infrared fixed point and that the Higgs branch of one theory is identified with the infrared quantum corrected Coulomb branch of the dual theory \cite{Intriligator:1996ex}. Since the HB and CB chiral rings are independent of the RG flow, we obtain the prediction that the HB chiral ring of one theory must match the CB chiral ring of the mirror dual theory. The duality swaps the $SU(2)_C$ and $SU(2)_H$ R-symmetry actions, the infrared enhanced topological symmetry $G_C$ and the flavor symmetry $G_F$, and the mass and FI deformation parameters.

The $T[SU(2)]$ theory is known to be self-dual under mirror symmetry, therefore its Higgs branch and Coulomb branch operators and ring relations should be mapped under the duality. In this simple theory, the map is easily found to be
\be\ba
\varphi \  &\leftrightarrow Z^{2}{}_2 (=-Z^1{}_1) \,, \cr
u^+ &\leftrightarrow Z^2{}_1 \,, \cr
u^- & \leftrightarrow Z^1{}_2 \,.
\label{TSU2MirrorMap}
\ea\ee
This identification maps the  CB quantum relation $u^+u^- = -\varphi^2$ to the HB trivial relation $Z^2{}_1Z^1{}_2 = Z^{2}{}_2Z^1{}_1$.
\medskip

The point of view that we develop in this paper is that the mirror map of operators can be found directly from the brane realizations. Indeed it is now well-known that mirror symmetry of the three dimensional theory follows from the action of type IIB S-duality on the brane realizations. In order to find the mirror map between operators, one can start with the brane setup realizing a local operator, then act with S-duality on the configuration and read off the mirror operator from the resulting brane setup. 

The action of S-duality on the various branes involved is described in Table \ref{tab:Sduality}. Here we combine S-duality with the space rotation $(x^{4,5,6}, x^{7,8,9}) \to (x^{7,8,9}, -x^{4,5,6})$, so that the branes always have orientations as in Table \ref{tab:orientationsAll}.
\begin{table}[h]
\begin{center}
\begin{tabular}{|c|ccccccc|}
  \hline
   Brane   & D3 & D5 & NS5 & F1 & D1 & D3' & D3"  \\  \hline
 S-dual brane  & D3 & NS5 & D5 & D1 & F1  & D3"  & D3'    \\ \hline
\end{tabular}
\caption{\footnotesize S-duality action on branes and strings.}
\label{tab:Sduality}
\end{center}
\end{table}

\begin{figure}[th]
\centering
\includegraphics[scale=0.7]{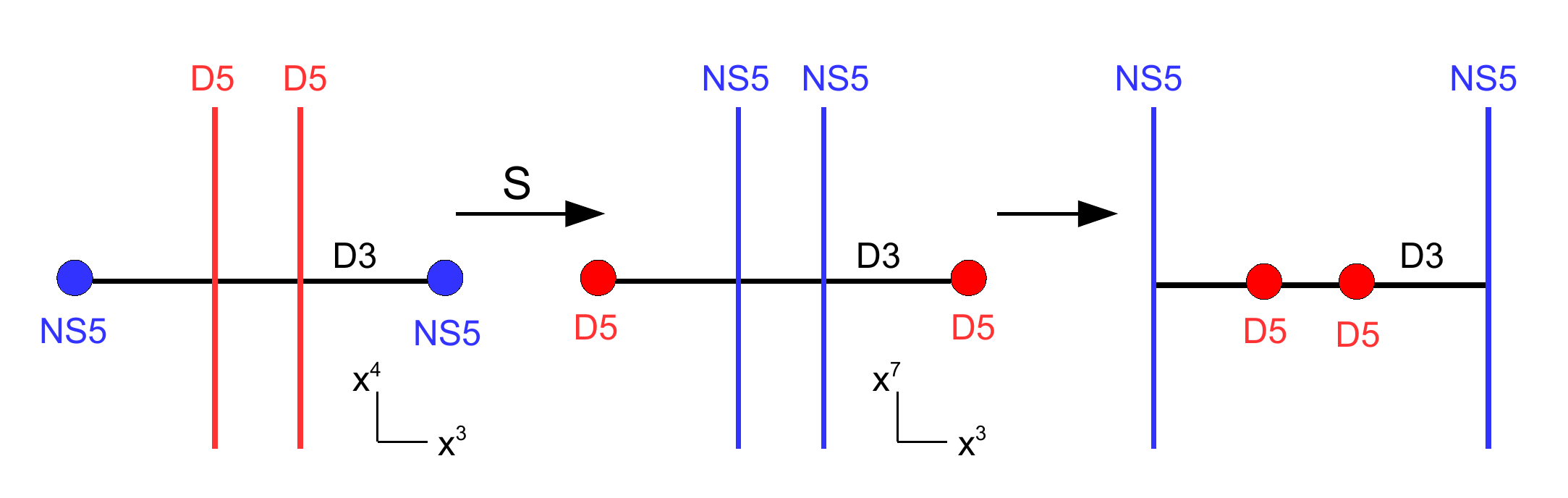}
\vspace{-0.5cm}
\caption{\footnotesize{Brane configuration of $T[SU(2)]$. After acting with the S transformation and moving the D5s to the middle, one recovers the initial configuration.}}
\label{TSU2Sduality}
\end{figure}

Let us see how this works for the $T[SU(2)]$ theory. In  the absence of local operator insertion the theory is realized by the brane configuration of Figure \ref{TSU2}. Applying S-duality, we obtain a configuration with a D3 stretched bewteen two D5s and crossing two NS5s. We then need to move the D5s to the middle of the configurations, so that each D5 passes across an NS5. Taking into account the HW brane creation effect (see Section \ref{ssec:BranesLocalOp}) we recover the initial configuration, confirming that the $T[SU(2)]$ theory is self-dual under mirror symmetry. This process is depicted in Figure \ref{TSU2Sduality}.
\medskip

\begin{figure}[th]
\centering
\includegraphics[scale=0.67]{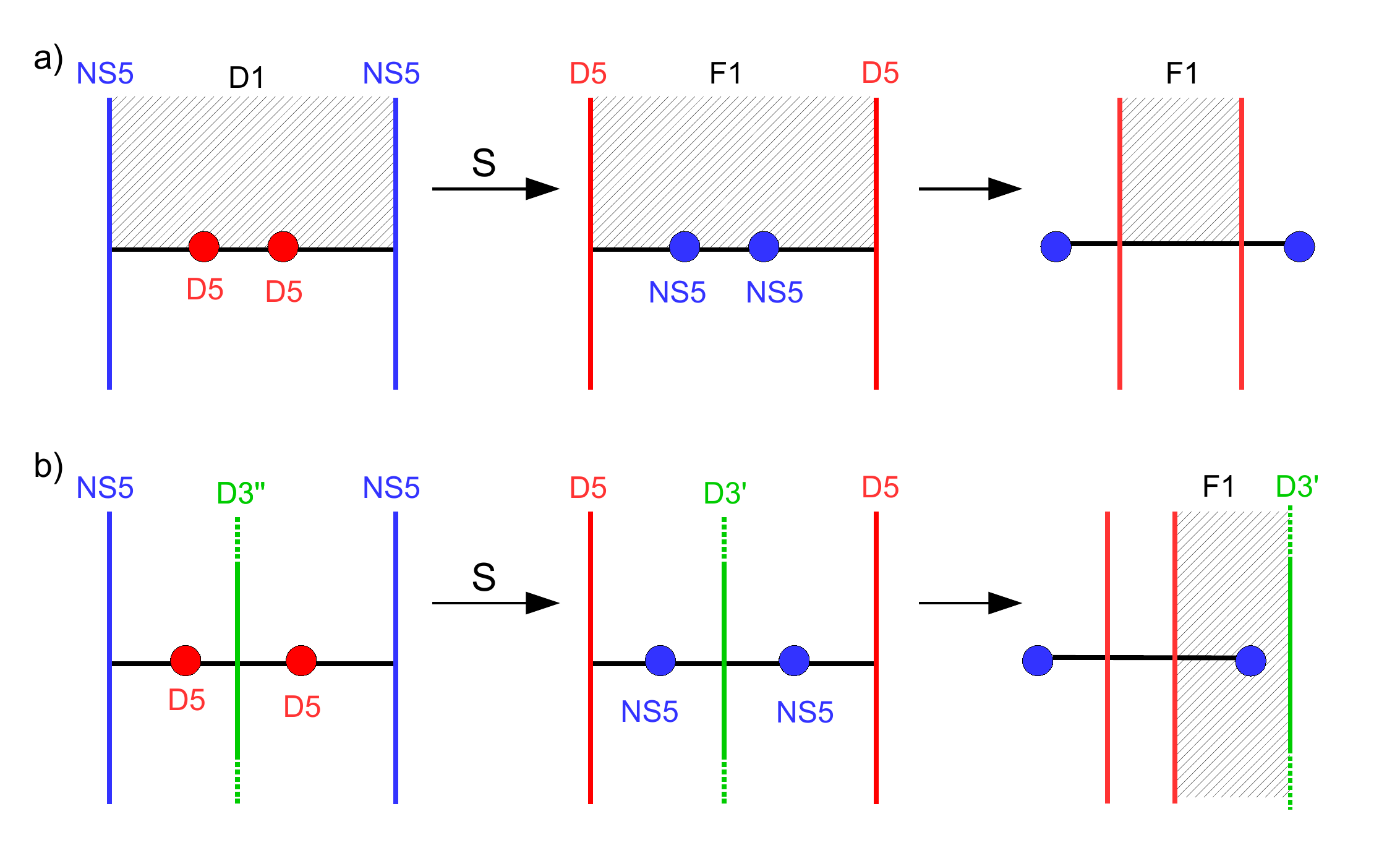}
\vspace{-0.5cm}
\caption{\footnotesize{a) Starting from the $u^+$ insertion setup and applying S-duality leads the $Z^2{}_1$ insertion setup, after brane rearrangements. b) Starting from the $\varphi$ insertion setup  and applying S-duality leads to the $Z^2{}_2$ insertion setup, after brane rearrangements. }}
\label{D1intoF1}
\end{figure}

 The brane setup realizing the monopole operator $u^+$ is that of Figure \ref{SemiD1}-a. Acting with the S-transformation and moving the D5s to the middle, as in Figure \ref{D1intoF1}-a, we directly obtain the setup realizing the meson operator $Z^2{}_1$. Similarly starting from the setup of Figure \ref{SemiD1}-b, realizing $-u^-$, we find that the S-dual setup is that of the operator $-Z^1{}_2$. We therefore obtain immediately the mirror map involving the monopole operators. 
 For the insertion of the scalar operator $\varphi$, we start with the configuration of Figure \ref{SemiD1}-c, which has a D3" brane crossing the D3. Acting with S-duality and moving the D5s we obtain the configuration of Figure \ref{D1intoF1}-b, which has a D3' crossing the D3. As argued in section \ref{ssec:Relations}, this setup realizes the insertion of the meson operator $Z^2{}_2$ or equivalently $-Z^1{}_1$. This can be seen by moving the D3' to the right of the configuration, leading to the creation of an F1 string ending on $D5_{(2)}$ (Figure \ref{SemiF1_2}-b and \ref{D1intoF1}-b), or to the left, leading to the creation of an F1 string ending on $D5_{(1)}$ (Figure \ref{SemiF1_2}-a). This reproduces the mirror map \eqref{TSU2MirrorMap}.
In addition, we observe that the brane configurations with two interpretations leading to the HB and CB relations are S-dual to each other.

\subsection{Deformations}
\label{ssec:Deformations}

To complete the analysis of the $T[SU(2)]$ theory, we discuss the complex mass and Fayet-Iliopoulos deformations.
\medskip

The mass deformations are simpler to understand. In the field theory there are three real mass deformations for each hypermultiplet, transforming as a triplet of $SU(2)_C$. They can be seen as background values of scalars in vector multiplets gauging the flavor symmetries.
The mass deformations lift the Higgs branch and modify the Coulomb branch geometry. For each triplet of masses, two out of the three parameters combine into a complex mass $m_{\bC}$\footnote{This choice is correlated to the choice of complex structure on the Coulomb branch (see Section \ref{sec:ModSpaceReview})}. In the $T[SU(2)]$ theory there are two such complex masses $m_\alpha$, $\alpha=1,2$, for the two hyper-multiplets, and the CB relation becomes \cite{Bullimore:2015lsa}
\be
u^+ u^- = -(m_1 - \varphi)(m_2- \varphi) \,.
\label{DeformedCBRel}
\ee
Simultaneous shift of complex masses by the same constant can be re-absorbed into a redefinition of the complex scalar $\varphi$, so that only the parameter $m_1-m_2$ is physical.

In the brane setup the complex mass deformations are associated to D5 brane displacements along the $x^8$ and $x^9$ directions. Denoting $x^{8,9}_\alpha$, $\alpha=1,2$, the positions of the two D5s, we have $m_\alpha \equiv  \varphi^{(D5)}_\alpha = x^8_\alpha + i x^9_\alpha$. 
It is easy to correct the derivation of the CB relation from the brane picture. In the second interpretation of the setup of Figure \eqref{FullD1} (see Section \ref{ssec:Relations}), considering the D1 crossing the D3, the zero-dimensional fermion $\chi_{\alpha}$ living at the intersection of the D1 and D5$_{(\alpha)}$ branes now has mass $m_\alpha - \varphi$, corresponding to the distance between the D1 and the D5 in the $x^8$ and $x^9$ directions,
\be
S_{\rm 0d ,(\alpha)} \sim \overline{\chi}_{(\alpha)} ( m_\alpha - \varphi) \chi_{(\alpha)} \,, \quad \alpha=1,2 \,.
\ee
Integrating out the two fermions yields the operator insertion
\be
(m_1 - \varphi)(m_2- \varphi) \,,
\ee
leading to the deformed relation \eqref{DeformedCBRel}.

\bigskip

\begin{figure}[th]
\centering
\includegraphics[scale=0.7]{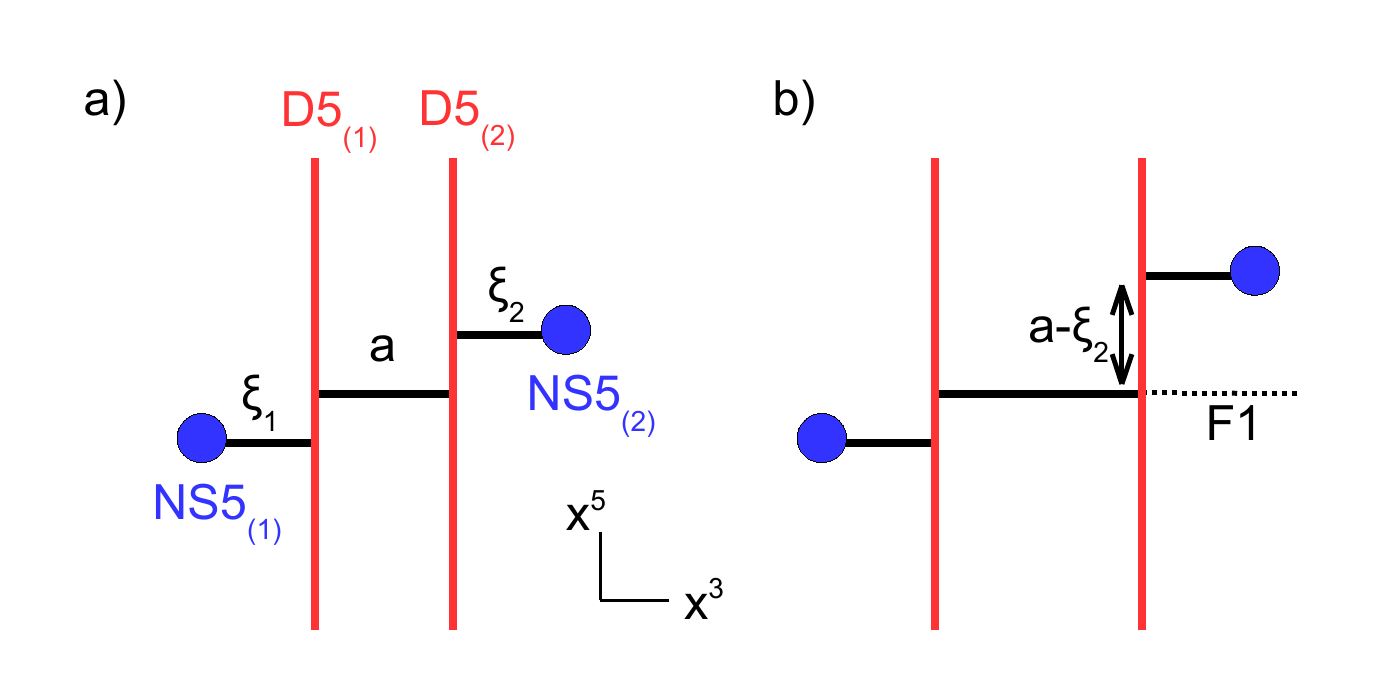}
\vspace{-0.5cm}
\caption{\footnotesize{a) Configuration associated to the FI deformation $\eta=\xi_1-\xi_2$ ($x^{3,5}$ plane). b) Brane setup inserting the operator $Y =Z^2{}_2 + \xi_2$.}}
\label{EtaDeform}
\end{figure}

Let us turn to the FI deformations. Those lift the Coulomb branch and deform the Higgs branch. They are parametrized by three real parameters, transforming as a triplet of $SU(2)_H$. Two out of the three deformations combine into a complex FI parameter $\eta_{\bC}$ which affects the F-term relations in the Higgs branch chiral ring. For the $T[SU(2)]$ theory there is a single complex FI parameter $\eta$ and the HB relations are
\be
\tr Z = \eta \,, \quad \det Z = 0 \,.
\label{DeformedHBRel}
\ee
 In the brane picture the $\eta$ deformation can be identified with displacements of the NS5s along the $x^5$ and $x^6$ directions. More precisely, denoting $x^{5,6}_i$, $i=1,2$, the positions of the two NS5s, we can define $\xi_i = x^5_i + i x^6_i$ and the complex FI parameter is $\eta = \xi_1 - \xi_2$. The modified setup is depicted in Figure \ref{EtaDeform}-a.

The modification of the brane analysis due to NS5 displacements is not straightforward to understand, as it is often the case with NS5 brane effects.
We can give a heuristic argument leading to the conclusion that the displacements affect the operator insertions, rather than the reading of the HB relations. One can imagine giving a vev to the meson operators $Z^2{}_2$ and $Z^1{}_1$. This corresponds in the brane picture to moving the D3 segment between the two D5s along the $x^{5,6}$ directions.  
In the absence of FI deformations, the D3 displacement along $x^{5,6}$ to the position $a = x^5 + i x^6$ corresponds to giving vevs $\vev{Z^2{}_2} = -\vev{Z^1{}_1} = a$ \footnote{This follows from the analysis in \cite{Gaiotto:2008sa}. It can also be seen as the mirror-dual operation of giving a vev $a$ to $\varphi$ by moving the D3 between the two NS5s along $x^{8,9}$. Indeed $\varphi$ is the mirror-dual operator to $Z^2{}_2 (= -Z^1{}_1)$ in the absence of FI deformation.}, namely giving a vev $a$ to the operator inserted by the setups of Figure \ref{SemiF1_2}-a and -b. The origin of the moduli space, $a = 0$, corresponds to the D3 segment in the middle being aligned with the two external D3 segments.

Let us focus on the brane setup of Figure \ref{EtaDeform}-b  and let us denote $Y$ the operator inserted by the F1 string ending on the right side of D5$_{(2)}$ and aligned with the middle D3 segment. Moving the middle D3 segment to a position $a$ is now associated to giving a vev to the operator $\scO$, $\vev{Y} =a$. When $a=\xi_2$, the middle D3 segment is aligned with the D3 segment stretched between D5$_{(2)}$ and NS5$_{(2)}$ (the NS5 on the left). Locally this situation is identical to the situation without FI deformation when we sit at the origin of the moduli space and must correspond to having $\vev{Z^2{}_2}= 0$, namely a vanishing vev for the meson sourced by D3-D5$_{(2)}$ open string modes. This suggests the identification 
\be
Z^2{}_2 = Y - \xi_2 \,.
\ee
A corresponding reasoning, applied to the brane setup of Figure \ref{SemiF1_2}-a, with the F1 string ending on D5$_{(1)}$, which also inserts the operator $Y$, leads to the identification
\be
-Z^1{}_1 = Y - \xi_1 \,.
\ee
The relations following from the brane realizations, as described in Section \ref{ssec:Relations}, with these deformed insertions are
\be
Z^2{}_2 +\xi_2 = -Z^1{}_1 +\xi_1 \equiv  Y  \,, \quad  -Z^2{}_1 Z^1{}_2 =  -Z^2{}_2 Z^1{}_1 \,,
\ee
where the second relation is the same as before. These relations match the deformed relations \eqref{DeformedHBRel}. 

There is an alternative and more robust way to understand these operator insertions by studying the brane configuration with a D3' brane between the two D5s, intersecting the D3 (middle figure in Figure \ref{D3pMove}), which is obtained from the above brane setup by HW moves. This setup was analyzed in Section \ref{ssec:Relations} using the results of \cite{Gaiotto:2008sa}, with the operator insertion following from integrating out a 0d fermion with complex mass $-Z^1{}_{1} = Z^2{}_{2}$. The FI deformation changes the analysis by modifying the boundary conditions on the scalar $\Phi$ to $\Phi(0) = \xi_1$ and $\Phi(L) = \xi_2$. Requiring the existence of a solution for $\Phi(x^3)$ imposes the modified F-term constraint $\xi_1 -Z^1{}_{1} - Z^2{}_{2} = \xi_2$, and the operator insertion is $\xi_1 -Z^1{}_{1} = \xi_2 + Z^2{}_{2} \equiv Y$, in agreement with our heuristic derivation.

\medskip

Mirror symmetry now related the $T[SU(2)]$ theory deformed by masses, with lifted Higgs branch and deformed Coulomb branch, to a dual $T[SU(2)]$ theory deformed by FI terms, with lifted Coulomb branch and deformed Higgs branch, with the map of operators
\be\ba
\varphi \  &\leftrightarrow Y \,, \cr
u^+ &\leftrightarrow Z^2{}_1 \,, \cr
u^- & \leftrightarrow Z^1{}_2 \,.
\label{DeformedMirrorMap}
\ea\ee
Note that this map is found after solving for the F-term relation on the HB side.
The second Higgs branch relation becomes
\be
Z^2{}_1 Z^1{}_2 = -(Y-\xi_1)(Y-\xi_2) \,,
\ee
which is mapped to the CB relation, through the usual map between masses and FI parameters $(m_1,m_2) \leftrightarrow (\xi_1,\xi_2)$.

\bigskip

We have now completed our study of the $T[SU(2)]$ theory. We have developed most of the tools needed to analyse moduli spaces from brane realizations and we are in a position to study more sophisticated theories.

\section{Abelian generalizations}
\label{sec:AbelGen}

In this section, we extend the analysis of the vacuum moduli spaces from brane configurations to more sophisticated abelian theories. We consider the abelian theory with $N_f$ fundamental hyper-multiplets, or $\N=4$ SQED, and its mirror dual theory, which is an abelian quiver theory. We show that the Coulomb and Higgs branch operators and the ring relations are correctly reproduced by following the brane reading rules found in the previous section and some new rules that we derive. We provide the mirror map of operators using S-duality of the brane picture. We illustrate our procedure in another couple of mirror dual theories.
Applications to arbitrary abelian quivers should then be straightforward. The final dictionary between brane setups and operator insertions, as well as the mirror map between HB and CB operators, are provided in Appendix \ref{app:Dictionary}.

\subsection{SQED}
\label{ssec:SQED}

\begin{figure}[th]
\centering
\includegraphics[scale=0.75]{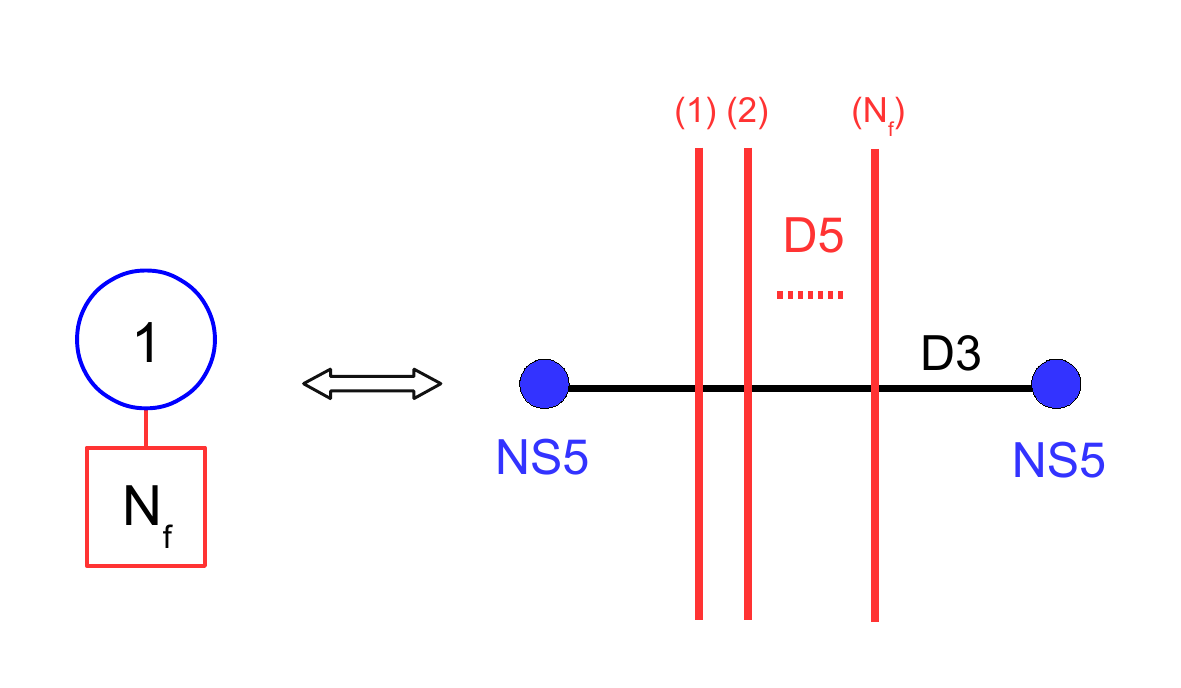}
\vspace{-0.5cm}
\caption{\footnotesize{Quiver and brane configuration of $\N=4$ SQED. There are $N_f$ D5 branes.}}
\label{SQED}
\end{figure}

We consider the $\N=4$ SQED theory. It has $U(1)$ gauge group and $N_f$ fundamental hyper-multiplets, with complex scalars $(Q^\alpha, \ti Q_\alpha)$, $\alpha=1,\cdots ,N_f$. The quiver diagram and the brane realization are shown in Figure \ref{SQED}. In particular, there are now $N_f$ D5s crossing the D3, sourcing the hyper-multiplets. We will denote them D5$_{(\alpha)}$ with $\alpha=1, \cdots, N_f$, labeling the branes from left to right.

\subsubsection{Higgs branch}
\label{sssec:SQEDHB}

The Higgs branch of the theory is  generated by the meson operators
\be
Z^\alpha{}_\beta = \ti Q_\beta Q^\alpha \,, \quad \alpha,\beta=1,\cdots , N_f \,,
\ee
which satisfy by definition
\be
Z^\alpha{}_\beta Z^\gamma{}_\delta - Z^\alpha{}_\delta Z^\gamma{}_\beta  = 0 \,, \quad \text{for all} \quad \alpha,\beta,\gamma,\delta \,,
\label{HBRelSQED1}
\ee
and are subject to the F-term constraint:
\be
\tr Z \equiv \sum_\alpha Z^\alpha{}_\alpha  = 0 \,.
\label{HBRelSQED2}
\ee
The relations \eqref{HBRelSQED1} can be recast as rank$(Z) \le 1$.
  \begin{figure}[th]
\centering
\includegraphics[scale=0.7]{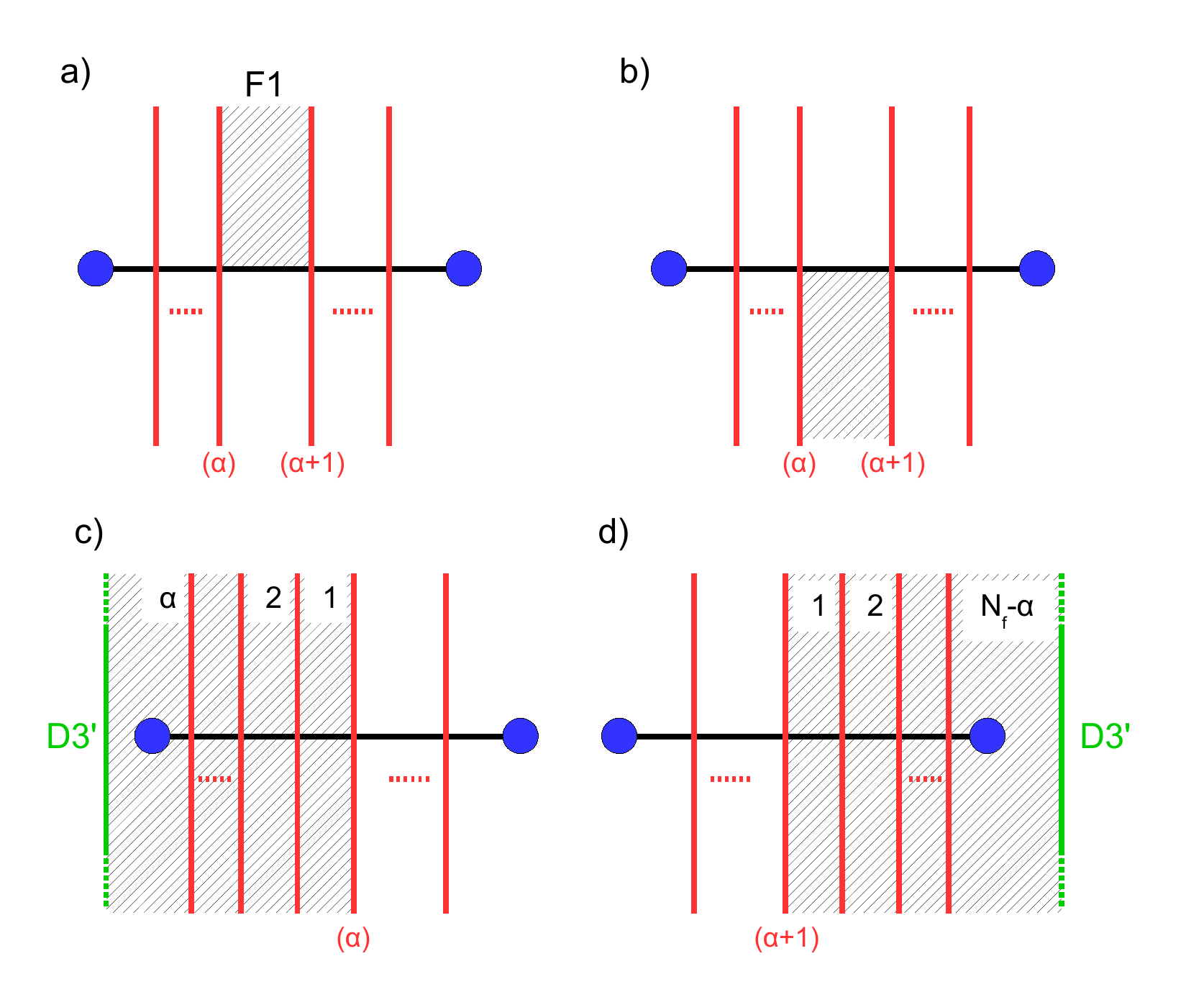}
\vspace{-0.5cm}
\caption{\footnotesize{Brane setups inserting the operators: a) $Z^{\alpha+1}{}_{\alpha}$; b) $-Z^\alpha{}_{\alpha+1}$; c) $-Z_{[1:\alpha]}$; d) $Z_{[\alpha+1:N_f]}$. The integers 1,2, ...,$\alpha$ indicate the number of superposed F1 strings in each region.}}
\label{SQEDSemiF1}
\end{figure}

It will be useful to introduce 
\be
Z_{[\alpha:\beta]} \equiv \sum_{\gamma=\alpha}^\beta Z^\gamma{}_\gamma \,.
\ee
The mesons $Z^\alpha{}_\alpha$ can be traded for the operators $Z_{[1:\alpha]}$ or $Z_{[\alpha+1:N_f]}$.
The mesons naturally realized in terms of brane setups are of the type $Z^{\alpha+1}{}_\alpha$, $Z^\alpha{}_{\alpha+1}$, $Z_{[1:\alpha]}$ and $Z_{[\alpha+1:N_f]}$ (no sum over $\alpha$), with the following dictionary:
\begin{itemize}
\item  The insertions of the meson operators $Z^{\alpha+1}{}_\alpha$ and $-Z^\alpha{}_{\alpha+1}$ are realized  by adding a semi-infinite F1 string stretched beween D5$_{(\alpha)}$ and D5$_{(\alpha+1)}$, and ending on the D3 from above and below respectively, as in Figure \ref{SQEDSemiF1}-a,b.
\item The insertion of the meson operator $-Z_{[1:\alpha]}$ is realized by adding a D3' brane on the left of the brane configuration and one F1-string stretched between the D3' and D5$_{(\beta)}$, for all $\beta\in [1,\alpha]$, so that there is a total of $\alpha$ F1s in the setup. This is described in Figure \ref{SQEDSemiF1}-c
\item The insertion of $Z_{[\alpha+1:N_f]}$ is realized by adding a D3' brane on the right of the brane configuration and one F1-string stretched between the D3' and D5$_{(\beta)}$, for all $\beta\in [\alpha+1,N_f]$, so that there is a total of $N_f-\alpha$ F1s. This is described in Figure \ref{SQEDSemiF1}-d.
\end{itemize}
These brane setups are simple generalizations of the one studied in Section \ref{ssec:HBop} for the $T[SU(2)]$ theory, which corresponds to the case $N_f=2$. \footnote{From our previous discussions, it is not obvious why the configuration \ref{SQEDSemiF1}-c (and similarly \ref{SQEDSemiF1}-d) realizes the insertion of the sum of meson operators $\sum_{\gamma=1}^\alpha Z^\gamma{}_\gamma$ and not the product of these mesons. However this turns out to be consistent with our general analysis. Moreover we will find in Section \ref{sec:OtherOp} different brane setups realizing the insertions of products of mesons.}

One important comment about these setups is that in each case there is a single way of interpreting the configuration, namely there is a single way to describe which brane ends on which other branes. This is obvious for the setups of Figure \ref{SQEDSemiF1}-a,b. For the setups of Figure \ref{SQEDSemiF1}-c,d, this follows from the {\it s-rule} which imposes that there is at most a single F1 string stretched between a D3' and a D5. This implies that the F1 cannot break into several pieces ending on both sides of some D5. 
Since there is a single way of interpreting these brane setups, they must insert operators which cannot be generated by products of operators, so they must belong to a basis of the HB chiral ring.

There is however a puzzle since the operators listed above are not enough to generate the whole Higgs branch ring, namely we are missing the operators $Z^\alpha{}_\beta$ with $|\alpha-\beta|>1$. We will see shortly that these operators appear in brane setups related to the HB ring relations.

\bigskip

\noindent{\bf Ring relations}:
\medskip

We observe immediately that the F-term relation \eqref{HBRelSQED2} follows from considering the configuration realizing $-Z_{[1:\alpha]}$, i.e. Figure \ref{SQEDSemiF1}-c, and moving the D3' brane from the left to the right in the configuration. Taking into account Hanany-Witten F1 creation effect, we end up with the configuration of Figure \ref{SQEDSemiF1}-d, realizing the $Z_{[\alpha+1:N_f]}$ insertion. We therefore obtain the relation
\be
-Z_{[1:\alpha]} = Z_{[\alpha+1:N_f]} \,,
\ee
which is nothing but $\tr Z =0$, for any chosen $\alpha$.

\medskip

\begin{figure}[th]
\centering
\includegraphics[scale=0.7]{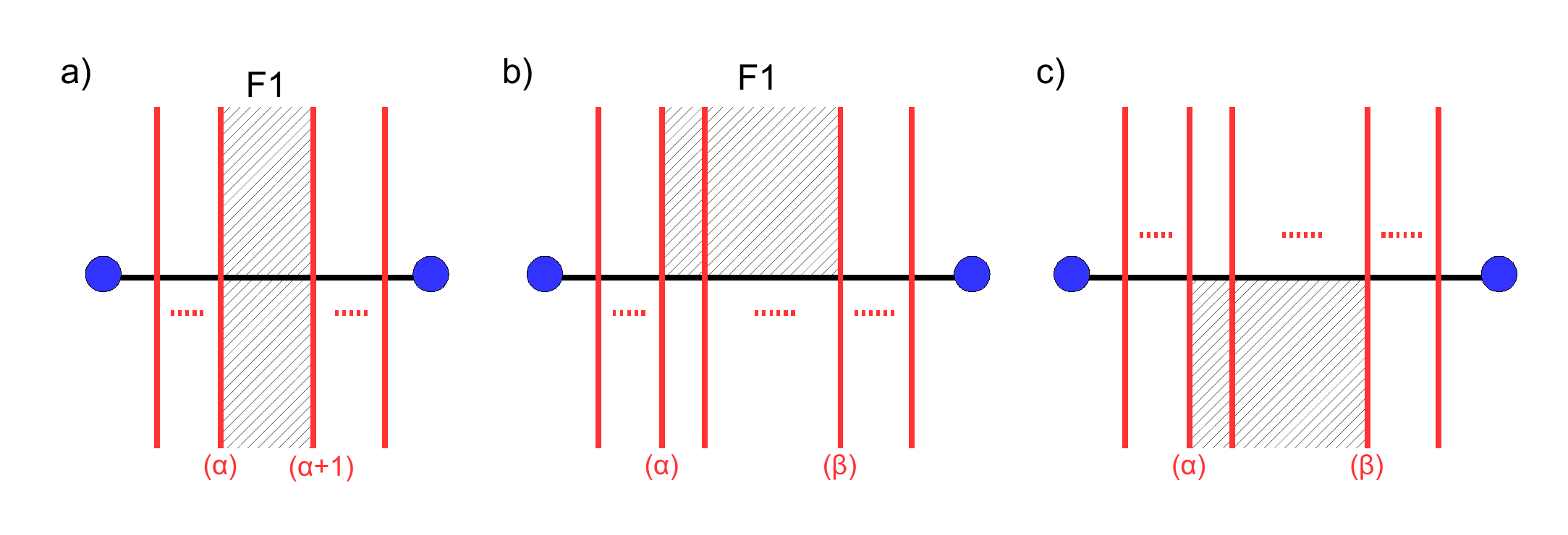}
\vspace{-0.5cm}
\caption{\footnotesize{a) Setup realizing $-Z^\alpha{}_{\alpha} Z^{\alpha+1}{}_{\alpha+1}$ or $-Z^{\alpha+1}{}_\alpha Z^{\alpha}{}_{\alpha+1}$. b) Setup realizing $Z^\beta{}_\alpha  \prod\limits_{\alpha<\gamma<\beta} Z^\gamma{}_\gamma$. c) Setup realizing $(-)^{\alpha+\beta+1} Z^\alpha{}_\beta  \prod\limits_{\alpha<\gamma<\beta} Z^\gamma{}_\gamma$. The F1 can split along the D5s, leading to alternative interpretations of the operator insertions.}}
\label{SQEDSemiF1_2}
\end{figure}

The other relations in the chiral ring \eqref{HBRelSQED2} follow from interpreting in several ways configurations with a semi-infinite or full F1 string stretched between two D5s, presented in Figure \ref{SQEDSemiF1_2}-a,b,c.

The brane setup of Figure \ref{SQEDSemiF1_2}-a is analogous to the one studied in Section \ref{ssec:Relations} and leads by the same reasoning to the relations
\be
Z^{\alpha+1}{}_\alpha Z^{\alpha}{}_{\alpha+1} = Z^\alpha{}_{\alpha} Z^{\alpha+1}{}_{\alpha+1}\,, 
\label{HBRel000}
\ee
 for all $\alpha$. These are part of the HB relations \eqref{HBRelSQED1}.

The setup of Figure \ref{SQEDSemiF1_2}-b has an F1 stretched between D5$_{\alpha}$ and D5$_{\beta}$, with $\alpha < \beta$, and ending on the D3 from above.
The analysis of the meson operator insertions in Section \ref{ssec:HBop} and the discussion of Section \ref{ssec:Relations} leading to the ``trivial" Higgs branch relation\footnote{We refer to the Higgs branch meson relations which follow from rearrangements of the elementary hyper-multiplet scalar fields as trivial. In terms of gauge invariant operators, these relations are not trivial at all.} can be adapted to the present situation. The setup of Figure \ref{SQEDSemiF1_2}-b can be interpreted as inserting the operator
\be
Z^\beta{}_\alpha  \prod\limits_{\alpha<\gamma<\beta} Z^\gamma{}_\gamma \,,
\ee
where the operator $Z^\beta{}_\alpha$ comes from the two F1 corners with D5$_{(\alpha)}$ and D5$_{(\beta)}$ respectively, and the operators $Z^{\gamma}{}_\gamma$, with $\alpha < \gamma <\beta$,  come from the F1 edges crossing the D5$_{(\gamma)}$ branes. We will say that $Z^\beta{}_\alpha$ is the contribution of an F1 stretched between D5$_{(\alpha)}$ and D5$_{(\beta)}$, ending on the D3 from above, and that $Z^{\gamma}{}_\gamma$ is the contribution of the D3-D5$_{(\gamma)}$ intersection with an F1 ending on the D3 from above.

Alternatively we can think of the setup as having $\beta -\alpha$ semi F1 strings, with one string stretched between D5$_{(\gamma)}$ and D5$_{(\gamma+1)}$, for $\gamma= \alpha, \cdots, \beta-1$. This corresponds to the insertion of the operator product
\be
\prod_{\gamma=\alpha}^{\beta-1} Z^{\gamma+1}{}_{\gamma} \,.
\ee
We therefore get the relation
\be
Z^\beta{}_\alpha  \prod\limits_{\alpha<\gamma<\beta} Z^\gamma{}_\gamma = \prod_{\gamma=\alpha}^{\beta-1} Z^{\gamma+1}{}_{\gamma} \,,
\label{HBRel0}
\ee
for each pair $(\alpha,\beta)$, with $\alpha<\beta$.
There are even more ways to read the setup of Figure \ref{SQEDSemiF1_2}-b, each corresponding to some splitting of the F1 string into several pieces ending on D5s. They lead to redundant relations.

The same considerations applied to the setup of Figure \ref{SQEDSemiF1_2}-c lead to the relations\footnote{In our conventions, each insertion comes with a minus sign and an overall factor $(-1)^{\alpha+\beta+1}$ drops from both sides of the relation.}
\be
 Z^\alpha{}_\beta  \prod\limits_{\alpha<\gamma<\beta} Z^\gamma{}_\gamma =  \prod_{\gamma=\alpha}^{\beta-1} Z^{\gamma}{}_{\gamma+1} \,,
\label{HBRel00}
\ee
for each pair $(\alpha,\beta)$, with $\alpha<\beta$.

Together, the relations \eqref{HBRel000}, \eqref{HBRel0} and \eqref{HBRel00} imply all the Higgs branch relations \eqref{HBRelSQED1}, as we show in Appendix \ref{app:SQEDHBRel}, up to one caveat. The caveat in the derivation of \eqref{HBRelSQED1} is that at some point in the computations we need to divide by products of $Z^\alpha{}_\alpha$ operators. This is valid only when these are non-zero, so strictly speaking we need to add this extra ingredient, or rule, to our derivation of the relations, saying that operators $Z^\alpha{}_\alpha$ appearing on both sides of a relation can be suppressed. This means that the resulting relations remain valid even at $Z^\alpha{}_\alpha=0$.  
To avoid confusions we will call the relations \eqref{HBRel000}, \eqref{HBRel0} and \eqref{HBRel00} the {\it pre-relations}, indicating that these are not yet the full set of ring relations, except in special cases (like in the $T[SU(2)]$ theory). From the pre-relations, the ring relations are uniquely determined by considering all the relations generated by the pre-relations and suppressing $Z^\alpha{}_\alpha$ operators appearing on both sides of a relation.
In the following we will derive the pre-relations from the brane setups and match them under mirror symmetry. This is equivalent to matching the full set of ring relations.

We have thus recovered the HB relations from the brane analysis. Again the ``trivial" relations follow from different readings of some brane setups, while the F-term relation follow from identifying brane setups after D3' moves.

\bigskip

As for the $T[SU(2)]$ theory, the Higgs branch can be deformed by FI terms. Turning on a FI term with complex parameter $\eta$, appropriate to the chosen complex structure on the Higgs branch, the F-term relation gets deformed to 
\be
\tr Z = \eta \,.
\ee
The counterpart in the brane picture is as in the $T[SU(2)]$ theory: the deformation corresponds to displacements of the NS5s along the $x^{5,6}$ directions, with $\eta = \xi_1 - \xi_2$ the difference between the two brane positions. The argumentation in Section \ref{ssec:Deformations}  leads in the present situation to a modification of the operator insertions for the brane setups of Figure \ref{SQEDSemiF1_2}-c and -d, which become $-Z_{[1:\alpha]} + \xi_1$ and $Z_{[\alpha+1:N_f]} + \xi_2$ respectively. The relation following from moving the D3' brane across the configuration becomes
\be
-Z_{[1:\alpha]} + \xi_1 = Z_{[\alpha+1:N_f]} + \xi_2  \equiv Y_\alpha \,,
\ee
reproducing the deformed F-term relation, for any $\alpha$.

The dictionary between brane setups and HB operator insertions is summarized in Appendix \ref{app:Dictionary}.

\subsubsection{Coulomb branch}
\label{sssec:SQEDCB}

The Coulomb branch of $\N=4$ SQED is simpler than the Higgs branch. The chiral ring is generated by the monopole operators $u^{\pm}$ of monopole charge $\pm 1$ respectively and the complex scalar $\varphi$, subject to the quantum relation
\footnote{Our conventions differ from those in \cite{Bullimore:2015lsa} by reversing the sign of the complex scalars $\varphi$ and complex masses $m_\alpha$.}
\be
u^+ u^- = - \prod_{\alpha=1}^{N_f} (m_\alpha - \varphi) \,,
\label{CBRelSQED}
\ee
where we have included the complex mass deformations $m_\alpha$ for the $N_f$ hyper-multiplets.

\medskip

\begin{figure}[th]
\centering
\includegraphics[scale=0.73]{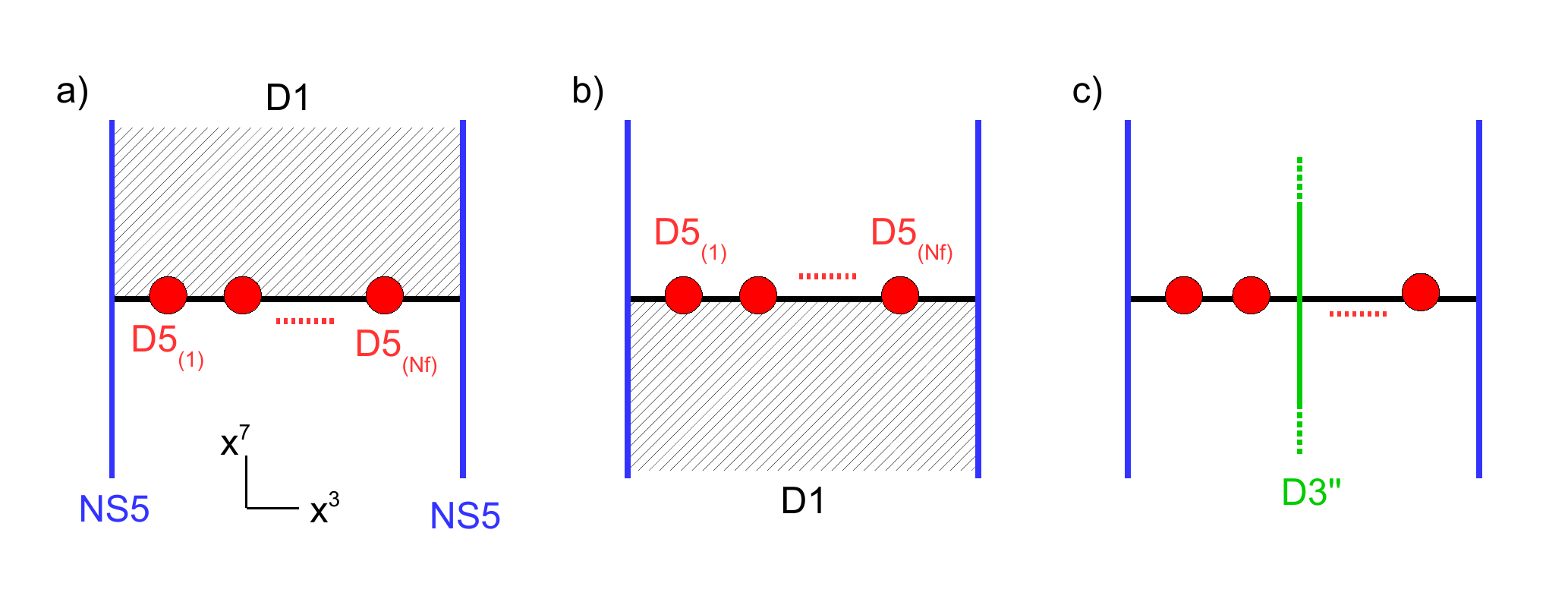}
\vspace{-0.8cm}
\caption{\footnotesize{a) Semi infinite D1 ending on the D3 from above, inserting the monopole operator $u^+$. b) Semi infinite D1 ending on the D3 from below, inserting the monopole operator $-u^-$. c) D3" brane crossing the D3 at an arbitrary position in $x^3$ between the two NS5s, inserting the complex scalar operator $\varphi$.}}
\label{SQEDSemiD1}
\end{figure}

The insertions of the operators $\pm u^{\pm}$ and $\varphi$ are realized as for the $T[SU(2)]$ theory, with the only difference that there are $N_f$ D5s instead of two D5s. The brane setups are shown in Figure \ref{SQEDSemiD1}-a,b,c. For the insertion of the operator $\varphi$ the position of the D3'' along $x^3$ (and with respect to the D5s) is irrelevant in the infrared limit.

\medskip

\begin{figure}[th]
\centering
\includegraphics[scale=0.75]{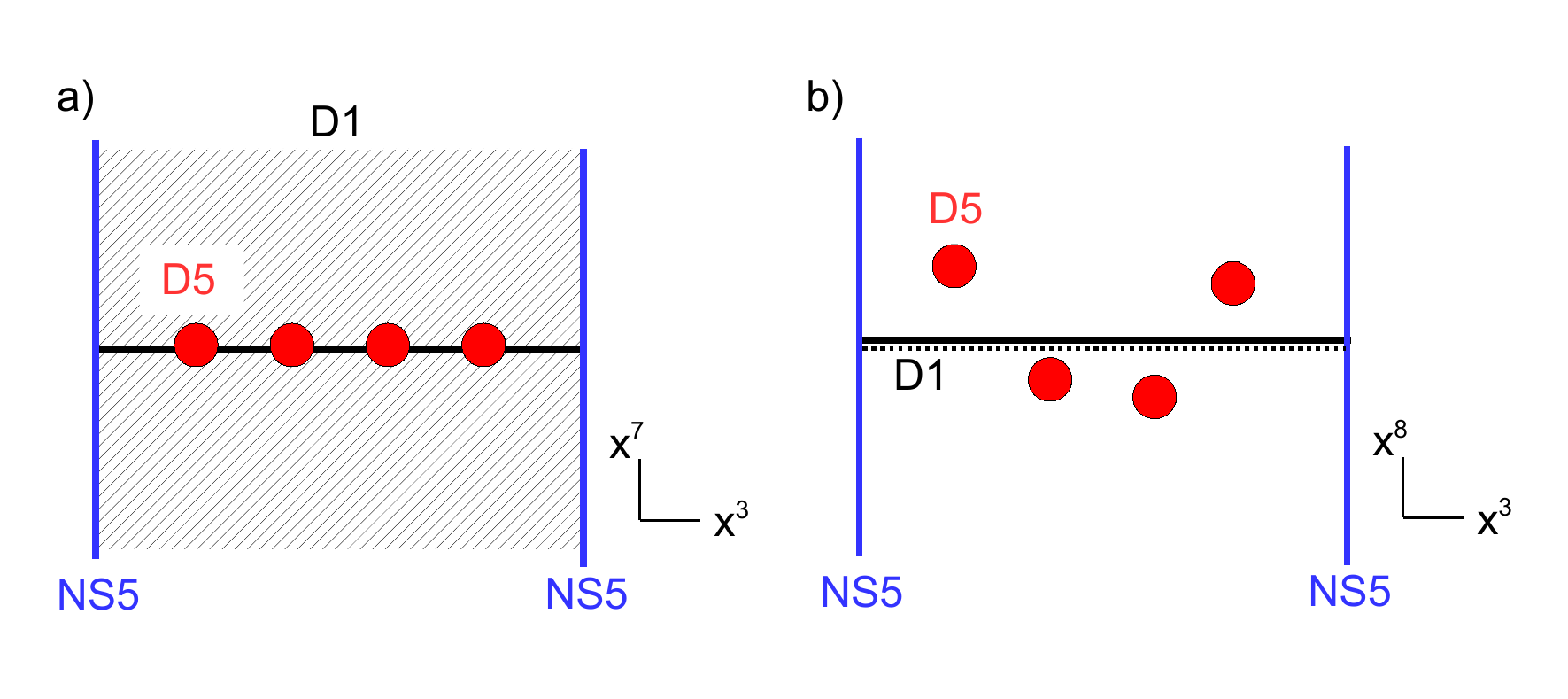}
\vspace{-0.8cm}
\caption{\footnotesize{a) Single D1 crossing the D3 ($x^{3,7}$ plane). b) The same configuration shown in the $x^{3,8}$ plane. The D1 appears on top of the D3. The D5s positions along $x^8$ correspond to the (real part of the) mass parameters $m_\alpha$.}}
\label{SQEDFullD1}
\end{figure}

The relation \eqref{CBRelSQED} is derived from the brane setup of Figure \ref{SQEDFullD1}, with a full D1 brane stretched between the two NS5s. This can be interpreted as inserting the product of monopole operators $- u^+ u^-$, with two semi-D1s ending on the D3 from above and from below. Alternatively it can be seen as a full D1 crossing the D3, with a one-dimensional theory living on its worldvolume in the low energy limit, together with a zero dimensional hyper-multiplet sourced by the D1-D3 strings and $N_f$ zero-dimensional fermions sourced by the D1-D5 strings.
The analysis of this system was done in Section \ref{ssec:Relations}. Integrating out the 1d theory and the hyper-multiplet yields a trivial factor. Integrating out the fermions produces the product of operators
\be
\prod_{\alpha=1}^{N_f} (m_\alpha - \varphi) \,,
\ee
where $m_\alpha - \varphi$ is the complex mass of the fermion sourced by the D5-D1$_{(\alpha)}$ open strings and corresponds to the distance along $x^8 + i x^9$ between the D5 and the D1. 
Identifying the two interpretations of the same brane setup gives the CB relation \eqref{CBRelSQED}.

Here we have obtained directly the CB relation from a brane setup. This is one of the special cases when the pre-relations -- the relations read from the brane setups -- directly match the full ring relations.

\bigskip

To check mirror symmetry we need to study the mirror theory, which is an abelian linear quiver.

\subsection{Abelian Quiver}
\label{ssec:AbQuiv}

\begin{figure}[th]
\centering
\includegraphics[scale=0.73]{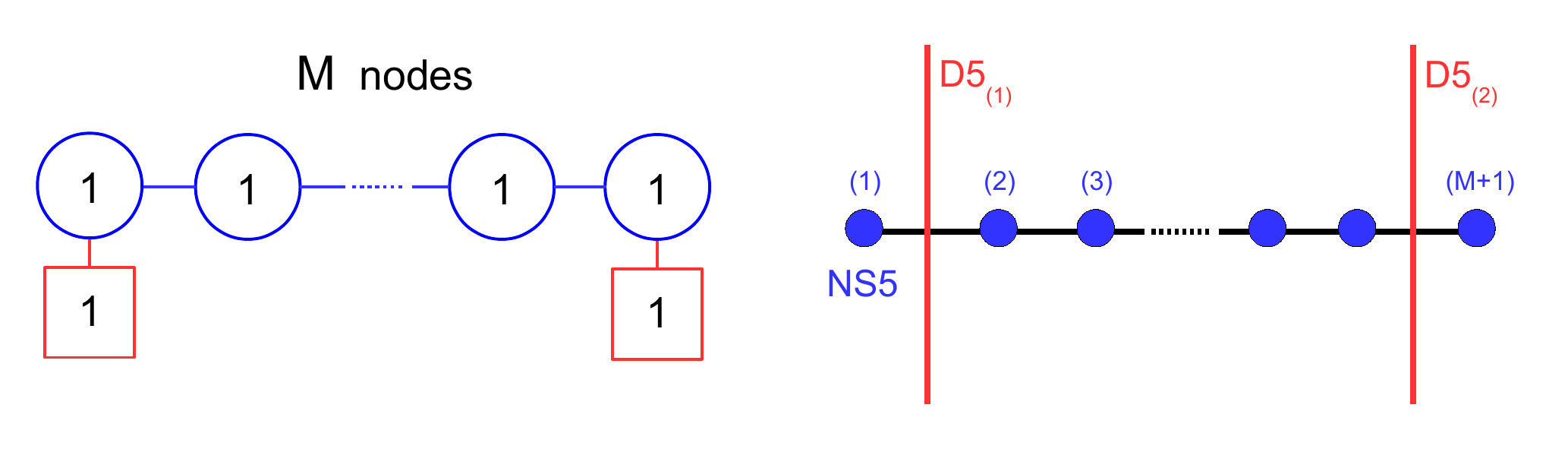}
\vspace{-0.5cm}
\caption{\footnotesize{Quiver and brane configuration of the abelian quiver $T_{\rm abel}$ with $M$ abelian nodes. There are $M+1$ NS5 branes in the brane configuration.}}
\label{Tabel}
\end{figure}

The extension to abelian quivers requires a few more efforts. We explain in detail the case of the abelian linear quiver with $M$ nodes and one fundamental hyper-multiplet in each exterior node, which is the mirror dual theory to SQED with $M+1$ flavor hyper-multiplets. In addition to the two fundamental hyper-multiplets, the matter content has bifundamental hyper-multiplets connecting the nodes in a linear fashion. We will call this theory $T_{\rm abel}$. The quiver diagram and the brane realization of $T_{\rm abel}$ are shown in Figure \ref{Tabel}.  The brane configuration has two D5 branes, which we denote $D5_{(1)}$ and $D5_{(2)}$, and $M+1$ NS5 branes, which we denote NS5$_{(i)}$ with $i=1, \cdots, M+1$, labeling the branes from left to right.

\subsubsection{Higgs branch}
\label{sssec:TabelHB}

The Higgs branch of $T_{\rm abel}$ is parametrized by gauge invariant combinations of the fundamental hyper-multiplet complex scalars $Q^1,\ti Q_1$ and $Q^2, \ti Q_2$ and the bifundamental hyper-multiplet complex scalars $q_i, \ti q_i$, $i=2, \cdots, M$ \footnote{The label $i$ is chosen to start at $i=2$, so that the scalars $q_i, \ti q_i$ are sourced by open string stretched across NS5$_{(i)}$.}. The HB chiral ring  is generated by the $M+1$ ``short" mesons\footnote{We include a minus sign in the definition of $X_1$, and a factor $(-1)^{M-1}$ in the definition of $\ti W$ below, for convenience.}
\be\ba
& X_1  = -\ti Q_1 Q^1 \,, \cr
& X_i =  \ti q_i q_i \,, \quad i=2, \cdots, M \,, \cr
& X_{M+1} = \ti Q_2 Q^2 \,,
\ea\ee
and the two ``long" mesons \footnote{In the notation of appendix \ref{app:Dictionary}, we have $X_1 \equiv -Z^1{}_{1}$, $X_{M+1} \equiv Z^2{}_{2}$, $W \equiv Z^2{}_{1}$ and $\ti W \equiv Z^1{}_{2}$. }
\be
W = \ti Q_1 q_2 q_3 \cdots q_{M} Q^2  \,, \quad \ti W =  (-1)^{M-1} Q^1 \ti q_2 \ti q_3 \cdots \ti q_{M} \ti Q_2 \,,
\ee
subject to the trivial relation
\be
W \ti W = (-1)^{M} \prod_{i=1}^{M+1} X_i \,,
\label{TabelHBRel1}
\ee
and the F-term relations
\be
X_i + \ti \xi_i = X_{i+1} + \ti \xi_{i+1} \,, \quad i = 1, \cdots, M,
\label{TabelHBRel2}
\ee
where we have included the deformations by  FI terms with complex parameter $\eta_i  = \ti \xi_{i} - \ti \xi_{i+1}$ for the $i$-th abelian node.

\medskip

\begin{figure}[th]
\centering
\includegraphics[scale=0.75]{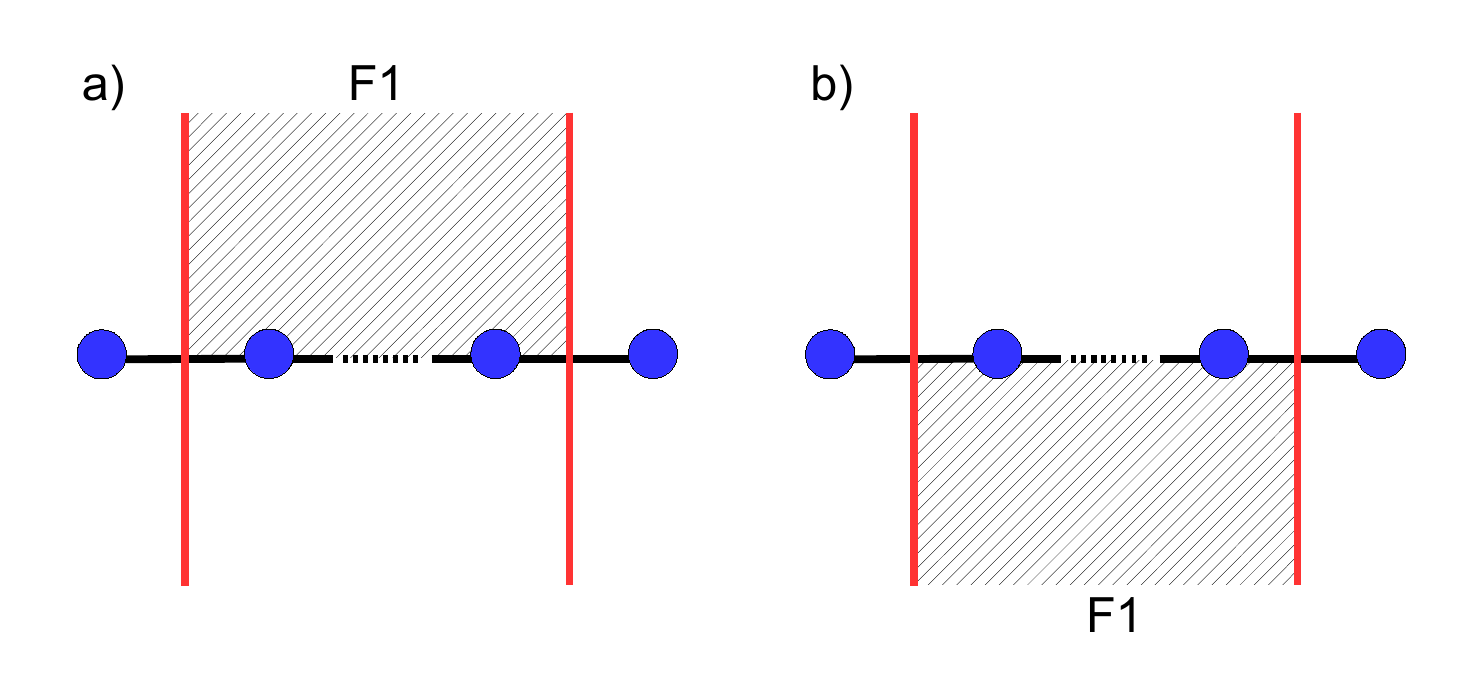}
\vspace{-0.5cm}
\caption{\footnotesize{Brane setups of the long mesons (a) $B$ and  (b) $-\ti B$.}}
\label{TabelHBop0}
\end{figure}

\begin{figure}[th]
\centering
\includegraphics[scale=0.75]{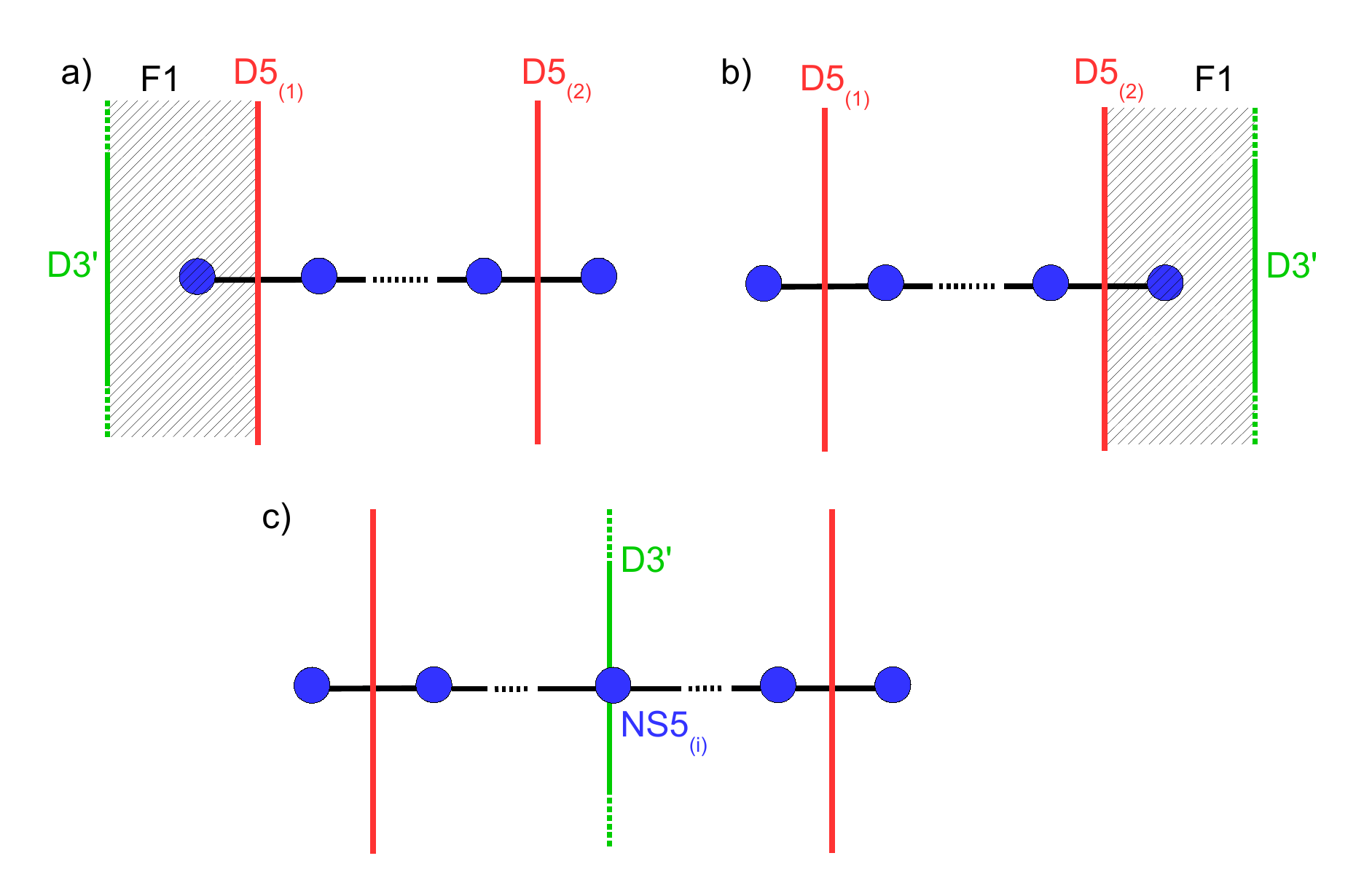}
\vspace{-0.5cm}
\caption{\footnotesize{Brane setups of the short mesons (a) $X'_1$, (b) $X'_{M+1}$ and (c) $X'_i$, $2 \le i \le M-1$. These brane configurations are all related by D3' brane moves along $x^3$ and are ultimately a single brane setup.}}
\label{TabelHBop}
\end{figure}

The brane realization of the corresponding operator insertions are
\begin{itemize}
\item The long mesons $W$ and $-\ti W$ are realized with a semi-infinite F1 string stretched between the two D5s and ending on the D3 segments from above and from below respectively, as in Figure \ref{TabelHBop0}-a,b.
\item The short meson $X'_1\equiv X_1 + \ti \xi_1$ is realized with one infinite F1 string extended on the left of the configuration and ending on D5$_{(1)}$, as shown in Figure \ref{TabelHBop}-a. The short meson $X'_{M+1} \equiv X_{M+1} +\ti \xi_{M+1}$ is realized with one infinite F1 string extended on the right of the configuration and ending on D5$_{(2)}$, as shown in Figure \ref{TabelHBop}-b. We can let the F1 strings end on a D3' brane away from the configuration, as in these figures.
\item The short meson $X'_i \equiv X_i +\ti \xi_i$, with $2 \le i \le  M$, is realized with a D3' brane crossing NS5$_{(i)}$ as shown in Figure \ref{TabelHBop}-c, with $\ti \xi_i$ the position of NS5$_{(i)}$ along $x^5 + i x^6$.
\end{itemize}

To justify the brane setup realizing the long meson $W$, we revisit the argument of Section \ref{ssec:HBop} for the insertion of meson operators, focusing on the part of the setup around one NS5 brane. In the region near NS5$_{(i)}$, there is an F1 string ending on the two D3 segments attached to the NS5 on the left and on the right. On the two D3 worldvolumes live two abelian vector multiplets for two adjacent nodes $U(1)_{i-1} \times U(1)_{i}$ of the quiver. Locally we know that the F1 inserts a half-BPS Wilson loop of charge one in each node:
\be
\exp \lp \int_{- \cdots }^{0} dx^3 \, ( i A^{(i-1)}_3 + \phi^{(i-1)}_4) \rp \, \exp \lp  \int_{0}^{+ \cdots} dx^3 \, ( i A^{(i)}_3 + \phi^{(i)}_4) \rp \,,
\ee
with NS5$_{(i)}$ sitting at $x^3=0$. This Wilson loop operator is not gauge invariant since the contours has a boundary at $x^3=0$. The minimal operator insertion which restores gauge invariance is the insertion of an extra bifundamental scalar $q_i(P)$, where $P$ is the point where the F1 meets the NS5 (in particular $x^3(P)=0$). We propose that the full operator insertion is 
\be
\exp \lp \int_{- (\cdots) }^{0} dx^3 \, ( i A^{(i-1)}_3 + \phi^{(i-1)}_4) \rp \, q_i(P) \exp \lp  \int_{0}^{+ (\cdots)} dx^3 \, ( i A^{(i)}_3 + \phi^{(i)}_4) \rp \,.
\ee
Taking into account the whole brane setup and the fact that in the low energy limit, with the Neumann boundary conditions along $x^3$, the Wilson loop factors trivialize, we obtain that the brane setup of Figure \ref{TabelHBop0}-a inserts  the product of operators\footnote{In the low energy limit all the insertion points of the scalars collapse to the same point $P$ in the $x^{0,1,2}$ 3d space.}
\be
\ti Q_1 q_2 q_3 \cdots q_{M} Q^2= W \,,
\ee
where the product of $q_i$ comes from the F1-NS5s regions and the factors $\ti Q_1, Q^2$ come from the F1-D3-D5 corners, as in previous sections. The insertion of the long meson $-\ti W$ from the setup of Figure \ref{TabelHBop0}-b follows the same logic,  with the insertion of the bifundamental scalar $-\ti q_i$ from the F1 ending from below on the D3-NS5$_{(i)}$ intersection. The minus is sign fixed for compatibility with the ring relations (that we derive below).


\smallskip

The setups corresponding to the insertion of the short mesons $X'_1$ and $X'_{M+1}$ in Figure \ref{TabelHBop} have already been explained in previous sections. 

\smallskip

The last setup, shown in Figure \ref{TabelHBop}-c,  is new. The D3' splits into two half-branes ending on NS5$_{(i)}$. In the absence of FI deformation ($\ti\xi_i=0$), the localized open strings low modes contain the 3d bifundamental hyper-multiplet with scalars $q_i, \ti q_i$ from D3-D3 strings across NS5$_{(i)}$, living on a $x^{0,1,2}$ slice, the 3d bifundamental hyper-multiplet with scalars $(q',\ti q')$ from D3'-D3' open strings across NS5$_{(i)}$, living on a $x^{7,8,9}$ slice, and two zero-dimensional fermions $\chi, \ti\chi$ from the D3-D3' open strings, living at the intersection point of the branes. The fermions $\chi$ and $\ti\chi$ have charges $(1,0,-1,0)$ and $(0,1,0,-1)$ under the $U(1)_{i-1}\times U(1)_{i} \times U(1)'_{\rm up} \times U(1)'_{\rm down}$ 3d gauge symmetry, where $U(1)'_{\rm up} \times U(1)'_{\rm down}$ refers to the symmetries gauged by the two half-D3' worldvolume fields, and which can be considered as a flavor symmetry in the infrared limit\footnote{Four-dimensional fields are frozen in the low-energy effective three-dimensional theory.}.

Our task is to integrate out the 0d fermions $\chi, \ti\chi$. These fermions can have quartic couplings with the hypermultiplet fields. From the supersymmetry of the brane configuration we know that these couplings must preserve the same four supercharges which annihilate the scalars $q_i, \ti q_i, q', \ti q'$. The gauge invariant quartic coupling between the four multiplets, compatible with this requirement\footnote{We remind the reader that the fermions $\chi, \ti\chi$ are invariant under preserved supersymmetry transformations, as are the complex scalars $q_i, \ti q_i, q', \ti q'$.}, is
\be
S_{\rm 0d} \sim  \overline{\chi} q_i \ti q' \ti\chi + \overline{\ti\chi} \ti q_i q \bar \chi \,.
\ee
There are also cubic couplings of the form $\bar\chi \phi \chi$ between the 0d fermions and the 4d complex scalars associated to the D3 (and D3') displacements along the directions $x^5+ i x^6$ but these scalars are set to zero by the Neumann boundary conditions on the 4d vector multiplets. 
Integrating out $\chi$ and $\ti\chi$ leads to the insertion of the operator
\be
q_i \ti q_i q' \ti q' \,, 
\ee
at the point where the D3' intersect the NS5 and the D3s.
This is not yet the insertion that we proposed above. After integrating out the 0d fermions, the fields $q'$ and $\ti q'$ are decoupled from the low-energy 3d theory living on the D3s and account only for an overall number, which we can set to one for convenience, since our analysis is not sensitive to such factors in the final operator insertions.
We end up with the insertion of the operator $\ti q_i q_i = X_i$, as predicted in the absence of FI deformation. 

The complex FI deformations are associated to the displacement of the NS5 branes along the direction $x^5+i x^6$. The displacement of NS5$_{(i)}$ to the position $\ti \xi_i$ is associated to a reconnection of the D3$_{(i-1)}$ and D3$_{(i)}$ branes. It is not obvious what is the corresponding deformation in terms of 0d couplings. On way to understand the effect of this deformation is to look at the brane setup inserting the meson $X'_{M+1}$ (Figure \ref{TabelHBop}-b). Moving D5$_{(2)}$ to the right of NS5$_{(M+1)}$, with D3 creation effect, and moving the D3' brane on top of NS5$_{(M+1)}$ lead to a configuration identical to those inserting the $X_i$ operators (a D3 and a D3' crossing an NS5). Indeed, in the absence of deformations, the operator inserted is $X_{M+1} =  \ti Q_2 Q^2$, where $(Q^2, \ti Q_2)$ are seen as bifundamentals of the $U(1)^{\rm gauge}_{M} \times U(1)^{\rm flavor}_{M+1}$ symmetry, in agreement with the above analysis. In the FI deformed theory, with NS5$_{(M+1)}$ at the position $\ti \xi_{M+1}$, we have argued that the operator inserted is $X'_{M+1}=X_{M+1} +\ti \xi_{M+1}$.
We deduce that the operator inserted is by the D3'-D3-NS5$_{(i)}$ intersection in the FI deformed theory is
\be
X'_i \equiv X_i + \ti \xi_i \,.
\ee
We could have derived this operator insertion solely from the knowledge of the $X'_{M+1}$ brane setup. The fact that we could understand it directly from the D3'-D3-NS5 intersection, in the absence of deformation, strengthen our conclusions.
Knowing the final operator insertion, we can deduce that the FI deformation corresponds in the 0d theory to giving masses $\pm \sqrt{\xi_i}$ to the two fermions,
\be
S_{\rm 0d} \sim  \sqrt{\ti\xi_i} (\overline{\chi} \chi - \overline{\ti\chi} \chi) +  \overline{\chi} q_i \ti q' \ti\chi + \overline{\ti\chi} \ti q_i q \bar \chi \,.
\ee
Integrating out the fermions produces the insertion $\ti\xi_i + q_i \ti q_i q' \ti q' \sim \ti\xi_i + X_i$.

\medskip

{\bf Ring relations}:
\medskip

\begin{figure}[th]
\centering
\includegraphics[scale=0.75]{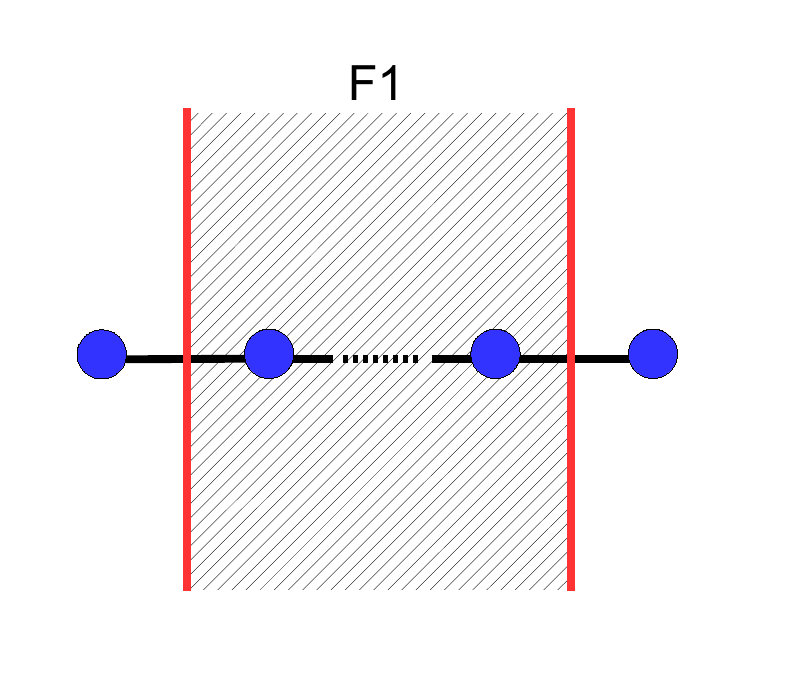}
\vspace{-0.5cm}
\caption{\footnotesize{A full F1 stretched between the two D5s and crossing the D3 segments, realizing $-W \ti W$ or $(-1)^{M+1} X_1 X_2 \cdots X_{M} X_{M+1}$.}}
\label{TabelBraneRel1}
\end{figure}
The trivial ring relation \eqref{TabelHBRel1} follows from considering the configuration of Figure \ref{TabelBraneRel1} , corresponding to the insertion of the product of long mesons $-W \ti W$ and reading it in an alternative way as a product of mesons $-X_1$ and $-X_{M+1}$, from the F1 string ending on D5$_{(1)}$ and D5$_{(2)}$ respectively, and of the mesons $-X_i$, for $i=2, \cdots, M$, from the full F1 string crossing each D3-NS5$_{(i)}$ intersection. This latter configuration is seen as the product of an F1 ending on the D3-NS5$_{(i)}$ intersection from above, inserting $q_i$ and an F1 ending on the D3-NS5$_{(i)}$ intersection from below, inserting $-\ti q_i$. The complete insertion is thus $\prod_{i=1}^{M+1} (-X_i)$ and the ring relation following from the two different readings is precisely \eqref{TabelHBRel1}.

It is not obvious, although it must be true, from the brane setup that the complex FI deformations do not affect this relation, namely that NS5s movements do not affect how we read the operator insertions.

The F-term relations \eqref{TabelHBRel2} follow from the D3' brane move from the left to the right of the quiver, connecting the brane setup inserting $X'_1$ (Figure \ref{TabelHBop}-a) to those inserting $X'_i$, for each $i$, (Figure \ref{TabelHBop}-c) and the brane setup inserting $X'_{M+1}$ (Figure \ref{TabelHBop}-b). There is therefore truely only a single brane setup for all these operators inserting the operator $Y= X_i + \ti\xi_i$, for any $i=1, \cdots, M+1$, solving for the F-term constraints.\footnote{It is also possible to analyse the related brane setups when the $D3'$ brane intersect a D3$_{(i)}$ segment away from the NS5s. Integrating out the zero-dimensional fermion living at the intersection then produces the expected meson operator insertion, using the analysis in \cite{Gaiotto:2008sa} of the 4d SYM theory on the D3 segments (see Section \ref{ssec:Relations}).}

\subsubsection{Coulomb branch}
\label{sssec:TabelCB}

The description of the Coulomb branch of the $T_{\rm abel}$ theory can be found in \cite{Bullimore:2015lsa}. As a hyper-K\"ahler maniflod the Coulomb branch admits a triplet of symplectic forms. The complex structure that is chosen to describe the Coulomb branch corresponds to a certain symplectic form, from which a Poisson bracket can be defined. As a Poisson algebra, the Coulomb branch is generated by the monopole operators with charge $\pm 1$ under a single node, $u_i^{\pm}$, for $i=1, \cdots, M$, and the complex scalars of the $U(1)$ vector multiplet $\varphi_i$, for $i=1, \cdots, M$.

As a simple ring, additional monopole operators are needed to generate the Coulomb branch. Those are monopoles with the same charge $+1$ or $-1$ in sequences of adjacent nodes, namely the operators $V^{\pm}_{[i:j]}$ with charges $+1$ and $-1$ respectively under each $U(1)_k$ with $i \le k \le j$ and zero charge under the other nodes. In the Poisson algebra description, these operators are generated by successive Poisson brackets of monopole operators, starting with $V^{\pm}_{[i:i]} = u^{\pm}_i$.

The ring relations, as derived in \cite{Bullimore:2015lsa}, are
\begin{align}
V_A V_B = V_{A+B} P_{A,B}(\varphi_i) \,,
\end{align}
with $V_A$ the monopole operator of charge $A\in \bZ^M$ under $U(1)^M$ and $P_{A,B}(\varphi_i)$ a certain polynomial of the $\varphi_i$, depending on the charges $A,B$, which is defined in Equation (3.13) of  \cite{Bullimore:2015lsa}. For most choices of charges $(A,B)$ the polynomial $P$ is trivial (equal to one) and the above relation can be used to eliminate the monopole $V_{A+B}$ from the description of the chiral ring. Ultimately the Coulomb branch can be described in terms of the monopole operators $V^\pm_{[i:j]}$, $i \le j$, and the scalars $\varphi_i$ with the following relations
\be\ba
& V^{\pm}_{[i:j]} V^{\pm}_{[j+1,k]} = V^{\pm}_{[i:k]} (\varphi_j - \varphi_{j+1}) \,, \cr
& V^{+}_{[i:j]} V^{-}_{[i,k]} = \left\lbrace
\begin{array}{c}
V^{-}_{[j+1:k]} (\varphi_{i-1} - \varphi_{i}) \,, \ \text{for} \  j < k \,, \\
V^{+}_{[k+1:j]} (\varphi_{i-1} - \varphi_{i}) \,, \ \text{for}\  k < j \,, 
\end{array} \right.  \cr
& V^{+}_{[i:j]} V^{-}_{[k,j]} = \left\lbrace
\begin{array}{c}
V^{+}_{[i:k-1]} (\varphi_{j} - \varphi_{j+1}) \,, \ \text{for} \  i < k \,, \\
V^{-}_{[k:i-1]} (\varphi_{j} - \varphi_{j+1}) \,, \ \text{for}\  k < i \,, 
\end{array} \right.  \cr
&  V^{+}_{[i:j]} V^{-}_{[i,j]} = (\varphi_{i-1} - \varphi_{i})(\varphi_j - \varphi_{j+1}) \,.
\ea\ee
with $\varphi_0 \equiv \ti m_1$ and $\varphi_{M+1} \equiv \ti m_2$, the complex masses of the fundamental hyper-multiplets of the $U(1)_1$ and $U(1)_{M}$ nodes respectively. 
In the following it will be useful to consider a subset of the ring relations, which we call {\it pre-relations}:
\begin{align}
& u^+_i u^-_i = (\varphi_{i-1} - \varphi_i)(\varphi_i - \varphi_{i+1}) \,, \quad i = 1, \cdots, M  \,, \label{TabelCBRel1} \\
& \prod_{k=i}^j u^{\pm}_k = V^{\pm}_{[i:j]} \prod_{k=i+1}^{j} (\varphi_{k-1} - \varphi_{k}) \,, \quad 1 \le i < j \le M \,, \label{TabelCBRel2}
\end{align}
where $u^{\pm}_i = V^{\pm}_{[i:i]}$. From the pre-relations \eqref{TabelCBRel1}, \eqref{TabelCBRel2}, one can generate the complete set of CB relations described above by manipulating them and allowing to suppress factors of $(\varphi_{i-1} - \varphi_i)$ which appear on both sides of a relation. This is the same procedure as described in Appendix \ref{app:SQEDHBRel} for the SQED Higgs branch relations, and, not surprisingly, we will show that the $T_{\rm abel}$ Coulomb branch pre-relations are mapped to the SQED Higgs branch pre-relations under mirror symmetry.

\medskip

\begin{figure}[th]
\centering
\includegraphics[scale=0.7]{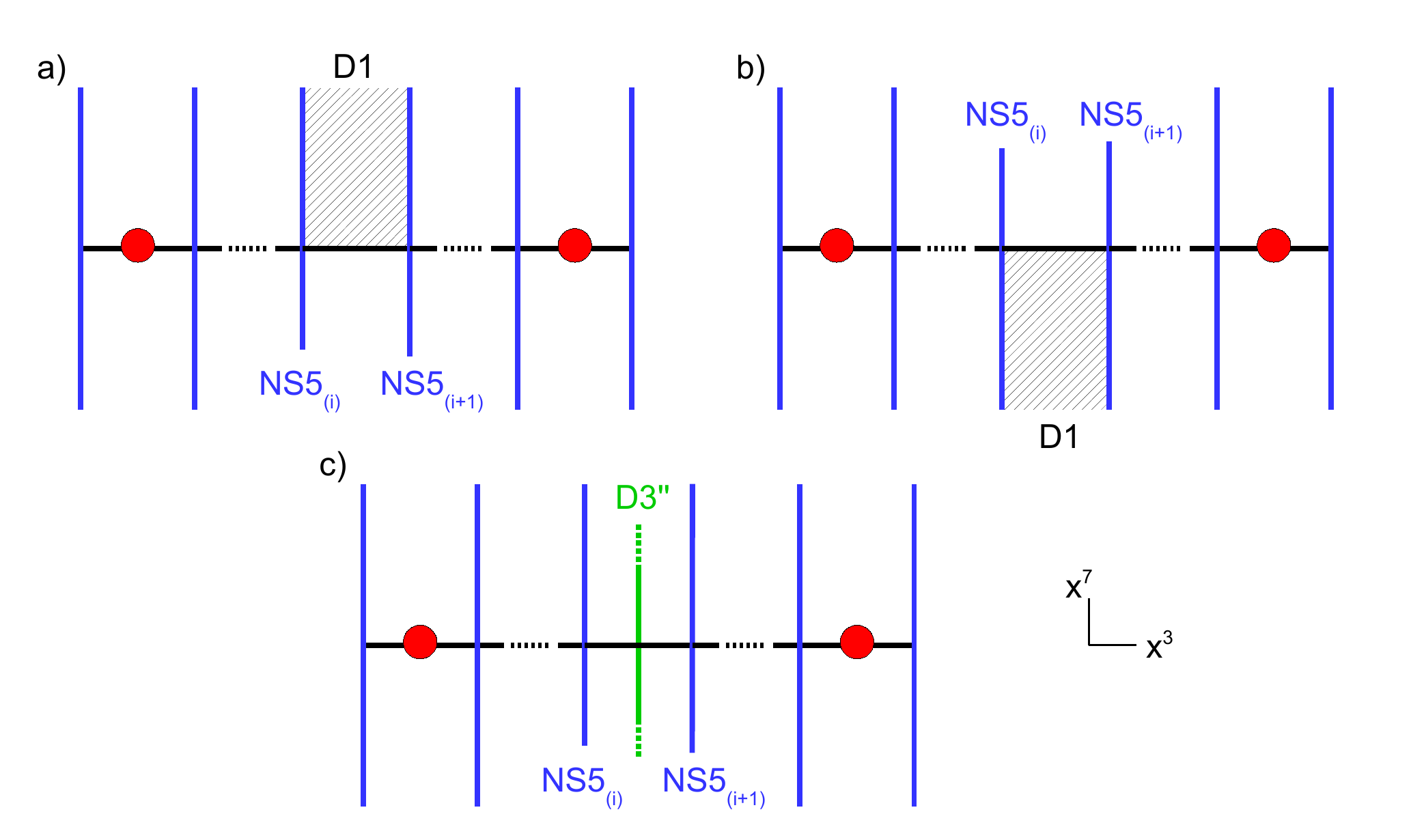}
\vspace{-0.5cm}
\caption{\footnotesize{Brane setups realizing the insertion of the CB operators: a) $u^+_i$, b) $-u^-_i$, c) $\varphi_i$.}}
\label{TabelSemiD1}
\end{figure}

In the dual brane description we find that only the generators of the Poisson algebra $u^{\pm}_i$ and $\varphi_i$ are naturally realized:
\begin{itemize}
\item The insertion of the monopole operators $u^{+}_i$ and $-u^{-}_i$, which have charge $1$ and $-1$ respectively under $U(1)_i$ and zero under the other $U(1)$s, are realized by adding a semi-infinite D1 stretched between NS5$_{(i)}$ and NS5$_{(i+1)}$ and ending on the D3$_{(i)}$ segment (the segment stretched between the two NS5s) from above and from below respectively. The setups are shown in Figure \ref{TabelSemiD1}-a and -b.
\item The insertion of the scalar operator $\varphi_i$ is realized by adding a D3" brane intersecting the D3$_{(i)}$ segment, which support the $U(1)_{(i)}$ vector multiplet, at a point, as in Figure \ref{TabelSemiD1}-c.
\end{itemize}
These operator insertions are immediate generalizations of the analysis of Section \ref{ssec:CBop}.

The other operators needed to generate of the Coulomb branch as a ring, namely the operators $V^{\pm}_{[i:j]}$, with $i<j$,  will appear from brane setups associated to the CB relations.

\bigskip

\noindent{\bf Ring relations}

\medskip

\begin{figure}[th]
\centering
\includegraphics[scale=0.7]{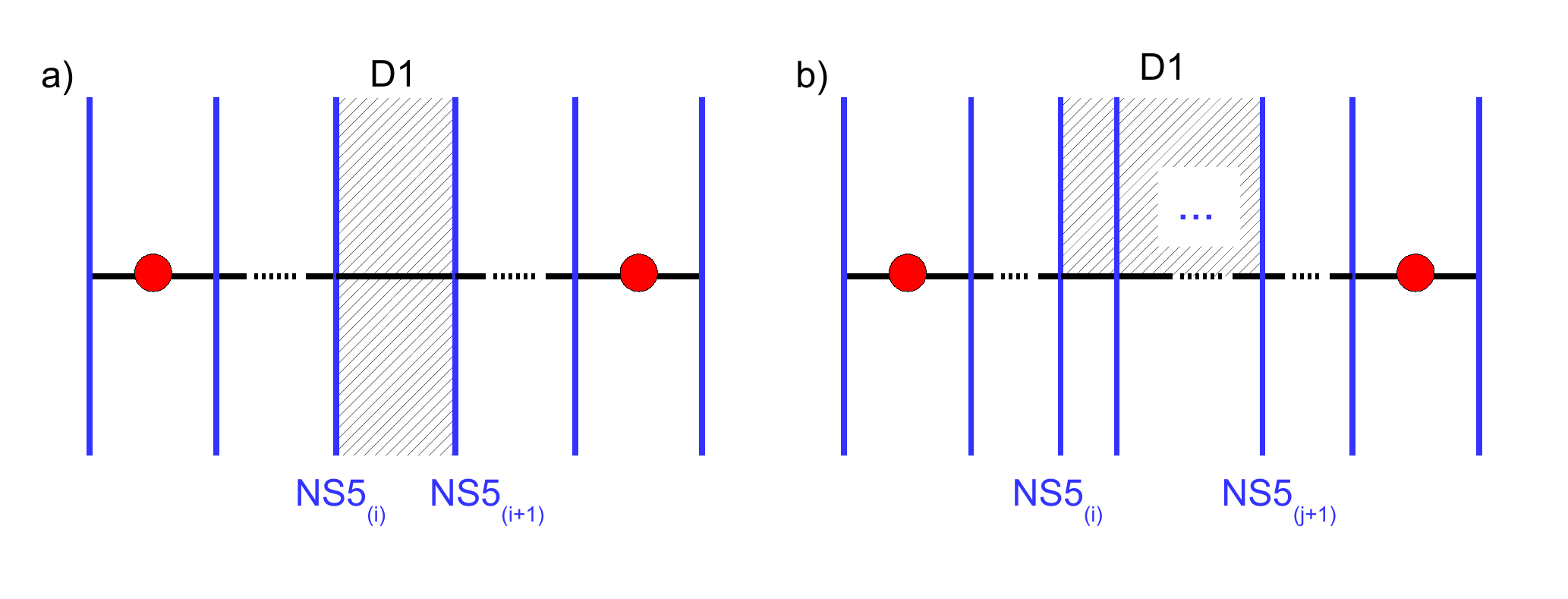}
\vspace{-0.5cm}
\caption{\footnotesize{The CB ring relations are derived from the configurations: (a) with a D1 stretched between two adjacent NS5s, realizing $-u^+_i u^-_i$ or $(\varphi_{i-1} - \varphi_{i})(\varphi_{i+1} - \varphi_{i})$, and (b) with a semi D1 stretched between two distant NS5s, ending on the D3 segments from above, realizing $\prod_{k=i}^j u^{+}_k$ or $V^{+}_{[i:j]} \prod_{k=i+1}^{j} (\varphi_{k-1} - \varphi_{k})$.}}
\label{TabelD1s}
\end{figure}

\begin{figure}[th]
\centering
\includegraphics[scale=0.75]{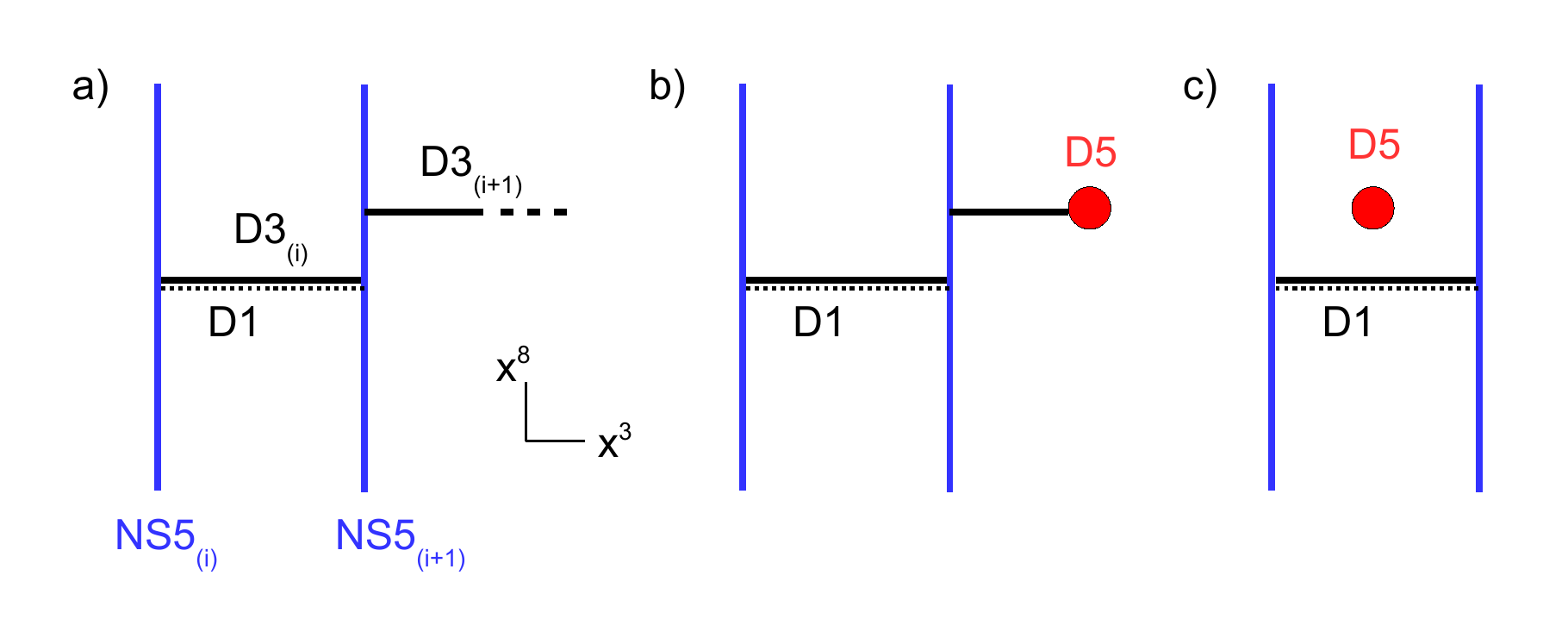}
\vspace{-0.5cm}
\caption{\footnotesize{The local setup (a) with an external D3 segment is related to the setup (c) with a D5 brane by letting the D3 segment end on a D5 (b) and moving it between the two NS5s. The distance between the D5 and the D3 along $x^8+ix^9$ is $\varphi_{i+1} -\varphi_i$.}}
\label{TabelHWmove}
\end{figure}

The ring relations are derived by considering the brane setups of Figure \ref{TabelD1s}.
To obtain the relation \eqref{TabelCBRel1}, we consider the setup of Figure \ref{TabelD1s}-a. Seen as two semi-infinite D1 branes ending on the D3$_{(i)}$ segment from above and from below, it corresponds to the insertion of the product of fundamental monopole operators $-u^+_i u^-_i$. Regarding it as a single D1 brane crossing the D3 segment instead, we can integrate out the 1d fields living on the D1 and the 0d fields sourced by the D1-D3$_{(k)}$, for $k\in \{i-1,i,i+1\}$, open string lowest modes. This configuration is actually very close to the one of Figure \ref{SQEDFullD1} studied in Section \ref{sssec:SQEDCB}. The only differences are the presence of D3 segments on both sides of the NS5s and the absence of D5s. However the configuration with D5s is related to the configuration with external D3 segments by HW brane moves. Starting with the setup of Figure \ref{TabelHWmove}-a with a D3 segment attached to the right of NS5$_{(i+1)}$, we can let the segment end on a D5 (Figure \ref{TabelHWmove}-b) and move the D5 between NS5$_{(i)}$ and NS5$_{(i+1)}$ (Figure \ref{TabelHWmove}-c). According to the analysis of Section \ref{sssec:SQEDCB} the final configuration realizes the insertion of the local operator $m - \varphi_i  = \varphi_{i+1}  - \varphi_{i}$, where $m \equiv \varphi_{i+1}$ is the mass of the hyper-multiplet sourced by the D3$_{(i)}$-D5 strings, or the D3$_{(i)}$-D3$_{(i+1)}$ strings in the original setup (corresponding to the position of D3$_{(i+1)}$ along $x^8+ix^9$).
The operator insertion corresponding to the brane setup of Figure \ref{TabelD1s}-a, seen as a single D1 crossing the D3$_{(i)}$ segment, is therefore the product $(\varphi_{i-1} - \varphi_{i})(\varphi_{i+1} - \varphi_{i})$ coming from integrating out the D3$_{(i-1)}$-D3$_{(i)}$ and D3$_{(i)}$-D3$_{(i+1)}$ modes. We obtain the CB relation \eqref{TabelCBRel1}.

\smallskip

The brane setup of figure \ref{TabelD1s}-b is less straightforward to analyse. It can be seen as $j-i+1$ semi-infinite D1s extended between neighboring NS5s, inserting the product of monopole operators $\prod_{k=i}^j u^{+}_k$. A second way to look at the same setup is to see it as a single semi-infinite D1 brane stretched between NS5$_{(i)}$ and NS5$_{(j+1)}$, ending on the D3 segments from above. The "single" D1 ending on the D3$_{(i)}$ to D3$_{(j)}$ segments corresponds to the insertion of the monopole operator $V^{+}_{[i:j]}$, which has monopole charge $+1$ in each $U(1)_k$ node from $k=i$ to $j$. In addition there are contributions from the D3-NS5 intersections with a D1 ending on the D3 from above (and crossing the NS5), whose stringy origin is unclear. This local setup is S-dual to the D3-D5 intersection with an F1 ending on the D3 from above (and crossing the D5), which inserts the same operator as the D3-D5 intersection with an F1 ending on the D5 from the right (and crossing the D3). Therefore, S-duality predicts that the D3-NS5 intersection with a D1 ending on the D3 from above (and crossing the NS5) inserts the same operator as the D3-NS5 intersection with a D1 ending on the NS5 from the right (and crossing the D3), which is the local setup we analyzed in the previous paragraph. The operator insertion is then $(\varphi_{i-1} - \varphi_{i})$ for the D3-NS5$_{(i)}$ intersection. The rules for such operator insertions are summarized in Appendix \ref{app:Dictionary}.
The contributions of the $j-i$ intersections, together with the monopole operator insertion, yield in total the insertion of the operator
$V^{+}_{[i:j]} \prod_{k=i+1}^{j} (\varphi_{k-1} - \varphi_{k})$. Identifying the two readings of the same brane setup we obtain the second set of relations \eqref{TabelCBRel2}. 

The other ways to interpret the setup of Figure \ref{TabelD1s}-b, splitting into various numbers of D1 pieces, lead to redundant relations.
\smallskip

The CB relations \eqref{TabelCBRel2} involving monopole operators with negative charges are obtained from the same brane setup as in Figure \ref{TabelD1s}-b, but with the semi D1 ending on the D3 segments from below. The two interpretations of the setup lead to the insertions of the operators $\prod_{k=i}^j (- u^{-}_k)$ and $(-V^{-}_{[i:j]}) \prod_{k=i+1}^{j} (\varphi_{k} - \varphi_{k-1})$ respectively, leading to the CB relations. Here again the negative charge monopole insertions come with a minus sign and $(\varphi_{k} - \varphi_{k-1})$ is the contribution from the D3-NS5$_{(k)}$ intersection with a D1 ending on the D3 from below, which is the same as the contribution from the D3-NS5$_{(k)}$ intersection with a D1 ending on the NS5 from the left, studied before.

\subsection{Mirror Symmetry}
\label{ssec:MirrorSymQuivers}

The SQED and $T_{\rm abel}$ theories are related by mirror symmetry with the identification $N_f = M+1$, relating the number of SQED fundamental hyper-multiplets to the number of nodes of $T_{\rm abel}$. The Coulomb branch of one theory should match the Higgs branch of the other. It is possible to find the mirror map of operators using the knowledge of the ring relations, but this is not completely straightforward. Instead, as we advocate in this paper, we can simply let the action of S-duality on the brane setups give us the answer.

\smallskip

In the absence of operator insertions, the two dual theories are realized by the brane setups of Figure \ref{SQED} and \ref{Tabel}, which are related by the action of type IIB S-duality, followed by the exchange of the external pairs of D5 and NS5, on the left and on the right of the brane configuration. The exchanges of 5-branes induce HW D3 brane creation effects, as explained in  Section \ref{ssec:BranesLocalOp}. In the presence of extra strings or/and branes inserting local operators, the identification of the brane setups proceeds in the same way, by acting with S-duality and exchanging the external D5-NS5 pairs.

\medskip

The SQED HB operators realized by the setups of Figure \ref{SQEDSemiF1}-a,b are mapped to the $T_{\rm abel}$ CB monopole operators realized by the S-dual setups of Figure \ref{TabelSemiD1}-a,b,
\be
Z^{i+1}{}_{i}\big|_{\rm SQED} \leftrightarrow u^+_i \big|_{T_{\rm abel}} \,, \quad 
Z^{i}{}_{i+1}\big|_{\rm SQED} \leftrightarrow u^-_i \big|_{T_{\rm abel}} \,.
\label{Map01}
\ee
The SQED HB operators realized by the setup of Figure \ref{SQEDSemiF1}-c,d are mapped to the $T_{\rm abel}$ Coulomb branch scalar operators realized setups of Figure \ref{TabelSemiD1}-c, by S-duality and Hanany-Witten D3'' move,
\be
Y_i \equiv Z_{[i+1:M+1]} + \xi_2 \, (= - Z_{[1:i]} + \xi_1) \, \big|_{\rm SQED} \leftrightarrow \varphi_i \big|_{T_{\rm abel}} \,, \quad i =1, \cdots, M\,,
\label{Map02}
\ee
where we have included the FI deformation in the SQED theory.
This provides the mirror map for operators generating the Higgs/Coulomb branches as Poisson algebras.

To obtain the mirror map for all the operators appearing in a basis of the chiral rings, we must consider the brane setups related to the ring relations and identify the various contributions to the operator insertions across S-duality. 
The setup of Figure \ref{SQEDSemiF1_2}-a is mapped to the setup of Figure \ref{TabelD1s}-a under S-duality. It implies the mapping of the ring relations \eqref{HBRel000} and \eqref{TabelCBRel1}
\be
Z^{i+1}{}_i Z^{i}{}_{i+1} = Z^i{}_{i} Z^{i+1}{}_{i+1} 
\ \leftrightarrow \ 
u^+_i u^-_i = (\varphi_{i-1} - \varphi_i)(\varphi_i - \varphi_{i+1})  \,, 
\ee
and the identification of the different contributions to the operator insertions leads to \eqref{Map01} and
\be
Z^i{}_{i} \, \big|_{\rm SQED} \leftrightarrow \varphi_{i-1} - \varphi_{i} \, \big|_{T_{\rm abel}}  \quad \text{for} \ i = 1 , \cdots, M+1 \,,
\label{Map03}
\ee
with $\varphi_0 \equiv \ti m_1$ and $\varphi_{M+1} = \ti m_2$, the mass parameters of the $T_{\rm abel}$ theory.
These latter identifications are equivalent to \eqref{Map02}, upon mapping the deformation parameters with
\be
(\xi_1, \xi_2) \big|_{\rm SQED} \leftrightarrow  (\ti m_1, \ti m_2)\big|_{T_{\rm abel}}  \,.
\label{ParameterMap1}
\ee 
The operator mapping \eqref{Map03} follows from identifying the contribution of a D3-D5 intersection with an F1 string ending on the D5, with the contribution of a D3-NS5 intersection with a D1 string ending on the NS5 (in the mirror theory).

The setup of Figure \ref{SQEDSemiF1_2}-b is mapped to the setup of Figure \ref{TabelD1s}-b under S-duality. It implies the mapping of the ring relations \eqref{HBRel0} and \eqref{TabelCBRel2} (with positive charge monopoles)
\be
 \prod_{k=i}^{j} Z^{k+1}{}_{k}  = Z^{j+1}{}_i  \prod\limits_{k=i+1}^{j} Z^k{}_k
\ \leftrightarrow \
\prod_{k=i}^j u^{+}_k = V^{+}_{[i:j]} \prod_{k=i+1}^{j} (\varphi_{k-1} - \varphi_{k}) \,,
\ee
and the identification of operator insertions yields, in addition to \eqref{Map03},  the mirror map
\be
Z^{j+1}{}_i  \, \big|_{\rm SQED} \leftrightarrow V^{+}_{[i:j]} \, \big|_{T_{\rm abel}} \,, \quad \text{for} \ 1 \le i \le j \le M \,.
\label{Map04}
\ee
Similarly the setup of Figure \ref{SQEDSemiF1_2}-c is mapped to a setup identical to Figure \ref{TabelD1s}-b but with the D1 branes ending on the D3 from below, implying the map of the ring relation \eqref{HBRel00} and \eqref{TabelCBRel2} (with negative charge monopoles) and the identification
\be
Z^{i}{}_{j+1}  \, \big|_{\rm SQED} \leftrightarrow V^{-}_{[i:j]} \, \big|_{T_{\rm abel}} \,, \quad \text{for} \ 1 \le i \le j \le M \,.
\label{Map05}
\ee
The operator map is summarized in Table \ref{tab:MirrorMap}.

\medskip

\begin{figure}[th]
\centering
\includegraphics[scale=0.67]{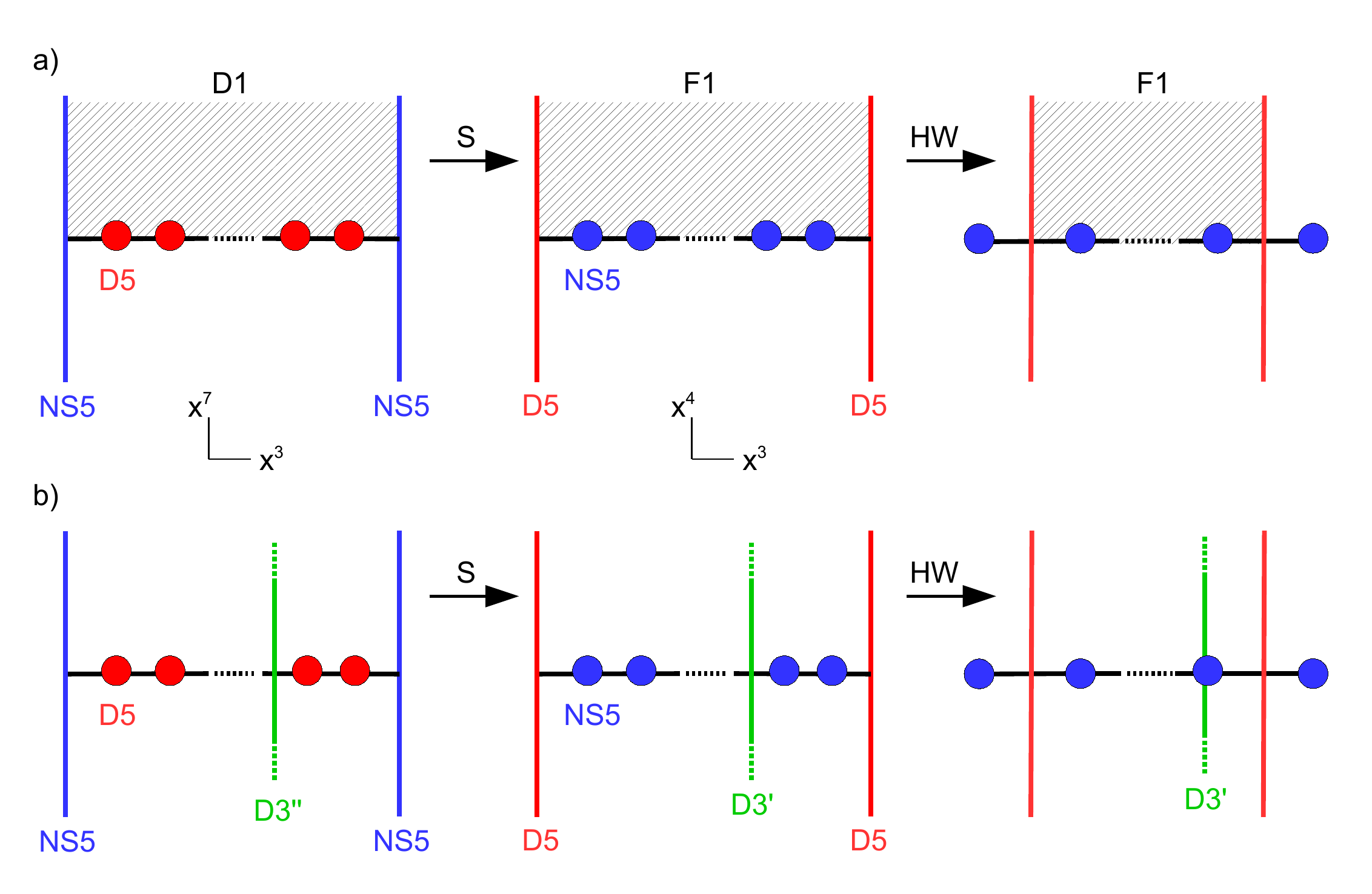}
\vspace{-0.5cm}
\caption{\footnotesize{a) From the brane setup realizing the SQED monopole operator $u^+$ to the setup realizing the insertion of the long meson $W$ in $T_{\rm abel}$. b)  From the brane setup realizing the SQED scalar operator $\varphi$ to the setup realizing the insertion of the meson $Y (= X'_i)$ in $T_{\rm abel}$. The letters $S$ and $HW$ denote the action of S-duality and Hanany-Witten brane moves respectively.}}
\label{MirrorMapBrane}
\end{figure}

It remains to compare the Coulomb branch of SQED and the Higgs branch of $T_{\rm abel}$. 
The setups of Figure \ref{SQEDSemiD1}-a,b inserting the SQED monopole operators are mapped to the setups of Figure \ref{TabelHBop0}-a,b inserting the long mesons, providing the map
\be
u^{+}  \, \big|_{\rm SQED} \leftrightarrow W \, \big|_{T_{\rm abel}} \,, \quad u^{-}  \, \big|_{\rm SQED} \leftrightarrow \ti W \, \big|_{T_{\rm abel}} \,.
\label{Map06}
\ee
The setup of Figure \ref{SQEDSemiD1}-c inserting the SQED scalar operator is mapped to the setups of Figure \ref{TabelHBop}, which are all equivalent, being related by D3' move along $x^3$, giving the map
\be
\varphi \, \big|_{\rm SQED} \leftrightarrow Y   \, \big|_{T_{\rm abel}}  \,,
\label{Map07}
\ee
with $ Y \equiv X'_i  \equiv X_i + \ti \xi_{i}$, for any $i \in [1, M+1]$. We illustrate these last two S-duality maps of brane setups in Figure \ref{MirrorMapBrane}.

Finally the brane setups of Figure \ref{SQEDFullD1} and \ref{TabelBraneRel1} are mapped under S-duality and correspond to the identification of the SQED CB relation \eqref{CBRelSQED} and $T_{\rm abel}$ HB relation \eqref{TabelHBRel1},
\be
u^+ u^- = - \prod_{i=1}^{M+1} (m_{i} - \varphi) \ \leftrightarrow \  W \ti W = - \prod_{i=1}^{M+1} (\ti \xi_i - Y) \, ,
\ee
provided we identify the FI and mass parameters as
\be
m_{i} \,  \big|_{\rm SQED} \ \leftrightarrow \  \ti \xi_i  \, \big|_{T_{\rm abel}} \,, \quad i=1, \cdots, M+1 \,.
\label{ParameterMap2}
\ee

The operator and parameter maps are summarized in Table \ref{tab:MirrorMap}.
\begin{table}[h]
\begin{center}
\setlength\extrarowheight{3pt}
\begin{tabular}{|c|c|c|c|c|}
  \cline{1-2} \cline{4-5}
   SQED HB  &  $T_{\rm abel}$ CB & &  SQED CB & $T_{\rm abel}$ CB \\    
   \cline{1-2} \cline{4-5}
 $Z^i{}_{i}$ \, $(i=1, \cdots, M+1)$  & $\varphi_{i-1} - \varphi_{i}$  & & $u^+$ & $W$ \\ [2pt]
      \cline{1-2} \cline{4-5} 
 $Z^{j+1}{}_{i}$ \, $(i\le j)$ & $V^+_{[i:j]}$  & & $u^-$ & $\ti W$ \\ [2pt]
     \cline{1-2} \cline{4-5}
  $Z^{i}{}_{j+1}$ \, $(i\le j)$ & $V^-_{[i:j]}$  & & $\varphi$ & $Y $  \\  [2pt]
     \cline{1-2} \cline{4-5} 
      \multicolumn{5}{c}{} \\
      \cline{1-2} \cline{3-4} 
    \multicolumn{2}{|c|}{SQED parameters}  &  \multicolumn{2}{c|}{$T_{\rm abel}$ parameters} & \multicolumn{1}{c}{} \\ 
        \cline{1-2} \cline{3-4} 
   \multicolumn{2}{|c|}{$\xi_1$, $\xi_2$ }  &  \multicolumn{2}{c|}{$\ti m_1$, $\ti m_2$} &\multicolumn{1}{c}{}   \\ 
        \cline{1-2} \cline{3-4} 
   \multicolumn{2}{|c|}{$m_{i}$ \, $(i=1, \cdots, M+1)$}  &  \multicolumn{2}{c|}{$\ti\xi_i$ } & \multicolumn{1}{c}{}  \\ 
        \cline{1-2} \cline{3-4} 
\end{tabular}
\caption{\footnotesize Mirror map of HB operators and CB operators between the SQED and $T_{\rm abel}$ theories, and mirror map of FI and mass parameters, with $\varphi_0 \equiv \ti m_1$ and $\varphi_{M+1} \equiv \ti m_2$.}
\label{tab:MirrorMap}
\end{center}
\end{table}

\subsection{Another example}
\label{ssec:Example}

We have worked out all the rules for the insertions of chiral operators, the derivation of the ring relations on the Higgs and Coulomb branches, and the mirror map between operators in abelian quiver theories. In this section we illustrate our method on one more example, extracting the results from the brane setups, without any computation, simply by applying the rules justified in the previous sections (and summarized in Appendix \ref{app:Dictionary}).
\medskip

Consider as theory A, or $T_A$, a linear quiver with two abelian nodes, with one fundamental hyper-multiplet of mass $m_1$ in the first node and three fundamental hyper-multiplets of masses $m_2,m_3$ and $m_4$ in the second node. The FI parameters are $\xi_1 - \xi_2$ and $\xi_2 - \xi_3$.
The mirror dual theory is theory B, or $T_B$, with three nodes, with two fundamental hyper-multiplets of masses $\ti m_1, \ti m_2$ in the first node and one fundamental hyper-multiplet of mass $\ti m_3$ in the third node. The FI parameters are $\ti\xi_{i} - \ti\xi_{i+1}$, for $i=1,2,3$.

\begin{figure}[th]
\centering
\includegraphics[scale=0.65]{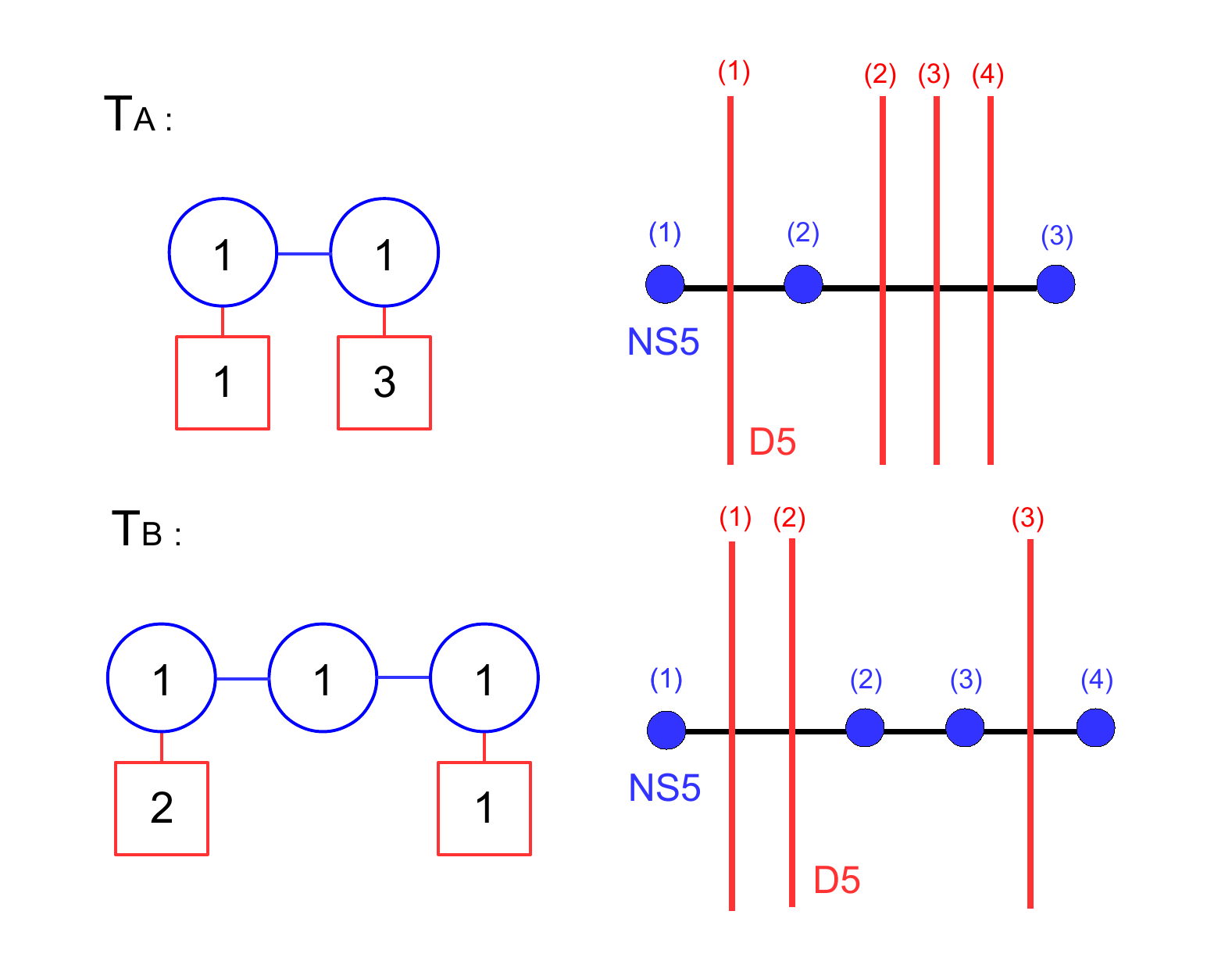}
\vspace{-0.5cm}
\caption{\footnotesize{Quiver diagrams and brane configurations of the mirror dual theories $T_A$ and $T_B$.}}
\label{TATB}
\end{figure}

The quiver diagrams and the type IIB brane configurations realizing these quiver theories are depicted in Figure \ref{TATB}. The D5 and NS5 branes are labeled D5$_{(\alpha)}$ and NS5$_{(i)}$, with $\alpha,i$ increasing from left to right. The brane configurations are related by S-duality and HW 5-brane moves, exchanging the external pairs of NS5 and D5 branes. 

In the following we analyse the Higgs branch of $T_A$, the Coulomb branch of $T_B$ and the mirror map between them. The analysis of the Coulomb branch of $T_A$ and the Higgs branch of $T_B$ is left as an exercise. The FI term deformations $\xi_1, \xi_2, \xi_3$ in theory $A$ will be turned on, and the masses set to zero, while the masses $\ti m_1, \ti m_2, \ti m_3$ in theory B will be turned on, with FI parameters set to zero. To make contact with the gauge theory language we denote by $(Q^\alpha, \ti Q_\alpha)$, $\alpha=1, \cdots 4$, the fundamental hyper-multiplet scalar and $(q, \ti q)$ the bi-fundamental hyper-multiplet scalars of the $T_A$ theory. We also denote by $V^\pm_{[i:j]}$, $1 \le i \le j \le 3$, the monopole operators of charge $\pm 1$ in the nodes $U(1)_k$, $i\le k \le j$, and $\varphi_i$, $i=1,2,3$, the vector multiplet complex scalars for the three nodes of the $T_B$ theory.

\begin{figure}[th]
\centering
\includegraphics[scale=0.7]{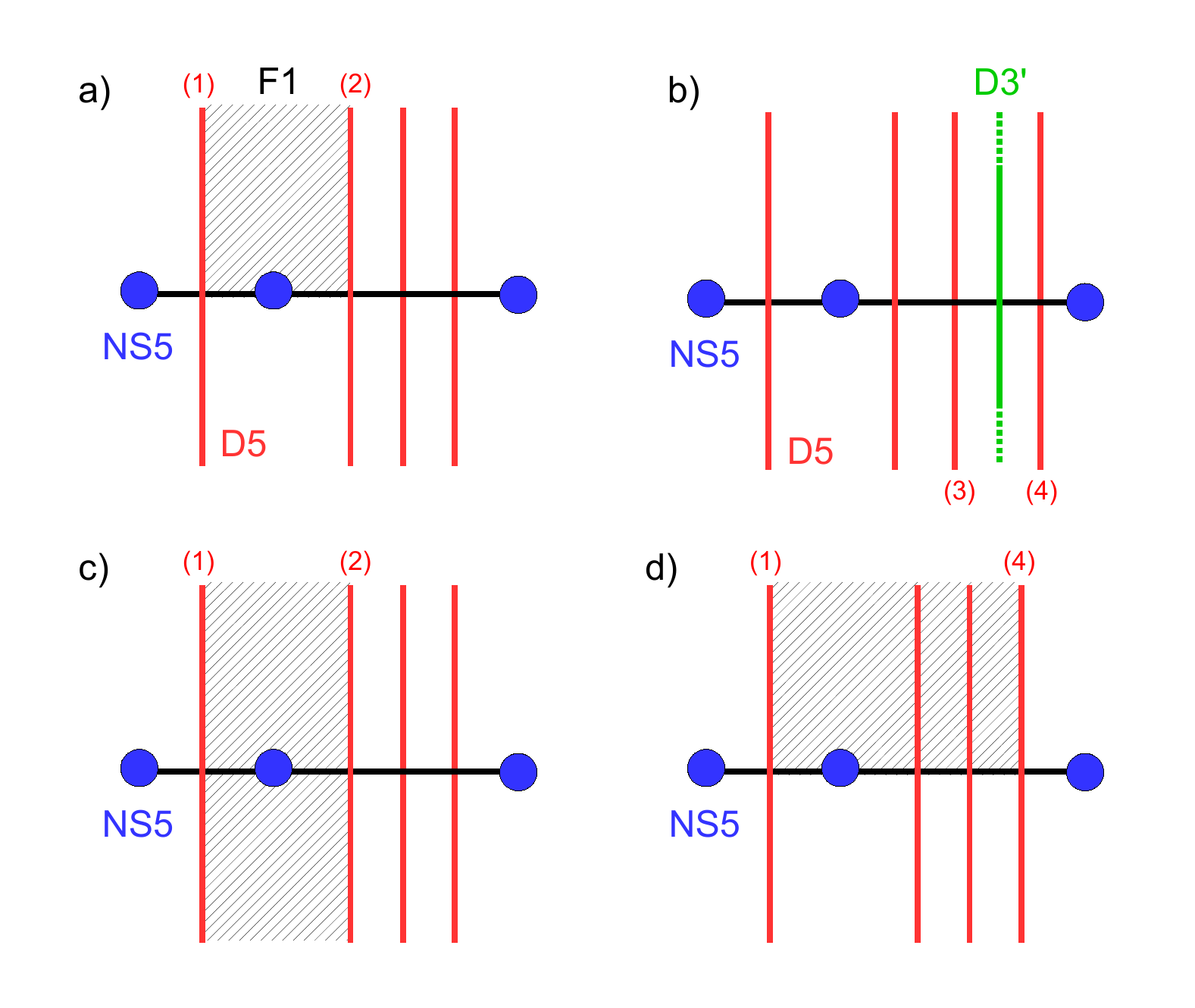}
\vspace{-0.5cm}
\caption{\footnotesize{Brane setups inserting the HB operators of theory A: a) $Z^2{}_{1}$; b) $Y_3$; c) $Z^2{}_{1}(-Z^1{}_2)$ or $(\xi_1 - Y_1)(\xi_2 - Y_1)(Y_2-Y_1)$; d) $Z^4{}_3 Z^3{}_2 Z^2{}_1$ or $Z^4{}_1 (Y_1 - Y_2)(Y_2-Y_3)$.}}
\label{TAHBopAndRel}
\end{figure}

The operators generating the Higgs branch of $T_A$  are of two kinds:
\begin{itemize}
\item For each pair of D5 branes, D5$_{(\alpha)}$-D5$_{(\beta)}$, with $1 \le \alpha < \beta \le 4$, there is a couple of  operators $(Z^{\alpha}{}_{\beta}, Z^{\beta}{}_{\alpha})$. The operator $Z^{\alpha}{}_{\beta}$ is associated to the contribution from a semi F1 string stretched between D5$_{(\alpha)}$ and D5$_{(\beta)}$, ending on the D3 segment from above. The operator $-Z^{\beta}{}_{\alpha}$ is associated to the contribution from a semi F1 string stretched between D5$_{(\alpha)}$ and D5$_{(\beta)}$, ending on the D3 segment from below. Figure \ref{TAHBopAndRel}-a presents the setup inserting the operator $Z^2{}_1$.\footnote{As explained in previous sections, there is no setup inserting the operators $Z^{\alpha}{}_{\beta}$, with $|\alpha - \beta| \ge 2$, alone.}
 In the gauge theory language they are meson operators $Z^{\alpha}{}_{\beta} = Q^\alpha \ti Q_\beta$ and $Z^\beta{}_{\alpha} = \ti Q_\alpha Q^\beta$, for $2 \le \alpha < \beta \le 4$, and $Z^{1}{}_{\beta} = -Q^1 \ti q \ti Q_\beta$ and $Z^\beta{}_{1} = \ti Q_1 q Q^\beta$, for $2 \le \beta \le 4$.
 \item For each pair of adjacent D5, D5$_{(\alpha)}$-D5$_{(\alpha+1)}$ with  $\alpha=1,2,3$, there is an operator $Y_{\alpha}$, whose insertion is achieved by adding a D3' brane between D5$_{(\alpha)}$ D5$_{(\alpha+1)}$ and crossing the D3 segment stretched between them. By moving the D3' brane along $x^3$, taking care of the F1 creation effects, we obtain various brane setups inserting the same operator $Y_\alpha$. Figure \ref{TAHBopAndRel}-b presents the setup inserting the operator $Y_3$. In the gauge theory language the operator $Y_{\alpha}$ corresponds to several combinations of meson operators, which are equal by the F-term constraints:
 \be
Y_{\alpha} = \xi_3 + \sum_{\beta=\alpha+1}^{4} Q^{\beta}\ti Q_{\beta}  
= \xi_2 + q \ti q - \sum_{\beta=2}^{\alpha} Q^{\beta}\ti Q_{\beta} 
 = \xi_1 - \sum_{\beta=1}^{\alpha} Q^{\beta}\ti Q_{\beta} \,, 
 \label{YZRel}
\ee
with $\alpha =1,2,3$. The first expression for $Y_\alpha$ follows from moving the D3' to the right of the configuration and reading the operator insertion, the second expression is obtained by moving the D3' on top of NS5$_{(2)}$, and the third expression is obtained by moving the D3' to the left of the configuration.
The operators $Y_\alpha$ are related to the standard meson operators $X = q \ti q$ and $Z^\alpha{}_{\alpha} = Q^\alpha \ti Q_\alpha$ by the equations \eqref{YZRel}, which can be recast as
\be
X =  Y_1 -\xi_2 \,, \qquad  Z^\alpha{}_{\alpha} = Y_{\alpha-1} - Y_\alpha \,, \ \alpha =1,2,3,4\,,
\ee
with $Y_0 \equiv \xi_1$ and $Y_4 = \xi_3$.
\end{itemize}

There are in total six operators $Z$ and three operators $Y$, so nine HB generators, after the F-term constraints have been imposed (this is already implemented in the brane setups). 

To read the HB relations constraining them, we consider two kinds of brane configurations. 
The first set of configurations have a full F1 stretched between two consecutive D5s, D5$_{(\alpha)}$-D5$_{(\alpha+1)}$, as in Figure \ref{TAHBopAndRel}-c for $\alpha=1$. Looking at these configurations as two semi-infinite F1 or a single F1 leads to the identities between the operator inserted,
\be
  Z^{\alpha+1}{}_{\alpha} (-Z^{\alpha}{}_{\alpha+1}) = \left\lbrace
\begin{array}{c}
Z^1{}_{1}(-X)(-Z^2{}_{2}) = (\xi_1 - Y_1)(\xi_2 - Y_1)(Y_2-Y_1)  \,,  \  \alpha=1 \,, \cr
(Z^\alpha{}_{\alpha})(-Z^{\alpha+1}{}_{\alpha+1}) = (Y_{\alpha - 1} - Y_\alpha)(Y_{\alpha+1} - Y_{\alpha})  \,,  \  \alpha=2,3\,,
\end{array}
\right.
\ee
where each factor in the product  corresponds to the contribution of a local brane ingredient (F1 ending on D3, F1 ending on D5, D3-NS5 intersection with F1, ...), as explained in detail in appendix \ref{app:Dictionary}.
We obtain the three relations
\be\ba
Z^2{}_{1}Z^1{}_2 &= (Y_1-Y_2)(\xi_1 - Y_1)(\xi_2 - Y_1) \,, \cr 
Z^3{}_{2}Z^2{}_3 &= (Y_1 - Y_2)(Y_2-Y_3) \,, \cr  
Z^4{}_{3}Z^3{}_4 &= (Y_2 - Y_3)(Y_3-\xi_4) \,.
\label{TAHBRel1}
\ea\ee
 
The second set of configurations have a semi F1 stretched between a pair of D5 branes, D5$_{(\alpha)}$-D5$_{(\beta)}$, with $\alpha +1 < \beta$, ending on the D3 segments from above or below. An example is shown in Figure \ref{TAHBopAndRel}-d. 
Looking at the setups as several semi F1 patches or a single semi F1 extended from D5$_{(\alpha)}$ to D5$_{(\beta)}$ leads to the second set of HB pre-relations. For instance for $(\alpha,\beta)=(1,3)$, we obtain the two relations
\be\ba
 Z^3{}_2 Z^2{}_1 &= Z^3{}_1 Z^2{}_2 = Z^3{}_1 (Y_1 - Y_2) \,, \cr
  (-Z^1{}_2) (-Z^2{}_3) &= (-Z^1{}_3)(-Z^2{}_2) = (-Z^1{}_3)(Y_2 - Y_1) \,.
\ea\ee
In total we get the six pre-relations
\be\ba
Z^3{}_2 Z^2{}_1 &= Z^3{}_1 (Y_1 - Y_2) \,, \quad Z^1{}_2 Z^2{}_3 = Z^1{}_3(Y_1 - Y_2) \,, \cr
Z^4{}_3 Z^3{}_2 &= Z^4{}_2 (Y_2 - Y_3) \,, \quad Z^2{}_3 Z^3{}_4 = Z^2{}_4(Y_2 - Y_3) \,, \cr
Z^4{}_3 Z^3{}_2 Z^2{}_1 &= Z^4{}_1 (Y_1 - Y_2)(Y_2-Y_3) \,, \quad Z^1{}_2 Z^2{}_3 Z^3{}_4 = Z^1{}_4(Y_1 - Y_2)(Y_2-Y_3)  \,,
\label{TAHBRel2}
\ea\ee
completing the three relations \eqref{TAHBRel1}.

\medskip

\begin{figure}[th]
\centering
\includegraphics[scale=0.7]{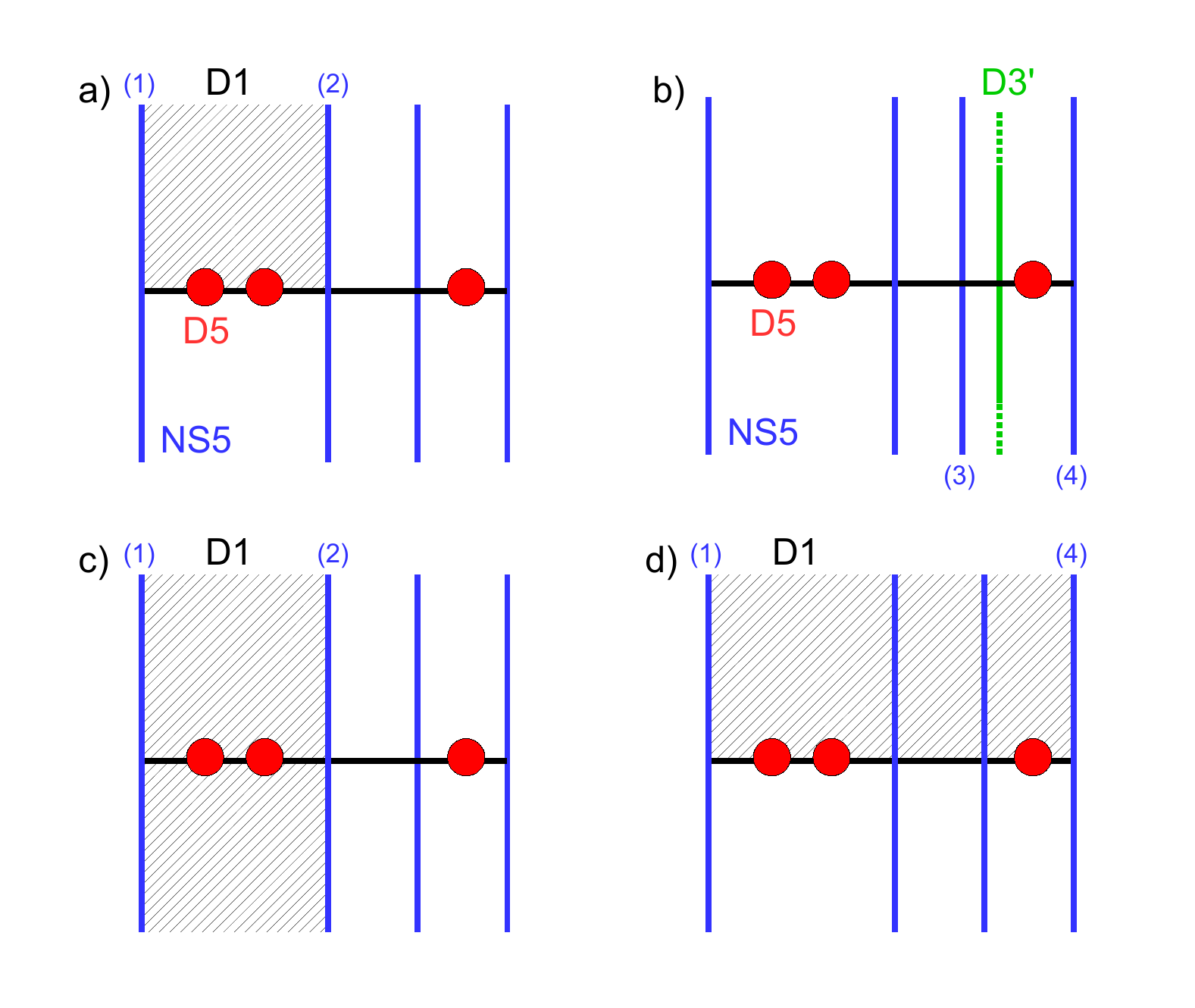}
\vspace{-0.5cm}
\caption{\footnotesize{Brane setups inserting the CB operators of theory B: a) $V^+_{[1:1]}$; b) $\varphi_3$; c) $V^+_{[1:1]}(-V^{-}_{[1:1]})$ or $(m_1-\varphi_1)(m_2-\varphi_2)(\varphi_2-\varphi_1)$; d) $V^+_{[1:1]}V^+_{[2:2]}V^+_{[3:3]}$ or $V^+_{[1:3]} (\varphi_1-\varphi_2)(\varphi_2-\varphi_3)$.}}
\label{TBCBopAndRel}
\end{figure}

The analysis of the Coulomb branch of the $T_B$ theory is completely analogous. A basis of CB generators is given by two sets of operators:
\begin{itemize}
\item For each pair of NS5 branes, NS5$_{(i)}$-NS5$_{(j+1)}$, with $1 \le i \le j \le 3$, there is a couple of operators $V^{\pm}_{[i:j]}$. The operator $V^{+}_{[i:j]}$ is associated to the contribution from a semi D1 string stretched between NS5$_{(i)}$ and NS5$_{(j+1)}$, ending on the D3 segments from above. The operator $-V^{-}_{[i:j]}$ is associated to the contribution from a semi D1 string stretched between NS5$_{(i)}$ and NS5$_{(j+1)}$, ending on the D3 segments from below. Figure \ref{TBCBopAndRel}-a presents the setup inserting the operator $V^{+}_{[1:1]}$.\footnote{As explained in previous sections, there is no setup inserting the operators $V^{\pm}_{[i:j]}$, with $|i-j|\ge1$, alone.}
 In the gauge theory language the operators $V^{\pm}_{[i:j]}$ are monopoles of charge $\pm 1$ under the $U(1)_{(k)}$ nodes for $i \le k\le j$, and vanishing charge under the other nodes.
 \item For each pair of adjacent NS5, NS5$_{(i)}$-NS5$_{(i+1)}$ for  $i=1,2,3$, there is an operator $\varphi_i$, corresponding to the vector multiplet complex scalar of the $i$th node, whose insertion is achieved by adding a D3'' brane between NS5$_{(i)}$ and NS5$_{(i+1)}$, crossing the D3 segment stretched between them. Figure \ref{TBCBopAndRel}-b presents the setup inserting the operator $\varphi_3$.
\end{itemize}

There are in total six operators $V^{\pm}$ and three operators $\varphi$, so nine CB generators. As for the Higgs branch, the CB pre-relations are read from two kinds of brane setups. 

The first setups consist of a full D1 stretched between two consecutive NS5s, NS5$_{(i)}$-NS5$_{(i+1)}$, as in Figure \ref{TBCBopAndRel}-c for $i=1$. Looking at these configurations as two semi-infinite D1, inserting the product of monopoles $V^+_{[i:i]}(-V^{-}_{[i:i]})$, or as a single D1 leads to the three identities
\be\ba
-V^+_{[1:1]}V^{-}_{[1:1]}  &= (\ti m_1-\varphi_1)(\ti m_2-\varphi_2)(\varphi_2-\varphi_1) \,, \cr
-V^+_{[2:2]}V^{-}_{[2:2]}  &= (\varphi_{1}-\varphi_2)(\varphi_{3}-\varphi_2) \,, \cr
-V^+_{[3:3]}V^{-}_{[3:3]}  &= (\varphi_{2}-\varphi_3)(\ti m_3-\varphi_3) \,.
\label{TBCBRel1}
\ea\ee
The right-hand side factors $(\ti m_i-\varphi_j)$ correspond to contributions from D1s crossing the D3$_{(j)}$ segment and intersecting D5$_{(i)}$ at a point, and the factors $\pm(\varphi_{i} - \varphi_{i-1}$) corresponds to the contributions of a NS5-D3 intersection with a D1 ending on the NS5 from the left or from the right.

The second kind of configurations have a semi D1 stretched between a pair of NS5 branes, NS5$_{(i)}$-NS5$_{(j+1)}$, with $i < j$, ending on the D3 segments from above or from below. The case $(i,j)=(1,3)$ is presented in Figure \ref{TBCBopAndRel}-d. 
Looking at the setups as several semi D1s inserting the product of monopole operators $\prod_{k=i}^{j} (\pm V^{\pm}_{[k:k]})$ or a single semi D1 extended from NS5$_{(i)}$ to NS5$_{(j+1)}$ leads to the six remaining CB pre-relations
\be\ba
 V^{\pm}_{[1:1]}  V^{\pm}_{[2:2]} &= V^{\pm}_{[1:2]} (\varphi_1-\varphi_2) \,, \cr
  V^{\pm}_{[2:2]}  V^{\pm}_{[3:3]} &= V^{\pm}_{[2:3]} (\varphi_2-\varphi_3) \,, \cr
 V^{\pm}_{[1:1]}  V^{\pm}_{[2:2]} V^{\pm}_{[3:3]} &= V^{\pm}_{[1:3]} (\varphi_1-\varphi_2)(\varphi_2-\varphi_3) \,.
 \label{TBCBRel2}
\ea\ee
The relations \eqref{TBCBRel1} and \eqref{TBCBRel2} are the pre-relations from which one can extract the full set of CB relations of the $T_B$ theory.

\bigskip

\noindent {\bf Mirror symmetry}:
\medskip

From our description of the HB and CB operators from brane setups, it is immediate to identify the mirror map. The setups realizing the HB operator insertions in theory A are S-dual to the setups realizing the CB operator insertions in theory B (Figures \ref{TAHBopAndRel}-a,b are S-dual to Figures \ref{TBCBopAndRel}-a,b), and the setups realizing the HB pre-relations in theory A are S-dual to the setups realizing pre-relations in theory B (Figures \ref{TAHBopAndRel}-c,d are S-dual to Figures \ref{TBCBopAndRel}-c,d). 
This leads to the mirror map of operators and parameters presented in Table \ref{tab:MirrorMap2} and the map between the pre-relations \eqref{TAHBRel1} and \eqref{TBCBRel1} on one side, and between the pre-relations \eqref{TAHBRel2} and \eqref{TBCBRel2} on the other side.
\begin{table}[h]
\begin{center}
\setlength\extrarowheight{3pt}
\begin{tabular}{|c|c|}
\hline
  Theory A: HB  &  Theory B : CB  \\    
\hline
 $Y_i$ \, $(i=1,2,3)$  & $\varphi_{i}$  \\ [2pt]
\hline
 $Z^{j+1}{}_{i}$ \, $(i\le j)$ & $V^+_{[i:j]}$  \\ [2pt]
\hline
  $Z^{i}{}_{j+1}$ \, $(i\le j)$ & $V^-_{[i:j]}$   \\  [2pt]
\hline
\hline
  Theory A: FI param. &    Theory B: mass param.  \\ 
\hline
$\xi_1, \xi_2, \xi_3$  &  $\ti m_1, \ti m_2, \ti m_3$   \\ 
\hline
\end{tabular}
\caption{\footnotesize Mirror map of HB operators of $T_A$ and CB operators of $T_B$, and mirror map of FI and mass parameters.}
\label{tab:MirrorMap2}
\end{center}
\end{table}

\section{Other chiral operators}
\label{sec:OtherOp}

So far the discussion has been focused solely on finding the brane setups inserting HB and CB operators entering in bases of the chiral rings. In the gauge theory description of the Coulomb branch it is natural to consider monopole operators of higher magnetic charges. Ultimately these higher charge monopoles are expressed through additional relations in terms of the monopole operators of lower charge forming a basis of the Coulomb branch. In the description of the Higgs branch these redundant operators are simply products of mesons. For completeness we discuss in this section the brane setups inserting some of these higher charge operators (or these products of operators). 

\begin{figure}[th]
\centering
\includegraphics[scale=0.7]{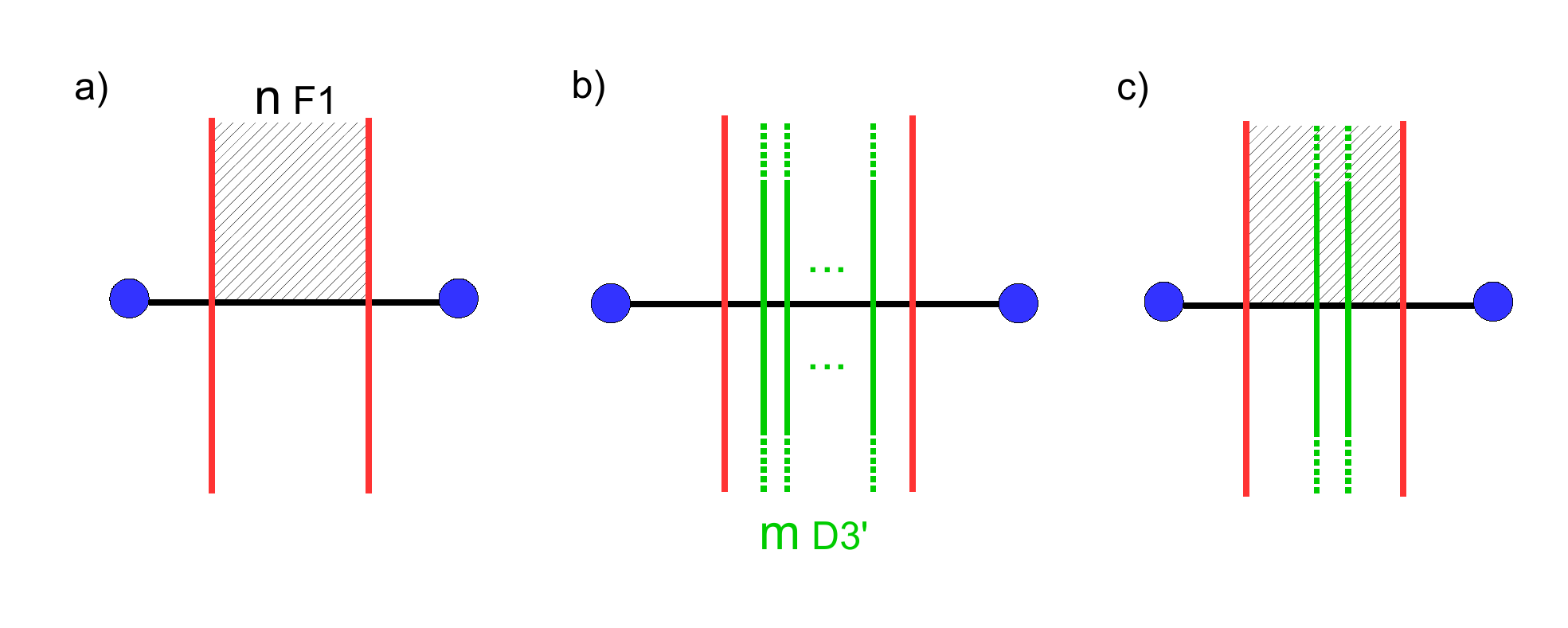}
\vspace{-0.5cm}
\caption{\footnotesize{Products of HB operators (mesons) are inserted by adding several F1 strings and D3' branes to the setups. a) $n$ semi F1s inserting $(Z^2{}_{1})^n$, b) $m$ D3's inserting $Y^m = (Z^2{}_2 - \xi_2)^m$, c) One semi F1 and two D3's inserting $Z^2{}_{1} Y^2$.}}
\label{OtherOpHB}
\end{figure}

On the Higgs branch side,  products of meson operators can be realized by brane setups with several F1 strings and/or several D3' branes. For instance, in the $T[SU(2)]$ theory studied in Section \ref{sec:TSU2} one can consider the setup with $n$ semi F1 strings stretched between the two D5s and ending on the D3 from above, as shown in Figure \ref{OtherOpHB}-a. Each semi F1 inserts a factor of the meson operator $Z^2{}_{1}$, leading to the total insertion of $(Z^2{}_{1})^n$.
We can also consider the setup with several $m$ D3' branes positioned between the two D5s, as in Figure \ref{OtherOpHB}-b, inserting the product of operators $Y^m \equiv (-Z^1{}_1 -\xi_1)^m = (Z^2{}_2 - \xi_2)^m$. A setup inserting the product $Z^2{}_{1} Y^2$ is shown in Figure \ref{OtherOpHB}-c.
Notice that the s-rule (see Section \ref{ssec:BranesLocalOp}) forbids to attach more that one F1 string between a D3' and a D5 brane, therefore the only way to have several strings attached to a D3' brane is that each of them ends on a different D5. This restricts the set of possible configurations. For the $T[SU(2)]$ theory, the various brane setups realize all the products of mesons $(Z^2{}_{1})^{n_1} (Z^{1}{}_{2})^{n_2} Y^{n_3}$, for $n_1, n_2,n_3 \ge 0$.
\medskip

\begin{figure}[th]
\centering
\includegraphics[scale=0.7]{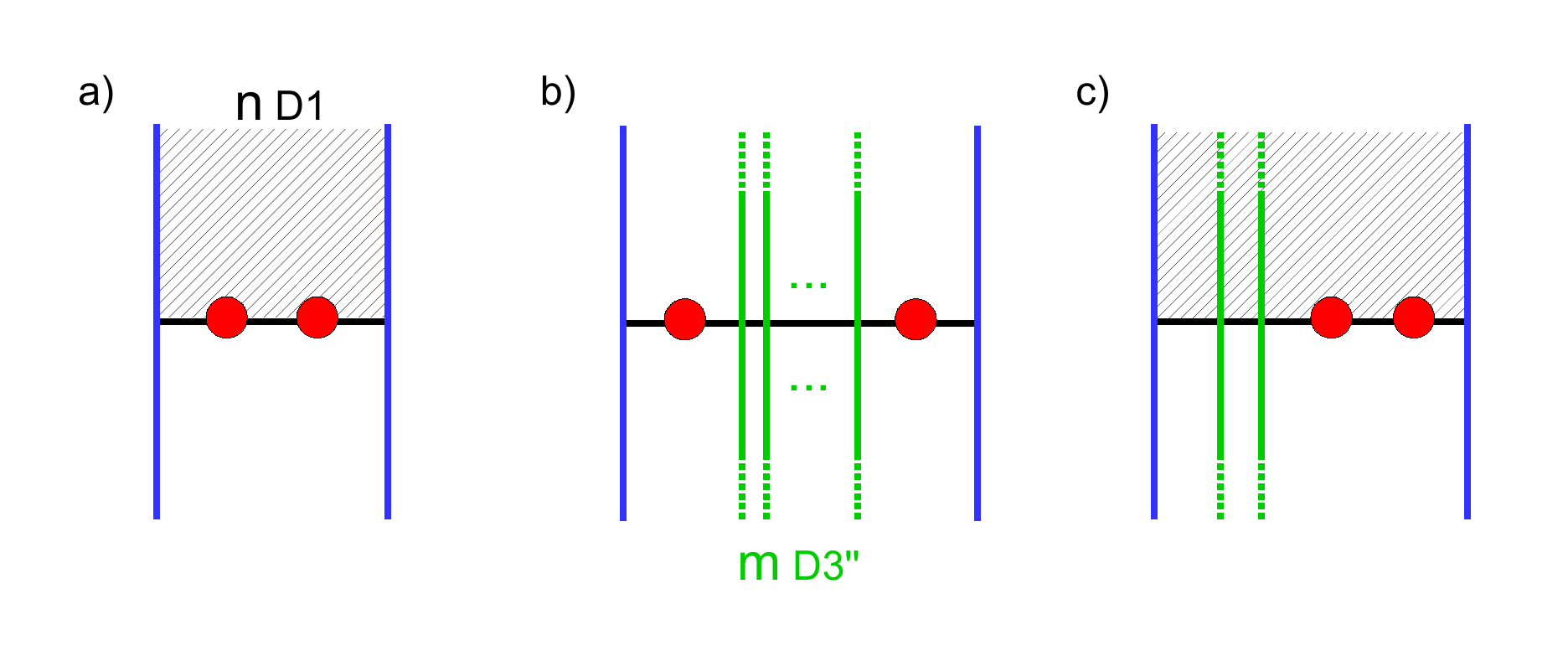}
\vspace{-0.5cm}
\caption{\footnotesize{Products of CB operators are inserted by adding several D1 strings and D3'' branes to the setups. a) $n$ semi D1s inserting $(u^+)^n$, b) $m$ D3''s inserting $\varphi^m$, c) One semi D1 and two D3''s inserting $u^+ \varphi^2$.}}
\label{OtherOpCB}
\end{figure}

On the Coulomb branch side, there are monopole operators $V_A$ of arbitrary magnetic charge $A \in \bZ^M$ under a $U(1)^M$ Cartan gauge symmetry, dressed by products of scalar operators $\varphi_i$ \cite{Bullimore:2015lsa}. These higher charge monopoles are realized by brane setups with several D1 strings and D3'' branes. 
In the simple case of the $T[SU(2)]$ theory, we can consider $n$ semi D1s stretched between the two NS5s and ending on the D3 segment from above, as in Figure \ref{OtherOpCB}-a. The insertion corresponds to a monopole operator $V_n$ of charge $n$. It can also be interpreted as the product of monopole operators $(u^+)^n$, leading to the CB relation $V_n = (u^+)^n$.
The setup of Figure \ref{OtherOpCB}-b with $m$ D3'' branes inserts the products of scalars $\varphi^m$. The setup of Figure \ref{OtherOpCB}-c inserts the operator $u^+ \varphi^2$. Here also the possible brane setups are restricted by the s-rule which imposes that there is at most a single D1 string stretched between a D3'' brane and an NS5 brane.

These three brane setups of Figure \ref{OtherOpCB} are S-dual to those of Figure \ref{OtherOpHB}, confirming the mirror map between operators $(u^+, \varphi) \leftrightarrow (Z^2{}_1, Y)$. 

\medskip

\begin{figure}[th]
\centering
\includegraphics[scale=0.75]{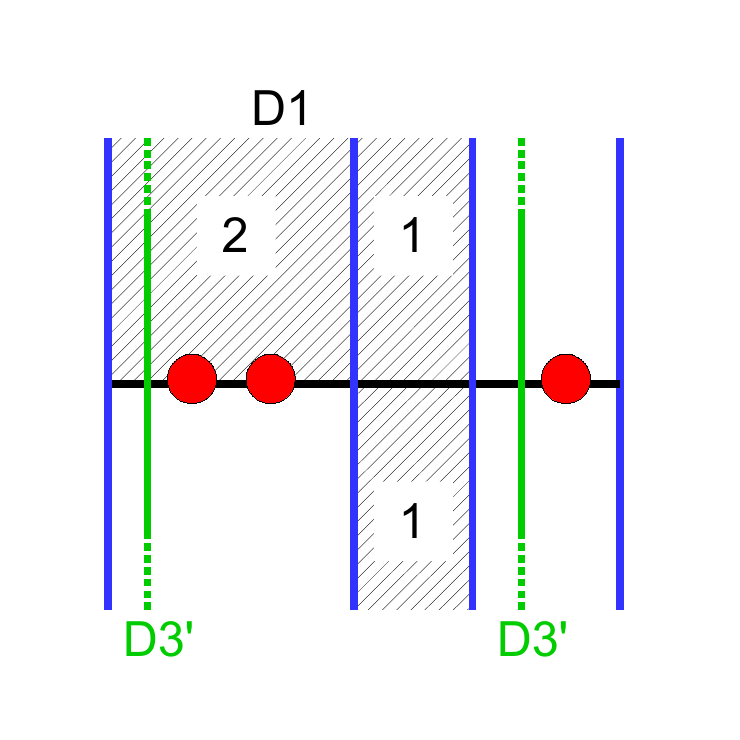}
\vspace{-0.5cm}
\caption{\footnotesize{Setup with D1s and D3''s. The numbers indicate the number of superposed D1s in each region. The setup can be interpreted as inserting for instance the operators $-V_{(2,0,0)} V_{(0,1,0)}  V_{(0,-1,0)}\varphi_1\varphi_3$ or $(V_{(1,0,0)})^2 (\varphi_1 - \varphi_2)(\varphi_3 - \varphi_2)\varphi_1\varphi_3$.}}
\label{OtherOpEx}
\end{figure}

In more sophisticated abelian theories one can consider richer setups which have many interpretations, corresponding to operator insertions identified by ring relations. Let us give one example. We consider the CB operator insertion in the theory B of Section \ref{ssec:Example} given by the brane setup of Figure \ref{OtherOpEx}. The presence of the two D3'' branes corresponds to the insertion of the product $\varphi_1\varphi_3$. The insertion due to the D1 branes can be interpreted in various ways. As four distinct semi D1s, they insert the products $-V_{(2,0,0)} V_{(0,1,0)}  V_{(0,-1,0)}  = -(V_{(1,0,0)})^2 V_{(0,1,0)}  V_{(0,-1,0)}$. Recombining the two semi D1 in the middle across the D3 segment leads to the insertion of $V_{(2,0,0)} (\varphi_1 - \varphi_2)(\varphi_3 - \varphi_2)  = (V_{(1,0,0)})^2 (\varphi_1 - \varphi_2)(\varphi_3 - \varphi_2)$. Recombining instead two D1s on the upper side across the NS5 leads to the insertion of $-V_{(2,1,0)} (\varphi_1-\varphi_2)  V_{(0,-1,0)}  = - V_{(1,0,0)} V_{(1,1,0)} (\varphi_1-\varphi_2)  V_{(0,-1,0)}$. Here we denoted $V_{(n_1,n_2,n_3)}$ the monopole operator of charge $(n_1,n_2,n_3)$. All these operators are equal by the CB relations. The relations that we obtain from such complicated setups are therefore correct but redundant. The operators $V_{(2,0,0)}$ and $V_{(2,1,0)}$ can be eliminated by the relations $V_{(2,0,0)} = (V_{(1,0,0)})^2$ and $V_{(2,1,0)} = V_{(1,0,0)} V_{(1,1,0)}$ (which follow from simpler brane setups).

\section{Non-abelian theories}
\label{sec:NATheories}

It would be very interesting to generalize this analysis to non-abelian theories and to find a set of simple rules to derive the non-abelian Coulomb branch relations and the mirror maps in a systematic way.
This generalization is not straightforward, in the sense that it is not obvious that brane setups can be found which insert each of the CB and HB operators in non-abelian theories with brane realizations.. However, by a simple analysis, we can derive the {\it abelianized relations} discussed in \cite{Bullimore:2015lsa}, from which one can extract the actual CB relations of the non-abelian theory by the method explained in that paper \footnote{The basic idea is the following. Each non-abelian operator can be expressed as a gauge invariant polynomial in the abelian operators ({\it abelianization map}) and the abelianized relations can be re-expressed as relations between the non-abelian operators, giving the CB relations of the non-abelian theory.}. These are the Coulomb branch relations of the abelian low-energy theory that one obtains by moving at a generic point of the Coulomb branch with corrections due to massive W-bosons. It is worth noting that the abelianized relations were postulated\footnote{Except for the Coulomb branch relation for $\N=4$ SQED which was derived in \cite{Borokhov:2002cg} using CFT methods.} in \cite{Bullimore:2015lsa} and that the present discussion constitutes a direct derivation of these relations relying on brane technology.

\begin{figure}[th]
\centering
\includegraphics[scale=0.7]{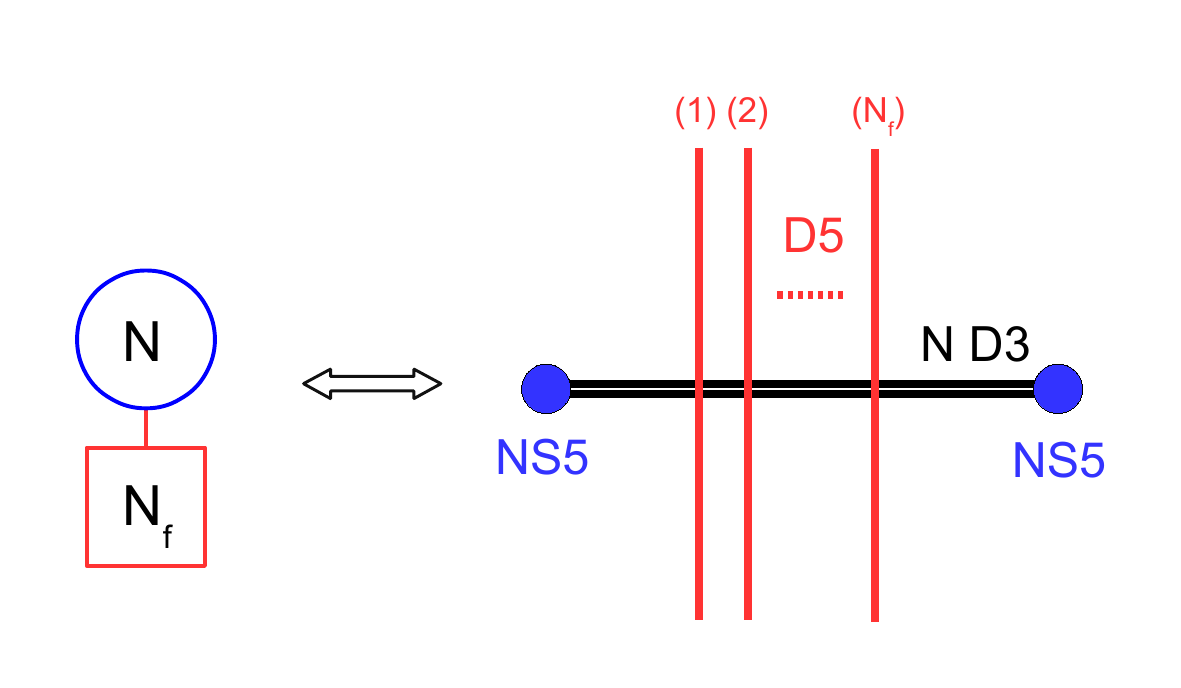}
\vspace{-0.5cm}
\caption{\footnotesize{Quiver diagram and brane realization of $U(N)$ SQCD.}}
\label{SQCD}
\end{figure}

Let us consider the simple example of $U(N)$ SQCD with $N_f$ fundamental hyper-multiplets. The gauge field $A_\mu$ and the complex scalar $\varphi$ belong to the $\mathfrak{u}(N)$ algebra and we can view them as $N\times N$ matrices. One can define a monopole operator $M_{A,p}$ by 
specifying Dirac monopole singularities with integer charges $A = (\mathfrak{n}_1,\mathfrak{n}_2, \cdots, \mathfrak{n}_N)$ for the diagonal components of the gauge field at a point in space. This breaks the gauge group $U(N)$ to a subgroup $H$, which is generically a maximal torus $U(1)^N$. The monopole operator is then dressed with an $H$-invariant polynomial $p(\varphi_H)$ of the complex scalar $\varphi$ restricted to $H$, which is generically a polynomial of the diagonal elements $\varphi_a$, $a=1, \cdots, N$. 
The abelianization map of \cite{Bullimore:2015lsa} maps a non-abelian monopole operator $M_{A,p}$ to a combination of monopole operator $V_A$ dressed with complex scalars $\varphi_a$ of the low-energy abelian theory of generic Coulomb branch loci. The Coulomb branch of the non-abelian theory is a sub-ring of the Coulomb branch of the abelian theory and the non-abelian ring relations can be extracted from ``abelianized relations" involving the operators of the abelian theory.

In the case of SQCD with $N_f$ flavors, the abelianized relations read\footnote{There is a sign difference with respect to the formula in \cite{Bullimore:2015lsa}, which can be absorbed in a redefinition of the vector multiplet complex scalars by a minus sign.}
\be
u^+_a u^-_a = - \frac{\prod\limits_{\alpha=1}^{N_f} (m_\alpha - \varphi_a)}{\prod\limits_{b=1 \atop b \neq a}^N (\varphi_b-\varphi_a)^2} \,, \quad  a=1, \cdots, N \,,
\label{SQCDAbRel}
\ee
where $u^{\pm}_a, \varphi_a$, $a=1, \cdots, N$, are the abelian monopole operators and complex scalar operators of the $U(1)^N$ low-energy abelian theory at a generic point on the Coulomb branch. The difference with the relations of actual abelian quiver theories is the presence of the factors $(\varphi_b-\varphi_a)^2$ in the denominator on the right hand side, due to the presence of massive W-bosons in the low-energy theory on the Coulomb branch. A basis of generators of the non-abelian Coulomb branch is given by Weyl invariant combinations of the abelian operators $u^{\pm}_a, \varphi_a$ with minimal magnetic charge, $M^{\pm}_{n} = \sum_{a=1}^N u^{\pm}_a \prod_{b_1\neq \cdots \neq b_n \neq a} \varphi_{b_1} \cdots \varphi_{b_n}$ and vanishing magnetic charge, $\Phi_n = \sum_{a_1\neq\cdots\neq a_n} \varphi_{a_1}\cdots\varphi_{a_n}$. The non-abelian CB relations obeyed by these operators amounts to rewriting the abelianized relations \eqref{SQCDAbRel} in terms the these Weyl invariant generators, as detailed in \cite{Bullimore:2015lsa}.
We propose to recover the (crucial) abelianized relations \eqref{SQCDAbRel} from a brane setup. 

\medskip

The brane configuration realizing $U(N)$ SQCD with $N_f$ hyper-multiplets as its low-energy limit consists of $N$ D3 segments stretched between two NS5 branes with $N_f$ D5 branes crossing the D3s ( Figure \ref{SQCD}).
In the brane picture, moving to a generic point on the Coulomb branch by giving vevs to the diagonal components $\varphi_a$ of  the complex scalar is achieved by moving the D3 segments along the $x^{8+i9}$ direction, with the position of the D3 segments corresponding to the abelian scalars vevs $\vev{\varphi_a}$, $a=1, \cdots, N$. 

\begin{figure}[th]
\centering
\includegraphics[scale=0.75]{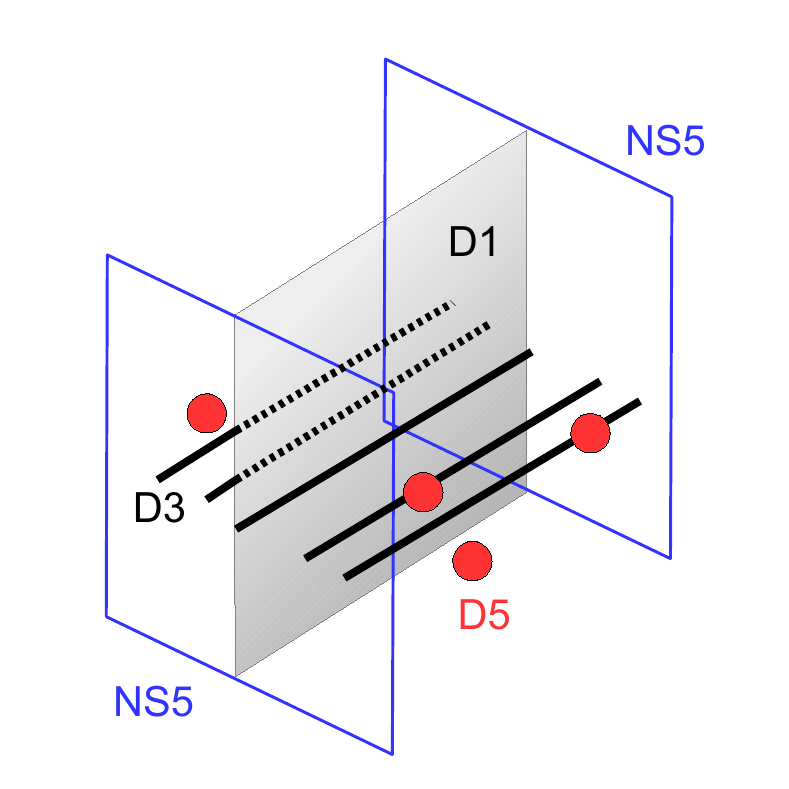}
\vspace{-0.5cm}
\caption{\footnotesize{Brane setup in the 3d space $(x^3,x^7, x^8)$ with a D1 string stretched between the two NS5s and intersecting a D3 brane. In this example there are $N=5$ D3s and $N_f=4$ D5s.}}
\label{SQCDBraneRel}
\end{figure}

To obtain the abelianized relations the relevant brane setup is that of Figure \ref{SQCDBraneRel}, where the D3 segments are separated. In addition the setup has a D1 brane stretched between the two NS5s and intersecting one of the D3 segments, say the D3$_{(a)}$ segment supporting the $U(1)_{(a)}$ Cartan gauge symmetry.
This setup can be interpreted as two semi D1 branes ending on D3$_{(a)}$ inserting the product of abelian monopole operators $-u^+_a u^-_a$ according to the rules explained in previous sections.\footnote{As in Section \ref{ssec:CBop}, we assume here that there is no contribution to the operator insertion from D1-D5 or D1-D3 string modes. This is can be heuristically understood by noting that a semi D1 is dissolved into the D3 segment (or forms a spike) so that there is no separate D1-D5 or D1-D3 strings, but only D3-D5 and D3-D3 strings.}

The second interpretation is that of a single D1 brane crossing the D3. This setup is essentially the same as that of Section \ref{ssec:Relations} and Figure \ref{FullD1}. The operator insertion comes from integrating out the low-energy fields living on the D1 and those living at the D1-D5 and D1-D3 intersections. 

The D1 brane theory is a 2d $\N=(8,8)$ $U(1)$ vector multiplet placed on an interval with half-BPS Neumann boundary conditions. These boundary conditions set to zero three out of the eight real scalars and the vector field component along $x^3$. In the low energy limit the massless modes make an 1d $\N=(4,4)$ vector multiplet, which has five scalars corresponding to the motions of the D1 along the directions $x^{0,1,2,7,8}$.
 
The D1-D3$_{b}$ open string low modes make a 1d hypermultiplet living on the interval between the two NS5s. The fields of the hyper-multiplet are $(q_r, \psi^{\dot r}_{\pm})$, $r=1,2$, $\dot r=1,2$, where $q_r$ denote the two complex scalars and $\psi^{\dot r}_{\pm}$ the four fermions, which have a single complex component in 1d. As is frequently the case when dealing with a supersymmetric theory on Euclidean space it will be useful to treat the conjugate fields $(\overline q^r, \overline\psi_{\pm\, \dot r})$ as independent fields to begin with.
The hyper-multiplet has charge $(1,-1)$ under the $U(1)_{\rm D1} \times U(1)_{\rm D3}$ gauge symmetries and therefore is coupled to the 2d and 3d vector multiplets living on the D1 and D3 respectively. The couplings are canonical couplings to 1d $N=(4,4)$ vector multiplets embedded inside the 2d and 3d vector multiplets. In particular the 2d and 3d vector multiplet scalars appear as masses for the 1d hyper-multiplet.\footnote{In this analysis the 2d and 3d fields can be treated as background fields.}

 The boundary conditions on this hyper-multiplet at the boundaries of the interval, compatible with the Neumann boundary conditions for the vector multiplets, are imposing $\p_3 q_r=0$. Since the boundary condition is half-BPS, half of the fermions are set to zero at the boundaries as well, while the other half obeys the Neumann condition $\p_3 \psi =0$. 
 To understand this in detail, let us write the low-energy effective action and the supersymmetry transformations, keeping all the fermions for the moment. In the low-energy limit the theory is zero dimensional and the only term in the action is a mass coupling to the vector multiplet scalars which are not set to zero by the boundary conditions,\footnote{In principle there should be additional Yakawa couplings involving the fermions of the 2d and 3d vector multiplets, which are the superpartners of $\Phi$. Here we perform the computation in the situation when these background fermions are set to zero, assuming that the final result is unchanged when they are turned on. The fact that we obtain in the end a supersymmetric chiral operator after integrating out the 0d theory and that these background fermions do not enter in chiral operators justifies this assumption.}
 \be
S_{\rm 0d} =  |\Phi |^2 \overline q^r q_r + \Phi \overline \psi_{+ \, \dot r} \psi^{\dot r}_-  + \overline\Phi \, \overline \psi_{- \, \dot r} \psi^{\dot r}_+ \,, 
\ee
where $\Phi = \varphi_b - \varphi_a$ is the  complex mass of the 0d hyper-multiplet, computed as the difference between the D3$_{(b)}$ position $\varphi_b$ and the D1 position $\varphi_a$ along $x^{8+i9}$. 
The 0d supersymmetry transformations obtained by dimensional reduction of the 2d $N=(4,4)$ transformations (or from 6d $N=(1,0)$) are
\be\ba
\delta q_r &=  \psi^{\dot r}_{-} \epsilon_{+ \, r \dot r} +  \psi^{\dot r}_+ \epsilon_{- \, r \dot r} \,, \quad  \delta \overline q^r =  - \overline\psi_{\dot r \, +} \epsilon_+^{r \dot r} -  \overline\psi_{\dot r \, -} \epsilon_-^{r \dot r} \,,  \cr
\delta \psi^{\dot r}_+ &= \Phi q_r \epsilon^{r \dot r}_- \,, \quad \delta \overline \psi_{+ \, \dot r} = \overline\Phi \overline q^r \epsilon_{+ \, r \dot r} \,, \cr
\delta \psi^{\dot r}_- &= \overline{\Phi} q_r \epsilon^{r \dot r}_+ \,, \quad \delta \overline \psi_{- \, \dot r} = \Phi  q^r \epsilon_{- \, r \dot r} \,, 
\ea\ee
where $\epsilon_{\pm \, r \dot r}$ parametrize eight supercharges. Out of these eight supercharges, four are broken by the boundary conditions on the interval, which set to zero half of the fermions. There are several choices of which fermions to set to zero, each preserving a certain subset of supercharges. Studying the different choices we find that there is a single choice leading to an operator insertion preserving the supercharges of a Coulomb branch operator, which are the supercharges preserved by the original brane configuration. The correct boundary condition is then
\be
\overline \psi_{+ \, \dot r} = \psi^{\dot r}_- = 0 \,,
\ee
preserving the four supercharges parametrized by $\epsilon_{- \, r \dot r}$, leading to a  0d  $\N=(4,0)$ theory.
This analysis indicates that the appropriate reality condition on the 0d fermions is not the usual Lorentzian reality condition $(\psi^{\dot r}_\pm)^\ast =  \pm \overline \psi_{\pm \, \dot r}$ and $(\epsilon_{\pm \, r \dot r})^\ast = \mp\epsilon^{r \dot r}_\mp$, which would lead to a real action, but instead we have $(\psi^{\dot r}_\pm)^\ast =  \pm \overline \psi_{\mp \, \dot r}$ and $(\epsilon_{\pm \, r \dot r})^\ast = \mp \epsilon^{r \dot r}_\pm$ and a complex action. 
After setting $\psi^{\dot r}_-=0$, we obtain
\be
S_{\rm 0d} =  |\Phi |^2 \overline q^r q_r  + \overline\Phi \, \overline \psi_{- \, \dot r} \psi^{\dot r}_+ \,.
\ee
Integrating out the two complex scalars $q^r$ and the two complex fermions $\psi^{\dot r}_+$ leads to the operator insertion
\be
\frac{ \overline\Phi ^2}{|\Phi |^4} = \frac{1}{\Phi^2} = \frac{1}{(\varphi_b - \varphi_a)^2} \,,
\ee
which is indeed a CB operator, confirming that we have correctly derived the boundary conditions on the hyper-multiplet fermions.

In the brane setup (Figure \ref{SQCDBraneRel}) there are $N$ D1-D3 hyper-multiplets, however one of them, the D1-D3$_{(a)}$ is massless, since the D1 is exactly crossing D3$_{(a)}$ and its contribution is diverging. The path integral must be regularized by removing the massless modes. This leaves the contribution of the $N-1$ massive hyper-multiplets
\be
\frac{1}{\prod \limits_{b\neq a} (\varphi_b - \varphi_a)^2} \,.
\ee
In addition there are $N_f$ zero-dimensional fermions with complex masses $m_\alpha - \varphi_a$ sourced by the D1-D5$_{(\alpha)}$ open strings modes, with $\alpha=1, \cdots, N_f$. As explained in Section \ref{ssec:Relations}, integrating out these fermions lead to the insertion of the product of CB operators $\prod_{\alpha=1}^{N_f} (m_\alpha - \varphi_a)$. The total operator insertion is then
\be
\frac{\prod_{\alpha=1}^{N_f} (m_\alpha - \varphi_a)}{\prod \limits_{b\neq a} (\varphi_b - \varphi_a)^2} \,.
\ee
Equating this insertion with the product of abelian monopoles $-u^+_a u^-_a$ obtained by the first interpretation of the brane setup reproduces the SQCD abelianized relation \eqref{SQCDAbRel}.

\medskip

Once again we have shown that the Coulomb branch relations can be derived by studying specific brane setups. This analysis can be generalized to deriving the abelianized ring relations of arbitrary non-abelian linear quiver theory with unitary nodes, from which the Coulomb branch relations of the non-abelian theory are derived following the approach of \cite{Bullimore:2015lsa}. We did not find a way to extract directly the non-abelian CB relations from the brane picture. It might be possible however to extract the mirror map of operators in a simple way. We leave this analysis for future work.

\medskip

Finally we can comment on the derivation of the Higgs branch relations in non-abelian theories. These relations are usually derived from the description of the Higgs branch as a hyper-K\"ahler quotient (or a holomorphic quotient), which follows from the Lagrangian description of the theory. However one may wonder how the Higgs branch would arise from the brane approach. It is beyond the scope of this paper to provide the algorithmic derivation of non-abelian Higgs branch relations from branes, but we can sketch how this would work.
As in the Coulomb branch analysis above, the idea would be to derive the equivalent of abelianized relations  from brane setups and to construct the non-abelian relations from them. The analogue of the low-energy abelian theory at generic points on the Coulomb branch is the low-energy free hypermultiplets theory at generic points on the Higgs branch\footnote{If the theory is {\it bad} in the sense of \cite{Gaiotto:2008sa} there is some residual interacting gauge theory as well.}, which corresponds to moving D3 segments stretched between D5s along the $x^{5,6}$ directions. In this ``free theory" the HB operators are inserted by F1 strings stretched between couples of D5s and ending on a given D3 segment, and by D3' branes placed between neighboring D5s and crossing a given D3 segment. These are the analogue the abelian operators $u^{\pm}_a,\varphi_a$ in the Coulomb branch analysis above. The HB operators in the non-abelian theory are then constructed as polynomials in the free theory operators which are invariant under permutations of the D3 segments (note that this is not the Weyl group of the non-abelian theory). The non-abelian HB relations would be obtained from a set of relations, which are the analogue of the abelianized relations for the Coulomb branch. These relations would follow from brane setups with full F1 strings stretched between two D5s and crossing a D3 segment. The details of the brane reading procedure would need a longer discussion and there might be additional brane setups to be considered to obtain the full set of Higgs branch relations in a generic non-abelian theory. The simplest situation where one could test this method would be for a non-abelian quiver theory whose brane realization has only two D5s (two fundamental hypermultiplets).

The main idea is that the brane analysis provides a derivation of the abelianized relations for the Coulomb branch and of an analogue set of relations for the Higgs branch, from which the CB and HB relations of the non-abelian theory can be derived following the method of \cite{Bullimore:2015lsa}.


\section*{Acknowledgments}

We wish to thank Costas Bachas, Stefano Cremonesi , Davide Gaiotto, Jaume Gomis and  Noppadol Mekareeya for enlightening discussions related to this work.
This research was supported in part by Perimeter Institute for Theoretical Physics. Research at Perimeter Institute is supported 
by the Government of Canada through the Department of Innovation, Science and Economic Development and by the Province of Ontario through the Ministry of Research, Innovation and Science.


\appendix

\section{SQED Higgs branch relations}
\label{app:SQEDHBRel}

In this appendix we show that the per-relations \eqref{HBRel000}, \eqref{HBRel0} and \eqref{HBRel00}, following from the brane setup analysis, imply the trivial Higgs branch relations \eqref{HBRelSQED1}.

Let us assume $\alpha < \beta < \gamma < \delta$. Using \eqref{HBRel0} and \eqref{HBRel00}, we have
\be\ba
& Z^{\delta}{}_{\alpha} \prod_{\epsilon = \alpha+1}^{\delta-1} Z^\epsilon{}_{\epsilon}  = \prod_{\epsilon = \alpha}^{\delta-1} Z^{\epsilon+1}{}_{\epsilon}  \
= Z^\beta{}_\alpha \big( \prod_{\epsilon = \alpha+1}^{\beta-1} Z^\epsilon{}_{\epsilon} \big) \big(  \prod_{\epsilon = \beta}^{\gamma-1} Z^{\epsilon+1}{}_{\epsilon} \big)  Z^\delta{}_{\gamma} \big( \prod_{\epsilon = \gamma+1}^{\delta-1} Z^\epsilon{}_{\epsilon} \big) \, \cr
 & Z^{\beta}{}_{\gamma} \prod_{\epsilon = \beta+1}^{\gamma-1} Z^\epsilon{}_{\epsilon}  = \prod_{\epsilon = \beta}^{\gamma-1} Z^{\epsilon}{}_{\epsilon+1} \,.
\ea\ee
Multiplying these two equations together we obtain
\be\ba
Z^{\delta}{}_{\alpha} Z^{\beta}{}_{\gamma} \prod_{\epsilon = \alpha+1}^{\delta-1} Z^\epsilon{}_{\epsilon} \prod_{\epsilon = \beta+1}^{\gamma-1} Z^\epsilon{}_{\epsilon}  
&= Z^\beta{}_\alpha  Z^\delta{}_{\gamma} \big( \prod_{\epsilon = \alpha+1}^{\beta-1} Z^\epsilon{}_{\epsilon} \big)
\big( \prod_{\epsilon = \gamma+1}^{\delta-1} Z^\epsilon{}_{\epsilon} \big) 
\big(  \prod_{\epsilon = \beta}^{\gamma-1} Z^{\epsilon+1}{}_{\epsilon} Z^{\epsilon}{}_{\epsilon+1} \big) \cr
&= Z^\beta{}_\alpha  Z^\delta{}_{\gamma} \big( \prod_{\epsilon = \alpha+1}^{\beta-1} Z^\epsilon{}_{\epsilon} \big)
\big( \prod_{\epsilon = \gamma+1}^{\delta-1} Z^\epsilon{}_{\epsilon} \big) 
\big(  \prod_{\epsilon = \beta}^{\gamma-1} Z^{\epsilon}{}_{\epsilon} Z^{\epsilon+1}{}_{\epsilon+1} \big) \cr
&= Z^\beta{}_\alpha  Z^\delta{}_{\gamma} \prod_{\epsilon = \alpha+1}^{\delta-1} Z^\epsilon{}_{\epsilon} \prod_{\epsilon = \beta+1}^{\gamma-1} Z^\epsilon{}_{\epsilon}   \,,
\ea\ee
where the second equality follows from using  \eqref{HBRel000}.
Dividing by $\prod_{\epsilon = \alpha+1}^{\delta-1} Z^\epsilon{}_{\epsilon} \prod_{\epsilon = \beta+1}^{\gamma-1} Z^\epsilon{}_{\epsilon} $ on both side yields
\be
Z^{\delta}{}_{\alpha} Z^{\beta}{}_{\gamma} = Z^\beta{}_\alpha  Z^\delta{}_{\gamma} \,,
\ee
which is the Higgs branch relation \eqref{HBRelSQED1}. This can be done when $\prod_{\alpha < \epsilon < \delta} Z^\epsilon{}_{\epsilon} \prod_{\beta < \epsilon < \gamma} Z^\epsilon{}_{\epsilon} \neq 0$. We know that the relation still holds even when this product vanishes, but, strickly speaking this does not follow from the relations \eqref{HBRel000}, \eqref{HBRel0} and \eqref{HBRel00}, and we must add this freedom to divide on both side of a relation by the $Z^\epsilon{}_{\epsilon}$ operators  to obtain the full set of HB relations. 

The relations with a different ordering between $\alpha,\beta,\gamma,\delta$, as well as the cases when some of these indices are equal, can be obtained with similar computations, without difficulty.

\section{Brane setup - Operator insertion dictionary}
\label{app:Dictionary}

In this appendix we summarize the rules for inserting/reading Higgs branch and Coulomb branch operators from brane setups and we provide the general mirror map of operators and deformation parameters.

The HB operator insertions require a little more effort to extract than the CB insertions. This is related to the fact that the brane setups automatically solve for the F-term conditions, so that the natural operator insertions are  sometimes combinations of the standard meson operators.

In the brane configuration realizing the abelian quiver theory with $M$ nodes and $K$ fundamental hyper-multiplets, we label the NS5 branes as NS5$_{(i)}$ with $i=1, \cdots, M+1$, increasing from left to right, and the D5 branes as D5$_{(\alpha)}$ with $\alpha=1, \cdots, K$, increasing from left to right as well.

\subsection{Higgs branch operators}

To each D5$_{(\alpha)}$-D3 intersection is associated a fundamental hyper-multiplet with complex scalars $(Q^\alpha, \ti Q_\alpha)$, sourced by the D5-D3 open strings. 
To each NS5$_{(i)}$-D3 intersection is associated a bifundamental hyper-multiplet with complex scalars $(q_i, \ti q_i)$, sourced by the D3$_{(i-1)}$-D3$_{(i)}$ open strings stretched across the NS5.

\begin{figure}[th]
\centering
\includegraphics[scale=0.75]{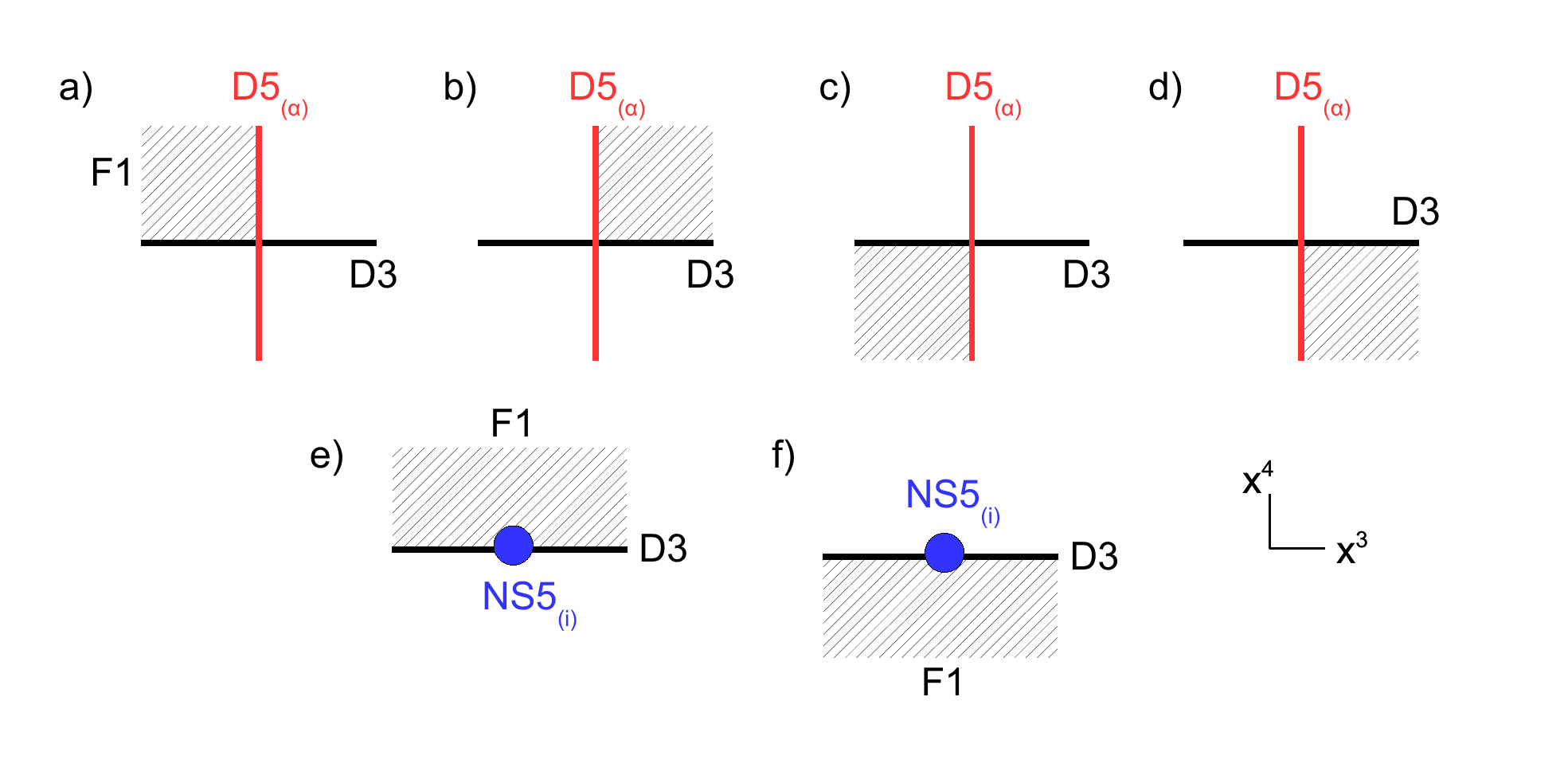}
\vspace{-1cm}
\caption{\footnotesize{Setups inserting hyper-multiplet scalar operators. The insertions are: a) $Q^a$, b) $\ti Q_a$, c) $-\ti Q_a$, d) $Q^a$, e) $q_i$, f) $-\ti q_i$.}}
\label{AppCornerF1}
\end{figure}

An F1 string ending on a D5$_{(\alpha)}$-D3 corner corresponds to the insertion of a scalar operator $Q^\alpha$, or $\pm \ti Q_{\alpha}$ as described in Figure \ref{AppCornerF1}-a,b,c,d, and an F1 string ending on the D3 of a NS5$_{(i)}$-D3 intersection corresponds to the insertion of a scalar operator $q_i$ or $-\ti q_i$, as described in Figure \ref{AppCornerF1}-e,f.

\begin{figure}[th]
\centering
\includegraphics[scale=0.72]{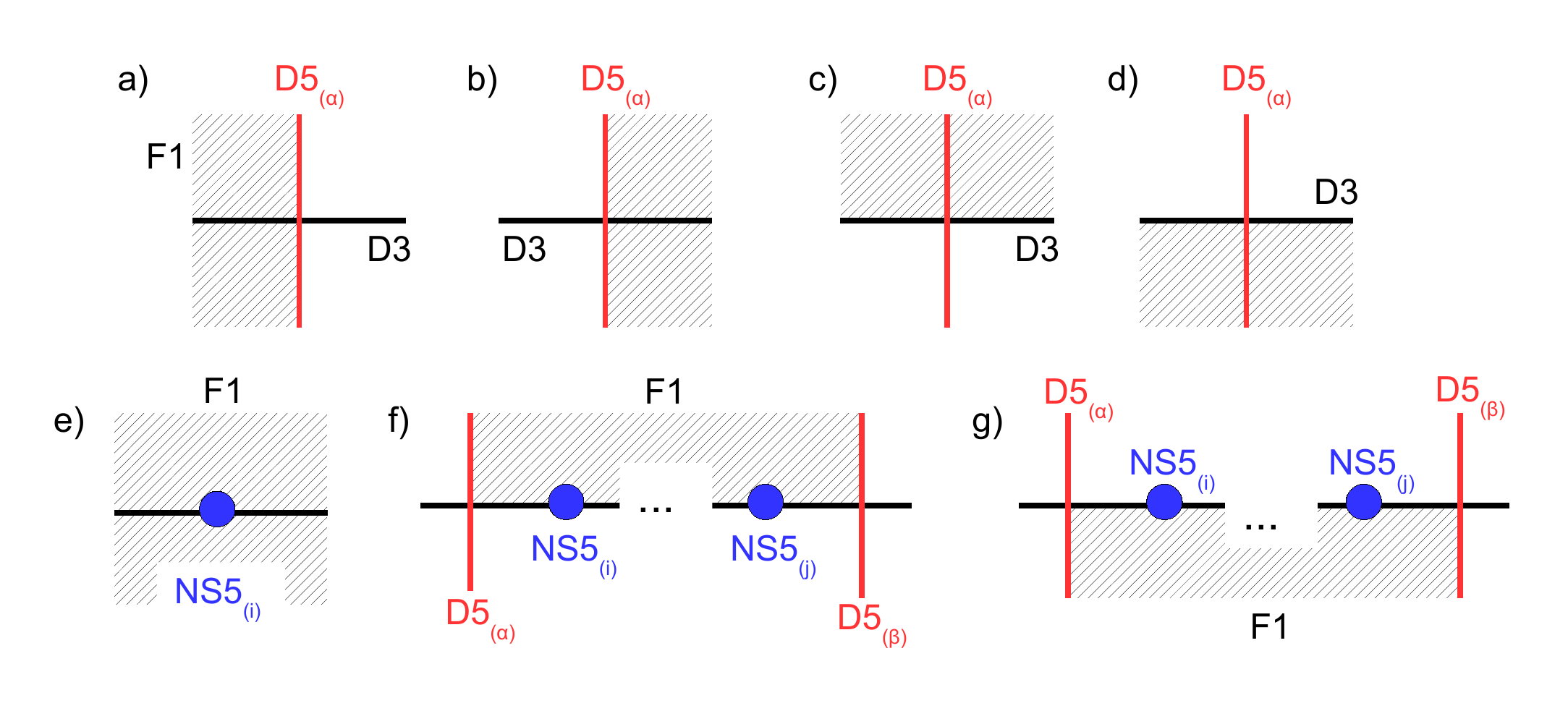}
\vspace{-1cm}
\caption{\footnotesize{Setups with F1 strings inserting HB operators (mesons). The insertions are: a) $-Z^\alpha{}_{\alpha}\equiv  -\ti Q_\alpha Q^\alpha$, b) $Z^\alpha{}_{\alpha}$, c) $Z^\alpha{}_{\alpha}$, d) $-Z^\alpha{}_{\alpha}$, e) $-X_i \equiv -q_i \ti q_i$, f) $Z^\beta{}_{\alpha} \equiv \ti Q_\alpha (\prod_{k=i}^j q_k) Q^\beta$, g) $-Z^\alpha{}_{\beta} \equiv  (-)^{i-j} Q^\alpha (\prod_{k=i}^j \ti q_k) \ti Q_\beta$.}}
\label{AppHBopSetups}
\end{figure}

\begin{figure}[th]
\centering
\includegraphics[scale=0.72]{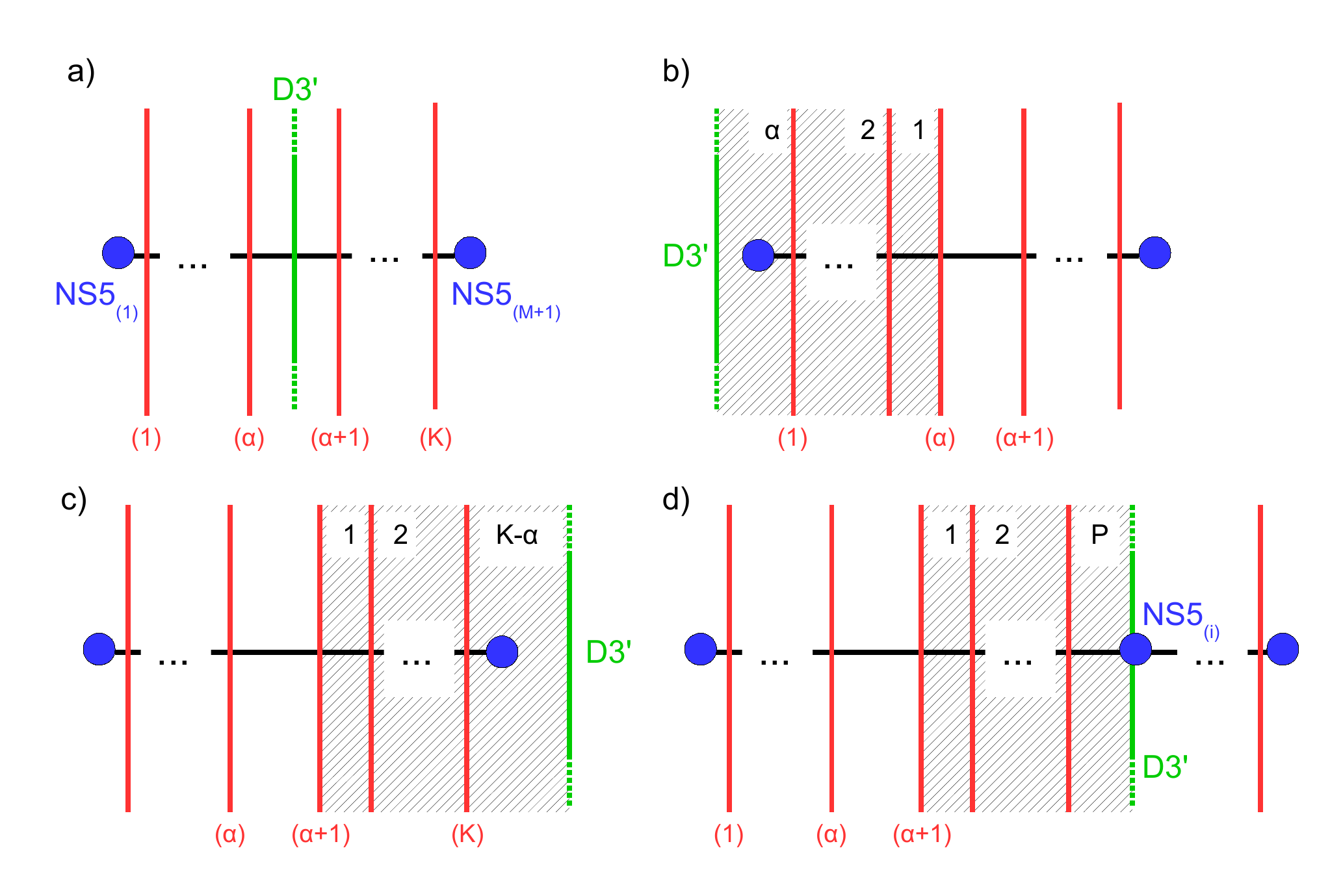}
\vspace{-0.5cm}
\caption{\footnotesize{Setups with D3' branes inserting HB operators. The insertion of the figures a,b,c and d are related by HW D3' move and insert the same operator $Y_{\alpha} \equiv \xi_{1} -\sum_{\gamma=1}^\alpha Z^\gamma{}_{\gamma} = \xi_{M+1} + \sum_{\gamma=\alpha+1}^K Z^\gamma{}_{\gamma} = \xi_i + X_i + \sum_{\gamma=\alpha+1}^{\alpha + P} Z^\gamma{}_{\gamma}$ (F-term constraints).}}
\label{AppHBopD3p}
\end{figure}

In terms of gauge invariant meson operators we have meson insertions according to the prescriptions of Figure \ref{AppHBopSetups}.
In addition there are setups involving D3' branes which also insert meson operators. Those insertions are affected by the FI term deformations $\eta_i = \xi_{i}-\xi_{i+1}$, with $\xi_i$ corresponding to the position of NS5$_{(i)}$. They are shown in Figure \ref{AppHBopD3p}. The mesons $Y_\alpha$ inserted by the setups with a D3' standing between D5$_{(\alpha)}$ and D5$_{(\alpha+1)}$ are related to the mesons $Z^\alpha{}_{\alpha}$ inserted by the setups of Figure \ref{AppHBopSetups}-a,b,c,d,e by the relation
\be\ba
Z^\alpha{}_{\alpha} &= Y_{\alpha-1} - Y_{\alpha} \,, \quad \alpha = 1, \cdots, K \,, \cr
X_i &= Y_{\delta(i)} - \xi_i \,, \quad  \text{with} \ \ \text{NS5}_{(i)} \in [\text{D5}_{(\delta(i))}, \text{D5}_{(\delta(i)+1)}] \,,
\ea\ee
where $Y_{-1} =\xi_1$ and $Y_{K}= \xi_{M+1}$, and for the second relation $\delta(i)$ is defined by the fact that NS5$_{(i)}$ sits between D5$_{(\delta(i))}$ and D5$_{(\delta(i)+1)}$ in the brane setup.
The meson operators inserted by the setups of Figure \ref{AppHBopSetups} and \ref{AppHBopD3p} form a basis of the HB chiral ring.

\newpage

\subsection{Coulomb branch operators}

\begin{figure}[th]
\centering
\includegraphics[scale=0.7]{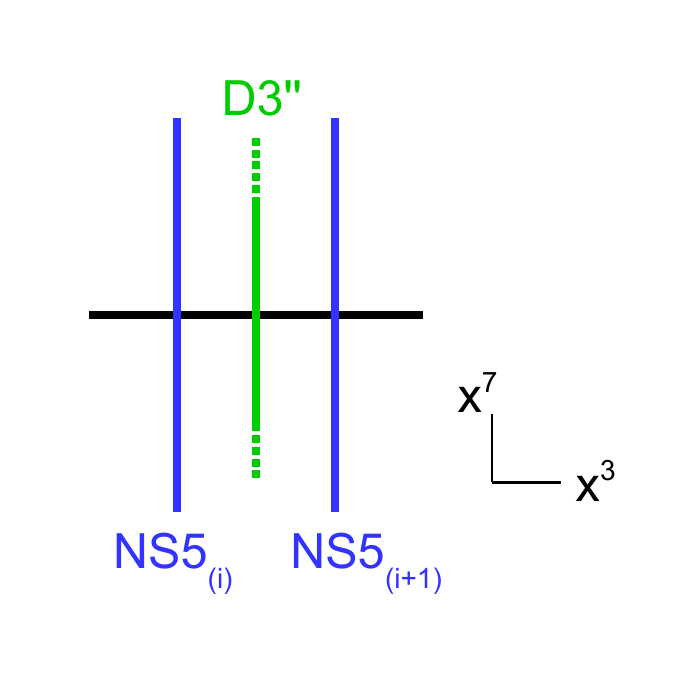}
\vspace{-0.5cm}
\caption{\footnotesize{Setup with a D3'' brane crossing the D3$_{(i)}$ segment, inserting $\varphi_i$.}}
\label{AppCBopD3pp}
\end{figure}

\begin{figure}[th]
\centering
\includegraphics[scale=0.7]{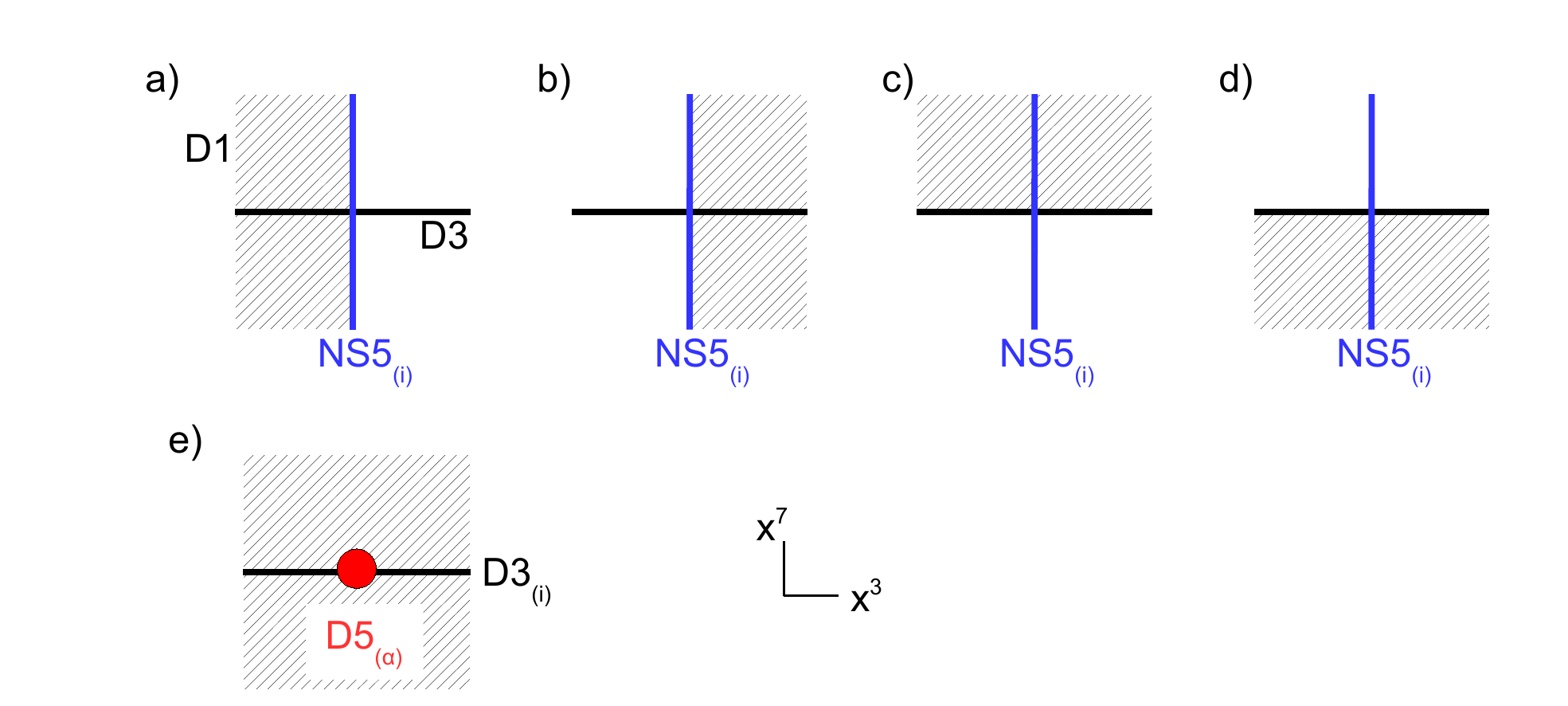}
\vspace{-0.5cm}
\caption{\footnotesize{Setups with D1 strings inserting scalar CB operators. The insertions are: a) $\varphi_{i} - \varphi_{i-1}$, b) $\varphi_{i-1} - \varphi_{i}$, c) $\varphi_{i} - \varphi_{i-1}$, d) $\varphi_{i-1} - \varphi_{i}$, e) $ m_\alpha - \varphi_i$.}}
\label{AppCBopD1}
\end{figure}

To a D3$_{(i)}$ segment stretched between  NS5$_{(i)}$ and NS5$_{(i+1)}$, is associated a U(1)$_{(i)}$ gauge symmetry and the vector multiplet complex scalar $\varphi_i$, whose insertion is realized by the brane setup of Figure \ref{AppCBopD3pp}, with a D3'' brane crossing the D3$_{(i)}$ segment.
In addition the setups with D1 and NS5$_{(i)}$ branes of Figure \ref{AppCBopD1}-a,b,c,d correspond to the insertions of the combinations of complex scalars $\pm (\varphi_{i-1} - \varphi_{i})$ and the setup with D1 and D5$_{(\alpha)}$ branes of Figure \ref{AppCBopD1}-e corresponds to the insertion of $(m_\alpha - \varphi_i )$, with $m_\alpha$ the complex mass of the hyper-multiplet sourced by the D5.

\begin{figure}[th]
\centering
\includegraphics[scale=0.7]{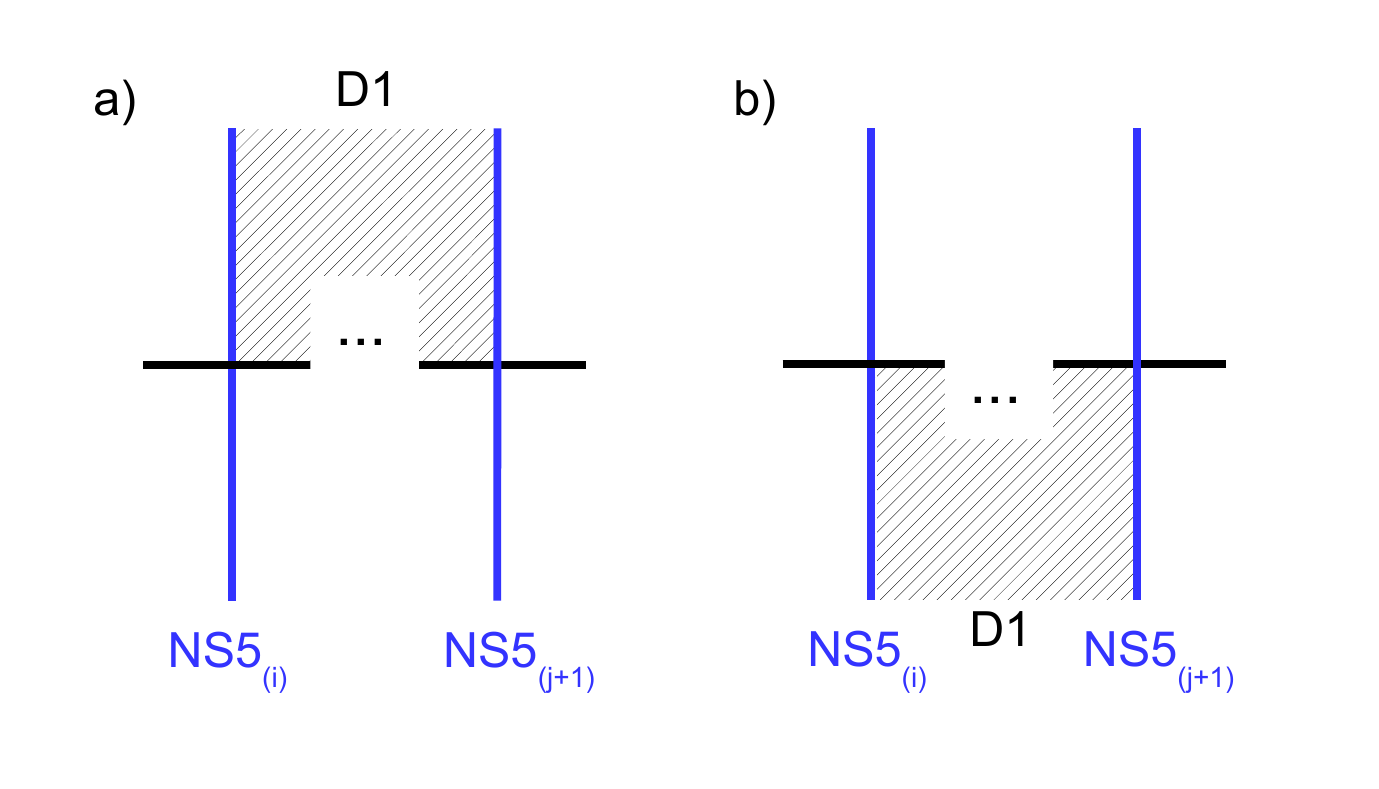}
\vspace{-0.5cm}
\caption{\footnotesize{Setups with D1 strings inserting monopole operators. The insertions are: a) $V^+_{[i:j]}$, b) $-V^-_{[i:j]}$. The branes which might stand in the place of the dots ``..." insert additional CB operators.}}
\label{AppCBopMonop}
\end{figure}

The insertion of monopole operators are related to the brane setups of Figure \ref{AppCBopMonop}. To a pair of NS5 branes, NS5$_{(i)}$-NS5$_{(j+1)}$ with $i \le j$, with a D1 string stretched between them and ending on the D3 segments,  is associated the monopole operators $V^{+}_{[i:j]}$ or $-V^{-}_{[i:j]}$, depending whether the D1 end on the D3 from above or from below, which have magnetic charges $+1$ or $-1$ under each $U(1)_{(k)}$ with $i \le k \le j$.

The CB operators inserted by the setups of Figure \ref{AppCBopD3pp} and \ref{AppCBopMonop} form a basis of the Coulomb branch chiral ring.


\newpage

\subsection{Mirror Map}

From the action of S-duality on the brane setups, one obtains the mirror map between HB and CB operators of a pair of dual theories A and B presented in Table \ref{tab:AppMirrorMap} and the  map of FI and mass parameters.
\begin{table}[h]
\begin{center}
\setlength\extrarowheight{3pt}
\begin{tabular}{|c|c|}
\hline
  Theory A: HB  &  Theory B : CB  \\    
\hline
 $Y_i$   & $\varphi_{i}$  \\ [2pt]
\hline
 $Z^i{}_i$   & $\varphi_{i-1}-\varphi_{i}$  \\ [2pt]
\hline
 $X_i$   & $ \varphi_{f(i)}- \ti m_i$  \\ [2pt]
\hline
 $Z^{j+1}{}_{i}$ \, $(i\le j)$ & $V^+_{[i:j]}$  \\ [2pt]
\hline
  $Z^{i}{}_{j+1}$ \, $(i\le j)$ & $V^-_{[i:j]}$   \\  [2pt]
\hline
\hline
  Theory A: FI param. &    Theory B: mass param.  \\ 
\hline
$\xi_i$  &  $\ti m_i$   \\ 
\hline
\end{tabular}
\caption{\footnotesize Mirror map of Higgs branch (HB) operators of $T_A$ and Coulomb branch (CB) operators of $T_B$, and mirror map of FI and mass parameters. $\varphi_{f(i)}$ denotes the scalar of the $U(1)_{f(i)}$ node under with the hyper-multiplet of mass $\ti m_i$ is charged.}
\label{tab:AppMirrorMap}
\end{center}
\end{table}


\bibliography{Benbib}
\bibliographystyle{JHEP}

\end{document}